\documentclass[english,prd,superscriptaddress,nofootinbib,twocolumn,floatfix,showpacs]{revtex4}
\usepackage{enumerate}
\usepackage{graphicx}
\usepackage{bm}
\usepackage{amsmath,amssymb}

\begin{document}
\font\sevenrm=cmr7\font\fiverm=cmr7
\def\dd{{\rm d}} \def\ds{\dd s} \def\e{{\rm e}} \def\etal{{\em et al}.}
\def\al{\alpha}\def\be{\beta}\def\ga{\gamma}\def\de{\delta}\def\ep{\epsilon}
\def\et{\eta}\def\th{\theta}\def\ph{\phi}\def\rh{\rho}\def\si{\sigma}
\def\ns#1{_{\hbox{\sevenrm #1}}}\def\Z#1{_{\lower2pt\hbox{$\scriptstyle#1$}}}
\def\Ns#1{\Z{\hbox{\sevenrm #1}}} \def\Ls#1{\Z{\hbox{\fiverm #1}}}
\def\goesas{\mathop{\sim}\limits}\def\w#1{\,\hbox{#1}}\def\chis{\chi_{i,s}^2}
\def\lsim{\mathop{\hbox{${\lower3.8pt\hbox{$<$}}\atop{\raise0.2pt\hbox{$
\sim$}}$}}} \def\ld{\ell_d}\def\bd{b_d}\def\sHi{\si\Z{H_i}}
\def\gsim{\mathop{\hbox{${\lower3.8pt\hbox{$>$}}\atop{\raise0.2pt\hbox{$
\sim$}}$}}} \def\sith{\si\Z\th} \def\Hth{H_{\al}}\def\bs{{\bar\si}}
\def\h{\,h^{-1}}\def\hm{\h\hbox{Mpc}}\def\deg{^\circ}\def\zhom{z\ns{hom}}
\def\OmMn{\Omega_{M0}}\def\OmRn{\Omega_{R0}}\def\OmLn{\Omega_{\Lambda0}}
\def\LCDM{$\Lambda$CDM}\def\Hm{H\Z{0}}\def\bH{\bar H}\def\Wi{W_{i\,\al}}
\def\kms{\w{km}\w{s}^{-1}}\def\kmsMpc{\kms\w{Mpc}^{-1}}\def\Wj{W_{j\,\al}}
\def\ave#1{\left\langle#1\right\rangle_s} \def\Be{{\boldsymbol\beta}}
\def\RV{R\Z V}\def\mr{{\bar r}}\def\bsa{\bs_\al} \def\dsp{\displaystyle}
\def\SumH#1{\sum_i\frac{\dsp#1}{\dsp\sHi^2}} \def\pt{\partial}
\def\PRL#1{Phys.\ Rev.\ Lett.\ {\bf#1}}\def\PR#1{Phys.\ Rev.\ {\bf#1}}
\def\ApJ#1{Astrophys.\ J.\ {\bf#1}}\def\AaA#1{Astron.\ Astrophys.\ {\bf#1}}
\def\MNRAS#1{Mon.\ Not.\ R.\ Astr.\ Soc.\ {\bf#1}}
\def\CQG#1{Class.\ Quantum Grav.\ {\bf#1}}
\def\GRG#1{Gen.\ Relativ.\ Grav.\ {\bf#1}}
\def\ApJs#1{Astrophys.\ J.\ Suppl.\ {\bf#1}}
\def\JCAP#1{J.\ Cosmol.\ Astropart.\ Phys.\ #1}
\def\beq{\begin{equation}} \def\eeq{\end{equation}}
\def\bea{\begin{eqnarray}} \def\eea{\end{eqnarray}}
\def\CS{{\textsc COMPOSITE}\ }
\def\latitudes{In all figures, the galactic longitudes $\ell=0\deg,180\deg,
360\deg$ are on the right edge, centre and left edge respectively.}

\def\Raa#1{Alnes and Amarzguioui \cite{#1}}
\def\Rablm#1{Abramowicz \etal\ \cite{#1}}
\def\Rakeke#1{Atrio-Barandela \etal\ \cite{#1}}
\def\Rbcmj#1{Bilicki \etal\ \cite{#1}}
\def\Rbennett#1{Bennett \etal\ \cite{#1}}
\def\Rbh#1{Bunn and Hogg \cite{#1}}
\def\Rdn#1{Davis and Nusser \cite{#1}}
\def\Rel#1{Erdo\u{g}du \etal\ \cite{#1}}
\def\Rfinkelman#1{Finkelman \etal\ \cite{#1}}
\def\Rfixsen#1{Fixsen \etal\ \cite{#1}}
\def\Rfreeman#1{Freeman \etal\ \cite{#1}}
\def\Rfwh#1{Feldman, Watkins and Hudson \cite{#1}}
\def\Rgls#1{Gordon, Land and Slosar \cite{#1}}
\def\Rhicken#1{Hicken \etal\ \cite{#1}}
\def\Rjrk#1{Jha, Riess and Kirshner \cite{#1}}
\def\Rkash#1{Kashlinsky \etal\ \cite{#1}}
\def\Rkeisler#1{Keisler \cite{#1}}
\def\Rkessler#1{Kessler \etal\ \cite{#1}}
\def\Rls#1{Li and Schwarz \cite{#1}}
\def\Rltmc#1{Lavaux \etal\ \cite{#1}}
\def\Rmd#1{McClure and Dyer \cite{#1}}
\def\Rms#1{Ma and Scott \cite{#1}}
\def\Rosborne#1{Osborne \etal\ \cite{#1}}
\def\Rrs#1{Rubart and Schwarz \cite{#1}}
\def\Rsdb#1{Scrimgeour \etal\ \cite{#1}}
\def\Rtully#1{Tully \etal\ \cite{#1}}
\def\Rturnbull#1{Turnbull \etal\ \cite{#1}}
\def\Rwfh#1{Watkins, Feldman and Hudson \cite{#1}}
\def\Rzrkd#1{Zehavi \etal\ \cite{#1}}
\def\tableline{\noalign{\hrule}}
\def\bigtable#1#2{\begingroup\squeezetable
\begin{table#1}\begin{ruledtabular}\begin{tabular}{#2}}
\def\enddata{\end{tabular}\end{ruledtabular}}
\title{Hubble flow variance and the cosmic rest frame}

\author{David~L.~Wiltshire}
\author{Peter~R.~Smale}
\author{Teppo~Mattsson}
\affiliation{Department of Physics \& Astronomy,
University of Canterbury, Private Bag 4800, Christchurch 8140, New Zealand}
\author{Richard~Watkins}
\affiliation{Department of Physics, Willamette University, Salem,
OR 97301, USA}

\begin{abstract}
We characterize the radial and angular variance of the Hubble flow in the
\CS sample of 4534 galaxies, on scales in which much of the flow is in
the nonlinear regime. With no cosmological assumptions other than the existence
of a suitably averaged linear Hubble law, we find with decisive Bayesian
evidence ($\ln B\gg5$) that the Hubble constant, when averaged in independent
spherical shells, is closer to its asymptotic value when referred to the
rest frame of the Local Group (LG), rather than the standard rest frame of the
Cosmic Microwave Background (CMB). An exception occurs for radial shells in the
range $40\h$ -- $60\hm$. Angular averages reveal a dipole structure in the
Hubble flow, whose amplitude changes markedly over the range $32\h$ -- $62\hm$.
Whereas the LG frame dipole is initially constant and then decreases
significantly, the CMB frame dipole initially decreases but then increases.
The map of angular Hubble flow variation in the LG rest frame is found to
coincide with that of the residual CMB temperature dipole, with correlation
coefficient $-0.92$. These results are difficult to reconcile with the
standard kinematic interpretation of the motion of the Local Group in response
to the clustering dipole, but are consistent with a foreground nonkinematic
anisotropy in the distance-redshift relation of 0.5\% on scales up to
$65\,\hm$. Effectively, the differential expansion of space produced by nearby
nonlinear structures of local voids and denser walls and filaments cannot be
reduced to a local boost. This hypothesis suggests a reinterpretation of bulk
flows, which may potentially impact on calibration of supernova distances,
anomalies associated with large angles in the CMB anisotropy spectrum, and
the dark flow inferred from the kinematic Sunyaev-Zel'dovich effect. It
is consistent with recent studies that find evidence for a
nonkinematic dipole in the distribution of distant radio sources.
\end{abstract}
\maketitle

\section{Introduction}
It is usually assumed that the cosmic microwave background (CMB) dipole
\cite{kogut93,fixsen96} is generated entirely by our own local peculiar
velocity. A local boost by the opposite velocity then defines the cosmic rest
frame in which we can be considered to be comoving observers in the background
geometry of homogeneous isotropic Friedmann-Lema\^{\i}tre-Robertson-Walker
(FLRW) model. Indeed measurements of cosmological redshifts are routinely
transformed to the CMB frame. According to the assumptions implicit in
such a transformation, the cosmic rest frame so defined should also be the
frame in which the Hubble flow is most uniform, with minimal statistical
variations as compared to other choices of the standard of rest.

Our understanding of the Hubble flow is, however, greatly complicated by the
fact that the universe is not completely homogeneous. Rather it appears to only
be homogeneous in some statistical sense, when one averages on scales $\gsim
100\hm$, the transition scale still being a matter of debate \cite{hogg05,%
sl09,sdb12,chr}. At scales below or comparable to the scale of statistical
homogeneity a complex pattern of variance in the Hubble flow is observed.
In the standard manner of thinking about the problem, Hubble flow
variance is interpreted as a field of peculiar velocities of galaxies with
respect to the expansion law of a FLRW model, which is linear on scales
up to $z\goesas0.1$, well above the scale of statistical homogeneity. The
CMB rest frame sets the standard of rest for a comoving observer at our
location in defining the leading order linear Hubble law. A great deal of
observational effort has gone into understanding the nearby peculiar motions
so derived; see, e.g., \cite{hudson04,ke06,springob07,tsk08,kkmt,ltmc,ap10,%
cmss,dks,wwwf,ic11}.

Some studies of peculiar velocities \cite{wfh09,fwh10,kash08,kash10}
have found results which indicate persistent bulk flows extending to
very large scales, and which are potentially in conflict with the
expectations of the perturbed FLRW model that underlies the standard
Lambda Cold Dark Matter (\LCDM) cosmology. Using the large \CS data set
of mainly non-SneIa galaxy distances \Rwfh{wfh09} report a large
bulk flow of $407\pm81\kms$ toward $\ell=287\deg\pm9\deg$, $b=8\deg\pm6\deg$,
with 90\% of the sample within a $107\hm$ sphere.

Different data sets and methods of analysis produce different, sometimes
contradictory, results. For example, \Rdn{dn11} and \Rms{ms13} analyse
samples which include large subsamples of the \CS sample, such as SF++
\cite{springob07}. Applying different methods they find bulk flows consistent
in direction with \Rwfh{wfh09,fwh10} but with different amplitudes, which are
consistent with \LCDM\ predictions. \Rdn{dn11} find a bulk flow $333\pm38
\kms$ towards $(\ell,b)=(276\deg,14\deg)\pm3\deg$ within a $40\hm$ sphere,
and $257\pm44\kms$ towards $(\ell,b)= (279\deg,10\deg)\pm3\deg$ within a
$100\hm$ sphere; \Rms{ms13} find amplitudes in the range $220$ -- $370\kms$ in
four different samples in an average direction $(\ell,b)=(280\deg\pm8\deg,5
\deg\pm6\deg)$ within a $80\hm$ sphere. \Rturnbull{turnbull11} have made
a study using 245 type Ia supernova (SneIa) distances on somewhat larger
scales $r\lsim190\hm$. They find a bulk flow $249\pm76\kms$ in the direction
$\ell=319\deg\pm18\deg$, $b=7\deg\pm14\deg$, which is consistent with the
predictions of the \LCDM\ model, but also marginally consistent with the
larger bulk flow of \Rwfh{wfh09,fwh10}. The results of \Rturnbull{turnbull11}
appear to be inconsistent, however, with the larger amplitude bulk flow of
$600$ -- $1,000\kms$ found by \Rkash{kash08,kash10} using the kinematic
Sunyaev-Zel'dovich effect, principally on larger scales $120\lsim r\lsim600
\hm$.

In the above papers and elsewhere in the literature, with very few exceptions
\cite{md07,ls08}, variance in the Hubble flow is generally attributed to
peculiar velocities whose radial components are defined as deviations from a
linear Hubble law
\beq
v\ns{pec}=cz-\Hm r\,\label{vpec}
\eeq
where $z$ is the redshift, $c$ the speed of light and $r$ an appropriate
distance measure. Such a definition implicitly makes a strong assumption about
spacetime geometry, namely on the scales of interest spatial curvature
can be neglected and the redshift associated with the Hubble expansion
can be treated in the manner of a recession velocity as in
special relativity.

{}From the point of view of general relativity, without any {\em a priori}
assumptions about the background geometry, such an assumption must
be questioned. In general relativity in an arbitrary spacetime background
the only velocities that are uniquely related to observables are those
corresponding to local boosts at a point. Given that the dust approximation
is not rigorously defined for the complex cosmic web of voids, walls and
filaments that constitute the present day universe on $\lsim100\hm$ scales
\cite{dust}, then the extension of the concept of a velocity in (\ref{vpec})
over the vast distances over which space is expanding is merely an ansatz,
whose validity remains to be justified. In any general relativistic framework
there must be some local peculiar velocities, which arise from the local
dynamics of galaxies within bound clusters. However, there is no {\em a
priori} reason for assuming that all redshifts on scales $\lsim100\hm$ can
be treated in terms of a simple Doppler shift in Euclidean space, which
is in practice the method of analysis adopted by most observationalists.

{}From the point of view of general relativity, variance in the Hubble flow
in its nonlinear regime is more naturally viewed as the differential expansion
of regions of different local densities, which have experienced different local
expansion histories over the billions of years that have elapsed since the
density field was close to uniform. While particular geometrical assumptions
would lead to the standard FLRW geometry with Newtonian perturbations, the
lack of convergence of bulk flows to the CMB dipole and the puzzle of several
possible anomalies associated both with bulk flows and the large angle
multipoles of the CMB anisotropy spectrum \cite{toh03,lm05,ebg07,heb09,kash08,%
kn10,Piso}, suggest that one should reconsider the problem from first
principles.

In this paper we will therefore reanalyse the largest available data set, the
\CS sample of \Rwfh{wfh09,fwh10}, from a fresh perspective. While the
particular analysis we adopt is one which is naturally suggested by the
cosmological model of Refs.~\cite{clocks,sol,equiv,obs}, our actual analysis is
independent of any cosmological model assumptions other than the most
elementary one that a suitably defined average linear Hubble law exists.

The plan of the paper is as follows. In Sec.~\ref{rad} we first consider
the spherical (monopole) variation of the Hubble law in independent radial
shells, obtain statistical bounds on the differences between the CMB and Local
Group (LG) frames, and provide an explanation of the systematic differences
between the two frames. In Sec.~\ref{ang} we consider angular averages, using a
Gaussian window function average to estimate the ratios of large angle
multipoles, and by fitting a dipole Hubble law in independent radial shells
for those shells with a dominant dipole. We determine the statistical
significance of the dipole and its amplitude
in relation to the monopole variations. We also identify those structures in
that might be responsible for the density gradient which induces both the
monopole and dipole variations. In Sec.~\ref{corr} we check that the Hubble
variance dipole correlates with the component of the CMB dipole usually
attributed to the motion of the LG, with very strong significance. In
Sec.~\ref{origin}, given that our combined results do not support a purely
kinematic response of the LG to the clustering dipole, we suggest a new
alternative mechanism for the generation of this contribution to the CMB
dipole. Finally in Sec.~\ref{dis} we discuss some potential implications of
our results.

\section{Spherical averages in radial shells}\label{rad}
We adopt the point of view that on scales of order $10\h$ -- $30\hm$ the
regional expansion history, and the regional average Hubble law should be
determined primarily by the relevant regional average density. From the point
of view of any observer, underdense voids will appear to be expanding faster
than denser wall regions, on account of the wall regions having decelerated
more. This is true independently of whether or not there is a homogeneous dark
energy which acts to accelerate the expansion by the same amount in all
regions.

The largest typical voids are shown by surveys to have a diameter
$\goesas30\hm$ \cite{hv02,hv04,pan11}. Our galaxy is located
in a filamentary sheet on the edge of a Local Void of at least this diameter,
formed from a complex of three smaller voids \cite{tsk08}.

Although the expansion rate just to the other side of a local void (wall) will
appear faster (slower) than average, whole sky spherical averages that include
many structures in different directions can be expected to have a reduced
variance as compared to measurements in particular directions.
Furthermore, once one also averages on radial scales a few times larger
than the largest typical structures then the variance of the Hubble parameter
averaged in spherical shells will reduce to a level consistent with
individual measurement uncertainties. (See Fig.~\ref{Hshells}.)
This provides an operational definition of the {\em scale of statistical
homogeneity} independent of any detailed cosmological model assumptions.

\begin{figure}[htb]
\vbox{\vskip\baselineskip\centerline{\scalebox{0.33
}{\includegraphics{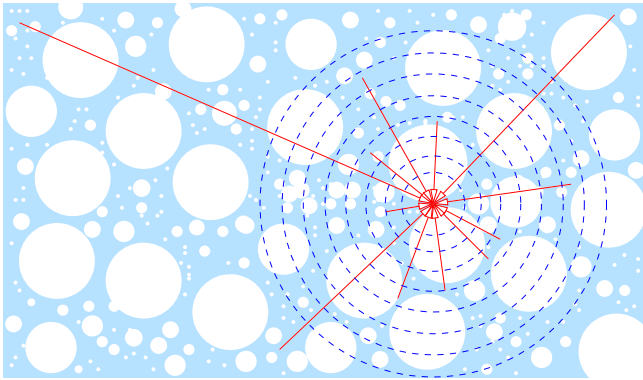}}}
\vskip\baselineskip
\caption{%
Schematic diagram of spherical averaging. The universe is described as ensemble
of filaments, walls and voids: expanding regions of different density which
have decelerated by different amounts and therefore experience different local
expansion rates at the present epoch. Observations show that the largest
typical nonlinear structures are voids of $\goesas30\hm$ \cite{hv02,hv04,%
pan11}, which occupy 40\% of the volume of the present epoch universe. If one
averages $cz/r$ in spherical shells (dotted lines) about a point then once the
shells are a few times larger than the typical nonlinear structures, an average
Hubble law with small statistical scatter is obtained (e.g., for longest null
geodesics represented by arrowed lines), whereas there are considerable
deviations for shells on scales comparable to the typical nonlinear structures.
This approach is model independent since it makes no assumptions about the
geometry of the universe on the scales in question.
\label{Hshells}}}
\end{figure}

Determining the scale of statistical homogeneity observationally is an
interesting challenge and a matter of debate \cite{hogg05,sl09}, one of
the issues being an appropriate definition of statistical homogeneity since
there will always be some cosmic variance on the largest of scales. Any
reasonable definition should encompass a notion of a transition between a
nonlinear regime and a linear regime of inhomogeneous perturbations relative
to an average almost homogeneous evolution, whether this is described by
a FLRW model or not \cite{dust}. Since the largest {\em typical} nonlinear
structures are voids of $30\hm$ diameter, which occupy of order 40\% of the
volume of the present universe \cite{hv02,hv04}, the scale of statistical
homogeneity should be at least a few times larger. On the other hand, since the
Baryon Acoustic Oscillation (BAO) feature is observed in galaxy clustering
statistics in accord with the expectations of linear perturbation theory on a
FLRW background, the scale of statistical homogeneity should necessarily be
of the same order or smaller, i.e., $\lsim110\hm$. This accords with
a recent measurement of the scale of statistical homogeneity by \Rsdb{sdb12}
in the WiggleZ survey.

A study of the spherically averaged Hubble flow, as a function of radial
distance was undertaken by \Rls{ls08} (henceforth LS08), using a subset of 54
distances from the Hubble Space Telescope (HST) Key project data
\cite{freedman01}. Fig.~2 of LS08 shows the radial variation $\de H(r)=(H(r)-
\Hm)/\Hm$ that results from such an analysis, with data restricted to the range
$r\ns{min}<r<r\ns{max}$, where $r\ns{min}=22.5\hm$ and $r\ns{max}=130\hm$.
Furthermore, their fit is computed for redshifts referred to a single linear
Hubble law $cz=H(r)r$ within a sphere of radius $r$, as $r$ is varied, and with
redshifts referred to the CMB frame.

Our first aim here is to perform a similar analysis to LS08 using the \CS
sample of cluster, group and galaxy distances compiled by \Rwfh{wfh09}
(henceforth WFH09) and slightly updated by \Rfwh{fwh10} (henceforth FWH10).
For each sample object, redshift, galactic latitude and longitude, distance,
and distance uncertainty are given. Distance uncertainties are about $15\%$ for
most individual galaxies. We include all 4,534 data points outside the
local group $r>2\hm$, which extends the range of distances considered in
LS08 to both smaller and larger values.

The data in the \CS sample combines nine independent, full-sky samples,
nearly all major peculiar velocity surveys published to date. Although each
survey uses a different distance measurement methodology, all of the surveys
were shown to be statistically consistent with each other \cite{wfh09}. The
survey of Ref.\ \cite{lp94} was not included, as it gave inconsistent results
\cite{wfh09}. WFH09 provide a detailed discussion of the issues involved
in combining subsamples with different characteristic depths and sky
coverages. For our analysis it is important to note that outside the Zone of
Avoidance (ZoA) the \CS sample has good all sky coverage, as is seen
from Fig.~1 of FWH10, and is further discussed in Sec.~\ref{meth} and
Appendix \ref{mc} below.

\subsection{Methodology}\label{meth}
Our analysis will feature two key differences to that of LS08. Firstly, rather
than simply performing the analysis in the CMB frame, we also perform the
analysis in the local group (LG) and local sheet (LS) frames. A comparison of
these frames is motivated by the Cosmological Equivalence Principle
\cite{equiv}: the LG frame corresponds to our ``finite infinity region''
\cite{clocks}, and should be close to the frame in which the variance in the
Hubble flow is minimized in the approach to cosmological averages advocated
in Refs.~\cite{clocks,sol,equiv}. The LG has a small peculiar velocity of
$66\pm24\kms$ relative to the LS \cite{tsk08} within which a local Hubble
flow is first defined.

Secondly, on account of the small number of data points, \Rls{ls08} included
all data within a radius $r$, as $r$ was varied in steps between $r\ns{min}$
and $r\ns{max}$. This has the effect that each binned data point shown in
their Fig.~2 is correlated with the previous data point. With 4,534
available data points in the \CS sample such correlations can be avoided
by the following technique: we will minimize the sum $\chi_s^2=\sum_i\left[\si
_i^{-1}(r_i-cz_i/H)\right]^2$ with respect to $H$, as a means of fitting a
Hubble law by a standard linear regression \cite{numrec},
in successive independent radial shells $r_s<r\le r_{s+1}$. We
consider the linear Hubble law with $r$ as a function of $z$ since all
uncertainties have been included as distance uncertainties\footnote{While
the measurement uncertainties in redshifts are negligible, using the standard
peculiar velocity framework a uniform velocity noise uncertainty was added in
quadrature to $\Hm\si_i$ in FWH10 in defining the maximum likelihood weights.
In the peculiar velocity framework galaxy motions are modeled using linear
theory. The velocity noise term then accounts for the fact that individual
peculiar velocities deviate from the local value of the linear peculiar
velocity field due to small-scale nonlinear motions. Here we are not using
linear theory to model deviations from a single global linear Hubble law, so
the addition of velocity noise to our analysis is unnecessary. In our
framework we would still have to take into account the noise due to peculiar
velocities of galaxies within gravitationally bound clusters. However, in
the \CS data set this has already been accounted for in
gravitationally bound systems by assigning distances and associated
uncertainties to clusters, rather than to the individual galaxies within
the clusters.} in the \CS sample. The value of the Hubble constant $H_s$
computed for the $s$th shell is then
\beq
H_s=\left(\sum_{i=1}^{N_s}{(cz_i)^2\over\si_i^2}\right)\left(\sum_{i=1}^{N_s}
{cz_i r_i\over\si_i^2}\right)^{-1}\,,\label{Hs}
\eeq
where $\si_i$ denote individual distance uncertainties in $\hm$.

The total uncertainty for $H_s$ in each shell\footnote{We use an overbar for
uncertainties in the Hubble constant obtained by either radial or angular
averages, to distinguish them from the distance uncertainties in individual
data points.}, $\bs_s$, is determined by
adding the following uncertainties in quadrature: (i) the uncertainty
determined by standard error propagation for the linear fit (\ref{Hs}) in the
$s$th shell
\beq
\bs\Z{1\,s}=\left(\sum_{i=1}^{N_s}{(cz_i)^2\over\si_i^2}\right)
^{3/2}\left(\sum_{i=1}^{N_s}{cz_i r_i\over\si_i^2}\right)^{-2}\,,\label{sigH}
\eeq
and (ii) a zero point uncertainty
\beq
\bs\Z{0\,s}=H_s{\si\Ns{0}\over\mr_s}\,,\label{sig0}
\eeq
where $\mr_s=\left(\sum_{i=1}^{N_s}{r_i\over\si_i^{2}}\right)\left(\sum_{i=1}^
{N_s}{1\over\si_i^{2}}\right)^{-1}$ is the weighted mean distance of the $N_s$
points in the $s$th shell and $\si\Ns{0}=0.201\hm$ is the distance uncertainty
arising from the $20\kms$ uncertainty in the heliocentric peculiar velocity of
both the Local Group and Local Sheet as given in Ref.\ \cite{tsk08} added in
quadrature to the 0.4\% uncertainty in the magnitude of the CMB
dipole\footnote{Very slightly different temperature dipoles are given by
\Rfixsen{fixsen96} and \Rbennett{bennett03}. Since much of the \CS data
set was determined before the \Rbennett{bennett03} result,
we assume that it has been normalized relative to the heliocentric frame
using the \Rfixsen{fixsen96} value of $v\Ls{CMB}=371\kms$ in a direction
$\ell=264.14\deg$, $b=48.26\deg$, which is the standard used in the NASA/IPAC
Extragalactic Database. Our heliocentric velocities of the LG and LS are
taken from Ref.\ \cite{tsk08} as $v\Ls{LG}=318.6\kms$ towards $\ell=106\deg$,
$b=-6\deg$, and $v\Ls{LS}=318.2\kms$ towards $\ell=95\deg$, $b=-1\deg$
respectively.} \cite{fixsen96}.

The uncertainty (\ref{sig0}) is included since the Hubble law is necessarily
determined by a linear fit through the origin for each shell. The local
velocity uncertainty when divided by $\Hm$ provides an additional distance
uncertainty in the mean distance of each shell relative to the origin, and the
related uncertainty (\ref{sig0}) in the mean slope $H_s$. This uncertainty
is significant for shells close to the origin, but much smaller
for shells at large radii for which the mean distance has a long lever arm.

In Eqs.~(\ref{sigH}), (\ref{sig0}) $H_0=100\,h\kmsMpc$ represents the
normalization used to convert velocity uncertainties to distance uncertainties
in the \CS data set. One other important issue is the asymptotic value
of the Hubble constant in each frame to which the variance in the Hubble
flow should be normalized, as this global value should have its own
uncertainty. To this end we have divided the data in $12.5\hm$ wide
shells out to those radial distances of order $150\hm$, for two different
choices of shells differing by the initial inner shell boundary, as shown
in Table~\ref{shell}. The penultimate shell, $10$ or $10'$, has been made
wider so that it contains a similar number of points to most inner shells.

\bigtable{*}{lrrrrrrrrrrr}
Shell $s$&1&2&3&4&5&6&7&8&9&10&11\\
$N_s$&92&505&514&731&819&562&414&304&222&280&91\\
$r_s$ ($h^{-1}$Mpc)&2.00&12.50&25.00&37.50&50.00&62.50&75.00&87.50&100.00&112.50&156.25\\
$\mr_s$ ($h^{-1}$Mpc)&5.43&16.33&30.18&44.48&55.12&69.24&81.06&93.75&105.04&126.27&182.59\\
$(H_s)\Ns{CMB}$&173.9&111.1&110.4&104.1&102.7&103.8&102.1&102.8&104.4&102.1&100.1\\
$(\bs_s)\Ns{CMB}$&6.8&1.5&1.1&0.7&0.7&0.8&0.9&0.9&1.0&0.8&1.7\\
$(Q_s)\Ns{CMB}$&0.000&0.000&0.000&0.037&0.985&0.997&1.000&1.000&1.000&0.979&0.999\\
$(\chi^2/\nu)\Ns{CMB}$&24.639&5.802&1.741&1.096&0.896&0.841&0.593&0.604&0.630&0.835&0.581\\
$(H_s)\Ns{LG}$&117.9&103.1&106.5&105.5&104.8&102.1&102.8&103.2&103.7&102.4&101.0\\
$(\bs_s)\Ns{LG}$&4.6&1.4&1.0&0.7&0.7&0.7&0.9&0.9&1.0&0.8&1.7\\
$(Q_s)\Ns{LG}$&0.000&0.000&0.000&0.000&0.998&0.940&1.000&1.000&1.000&0.993&0.999\\
$(\chi^2/\nu)\Ns{LG}$&23.656&7.767&2.185&1.419&0.864&0.909&0.594&0.542&0.622&0.803&0.590\\
$\ln B$ ($r\ge r_s$)&58.62&16.95&8.43&1.71&1.98&2.71&1.64&1.78&1.67&0.49&\\
\noalign{\smallskip}\tableline\noalign{\smallskip}
Shell $s$&1'&2'&3'&4'&5'&6'&7'&8'&9'&10'&11\\
$N_s$&321&513&553&893&681&485&343&273&164&206&91\\
$r_s$ ($h^{-1}$Mpc)&6.25&18.75&31.25&43.75&56.25&68.75&81.25&93.75&106.25&118.75&156.25\\
$\mr_s$ ($h^{-1}$Mpc)&12.26&23.46&37.61&49.11&61.74&73.92&87.15&99.12&111.95&131.49&182.59\\
$(H_s)\Ns{CMB}$&120.8&108.8&105.9&103.6&104.2&102.6&102.7&103.8&102.0&102.3&100.1\\
$(\bs_s)\Ns{CMB}$&2.1&1.2&0.9&0.7&0.8&0.8&0.9&0.9&1.0&0.9&1.7\\
$(Q_s)\Ns{CMB}$&0.000&0.000&0.000&0.555&0.959&1.000&1.000&1.000&0.992&0.997&0.999\\
$(\chi^2/\nu)\Ns{CMB}$&9.496&2.506&1.421&0.993&0.908&0.633&0.624&0.658&0.754&0.745&0.581\\
$(H_s)\Ns{LG}$&103.5&103.5&103.9&106.6&103.9&102.0&103.2&103.6&101.6&102.7&101.0\\
$(\bs_s)\Ns{LG}$&1.8&1.1&0.9&0.7&0.8&0.8&0.9&0.9&1.0&0.9&1.7\\
$(Q_s)\Ns{LG}$&0.000&0.000&0.000&0.031&0.960&1.000&1.000&1.000&0.996&0.999&0.999\\
$(\chi^2/\nu)\Ns{LG}$&11.427&3.246&1.792&1.090&0.907&0.701&0.592&0.608&0.728&0.711&0.590\\
$\ln B$ ($r\ge r_s$)&30.09&8.99&3.22&0.92&2.47&1.68&1.37&1.30&0.64&0.39&
\enddata
\caption{Hubble flow variation in radial shells in CMB and LG frames.
Spherical averages (\ref{Hs}) are computed for two different choices of shells,
$r_s<r\le r_{s+1}$, the second choice being labeled by primes. In each case we
tabulate the inner shell radius, $r_s$; the weighted mean distance, $\mr_s$;
the shell Hubble constants, $(H_s)\Ns{CMB}$ and $(H_s)\Ns{LG}$ in the CMB and
LG frames, and their uncertainties determined by linear regression within each
shell, together with its ``goodness of fit'' probability $Q_s$ and reduced
$\chi^2$ (for $\nu=N_s-1$); $\ln B$ where $B$ is the Bayes factor for the
relative probability that the LG frame has more uniform $\de H_s=0$ than the
CMB frame when $\chi^2$ is summed in all shells with $r>r_s$.
$H_s$ and $\bs_s$ are given in units \protect{$h\kmsMpc$}.}
\label{shell}\end{table*}\endgroup

\begin{figure*}[htb]
\vbox{\centerline{{\includegraphics{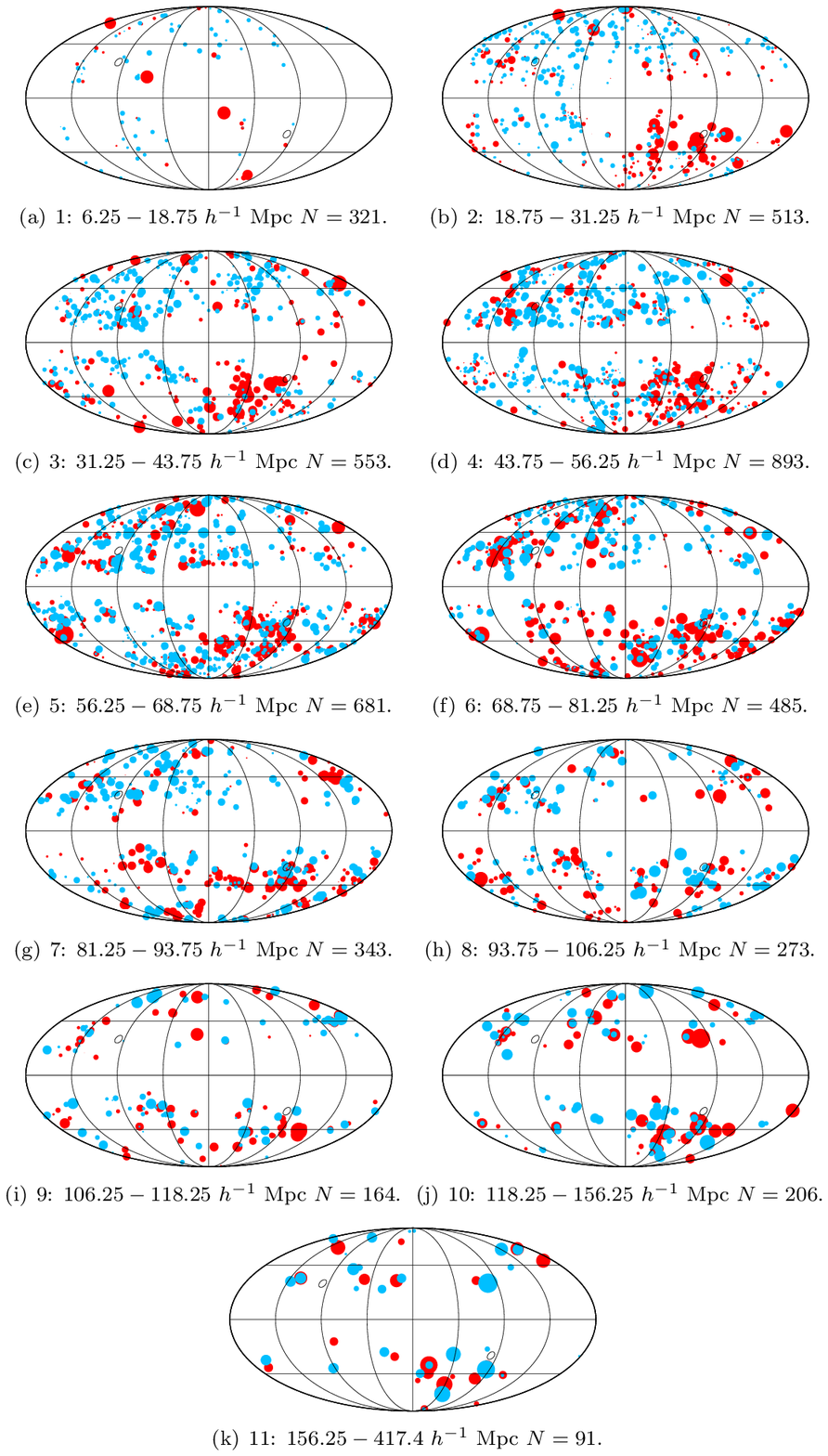}}}
\vskip\baselineskip
\caption{%
The angular distribution of the individual data points in the \CS sample is
plotted in a manner similar to Fig.~1 of Ref.\ \cite{fwh10}, but distributed
into the (unprimed) radial shells of Table~\ref{shell}. The radius of each
data point indicates the magnitude of the peculiar velocity relative to
$H_s r_i$, where $H_s$ is the value given for the LG frame in Table~\ref{shell}
in the unprimed case. Positive and negative peculiar velocities are
colour-coded red and blue (or darker and lighter in greyscale). In panels
(a)-(k) which show each of the unprimed shells, the galactic longitudes
$\ell=0\deg,180\deg,360\deg$ are on the right edge, centre and left edge
respectively. An open error circle marking the poles of the residual CMB dipole
at $\{(96.4\deg,-29.3\deg)\pm3.2\deg,\ (276.4\deg,29.3\deg)\pm3.2\deg\}$ in
the LG frame is shown for reference.
\label{skycover}}}
\end{figure*}

For both choices of shells 91 data points with $r>156.25\hm$ in shell 11 have
been used to determine the mean asymptotic value of the Hubble constant,
$\bH\Z0$, and its uncertainty. The inner boundary of this shell must be chosen
at a sufficiently large distance that it is greater than the scale of
statistical homogeneity. Thus we take the inner shell radius to be larger than
the baryon acoustic oscillation (BAO) scale, that being the largest scale which
could reasonably modify the gross features of the local Hubble flow\footnote{%
In seeking convergence of bulk flows to the CMB dipole, researchers working in
the peculiar velocity framework are currently considering the influence of the
Shapley Concentration on our local motion. Since Shapley is at a distance of
$138\hm$, such a large scale correlation would have to represent a very unusual
fluctuation relative to the statistical homogeneity scale if the standard
framework were correct. The standard framework has focused attention on the
largest overdense structures. However, the very underdense regions are equally
important in determining averages and are naturally incorporated in
the spherical averaging approach (c.f.~Fig.~\ref{Hshells}).}.
Furthermore, in the CMB frame the asymptotic Hubble constant should match the
$100\hm$ normalization used in the \CS data set. This is indeed
satisfied by our choice. We find $\bH\Ns{0}= (100.1\pm1.7)h\kmsMpc$ for the
CMB frame and $\bH\Ns{0}=(101.0\pm1.7)h\kmsMpc$ for the LG/LS frames. We
thus see that although the LG/LS value is 1\% larger than the CMB value, both
values agree within uncertainties, and also with the distance normalization
assumed in compiling the \CS sample.

We do not determine $\bH\Ns{0}$ from the whole \CS sample, since it
is dominated by points in the foreground, with a mean weighted distance
of $15.05\hm$. The fit of a single linear Hubble law to the
whole sample of 4,534 points gives $\bH\Ns{0}=(108.9\pm1.5)\,h\kmsMpc$ in the
CMB frame or $\bH\Ns{0}=(104.4\pm1.4)\,h\kmsMpc$ in the LG and LS frames. It is
precisely because voids dominate the volume of space that we expect
radial averages on scales comparable to the diameters of the largest typical
voids to skew the simple linear average to values
greater than the asymptotic global value. This is
confirmed by the full sample simple linear fit. Our purpose is to more
carefully quantify the foreground Hubble flow variance.

The key statistical point about the determination of the mean asymptotic
value, $\bH\Z0$, in each case is that its uncertainty provides a
significant contribution to the total uncertainty in the relative variation
of the Hubble parameter in the $s$th shell
\beq\de H_s=\left(H_s-\bH\Z0\right)/\bH\Z0\,.\eeq

We have checked that the angular sky coverage of the sample is consistent
in individual shells. This is important since we could get spurious
results if the data in any shell was limited to one side of the sky, and
potentially dominated by particular structures. In Fig.~\ref{skycover} we
plot figures similar to Fig.~1 of FWH10, which shows the sky coverage within
each of the unprimed shells of Sec.~\ref{rad}. We use a
Mollweide projection in galactic coordinates $(\ell,b)$ with $\ell=360\deg$ on
the extreme left and $\ell=0\deg$ on the extreme right. Additional peculiar
velocity information is encoded in the relative sizes and colours of the data
points.

We see that angular sky coverage is consistent in almost all shells, with some
large gaps only the innermost shell 1, $r<12.5\hm$. While shells 9 and 10
contain 222 and 280 points respectively, i.e., of order half the number in most
other shells, a Monte Carlo analysis discussed in Appendix~\ref{mc} establishes
that there is still sufficient data in these shells to support the existence of
a dipole feature in the CMB frame in at the 97.6\% and 99.7\% confidence levels
respectively. There are insufficient data points in the outermost shell 11 to
reliably distinguish any angular variations. However, this shell is only used
as a check on the asymptotic spherically averaged Hubble constant, for which
there are no statistical problems, the goodness of fit statistic being $0.999$.

For the primed shells, where the inner boundary is offset by $6.25\hm$ there is
no sky coverage problem, even in the innermost shell. We will retain the
(unprimed) shell 1 in our analysis, but our statistical conclusions will not
rely on it.

\begin{figure*}[htb]
\vbox{\centerline{\vbox{\halign{#\hfil\cr\scalebox{0.4}
{\includegraphics{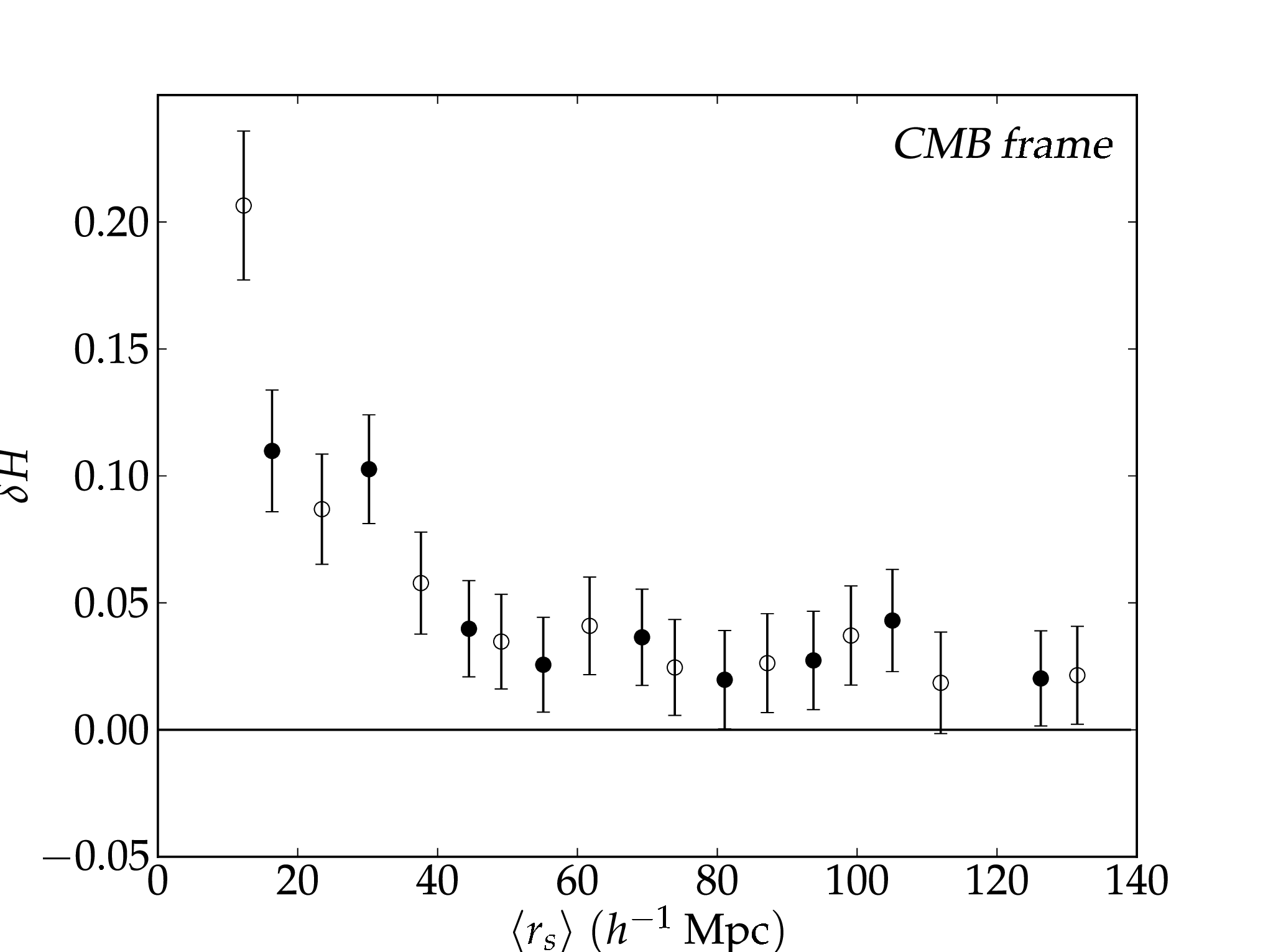}}\cr
\noalign{\vskip-5pt}\qquad{\bf(a)}\cr}}
\vbox{\halign{#\hfil\cr\scalebox{0.4}
{\includegraphics{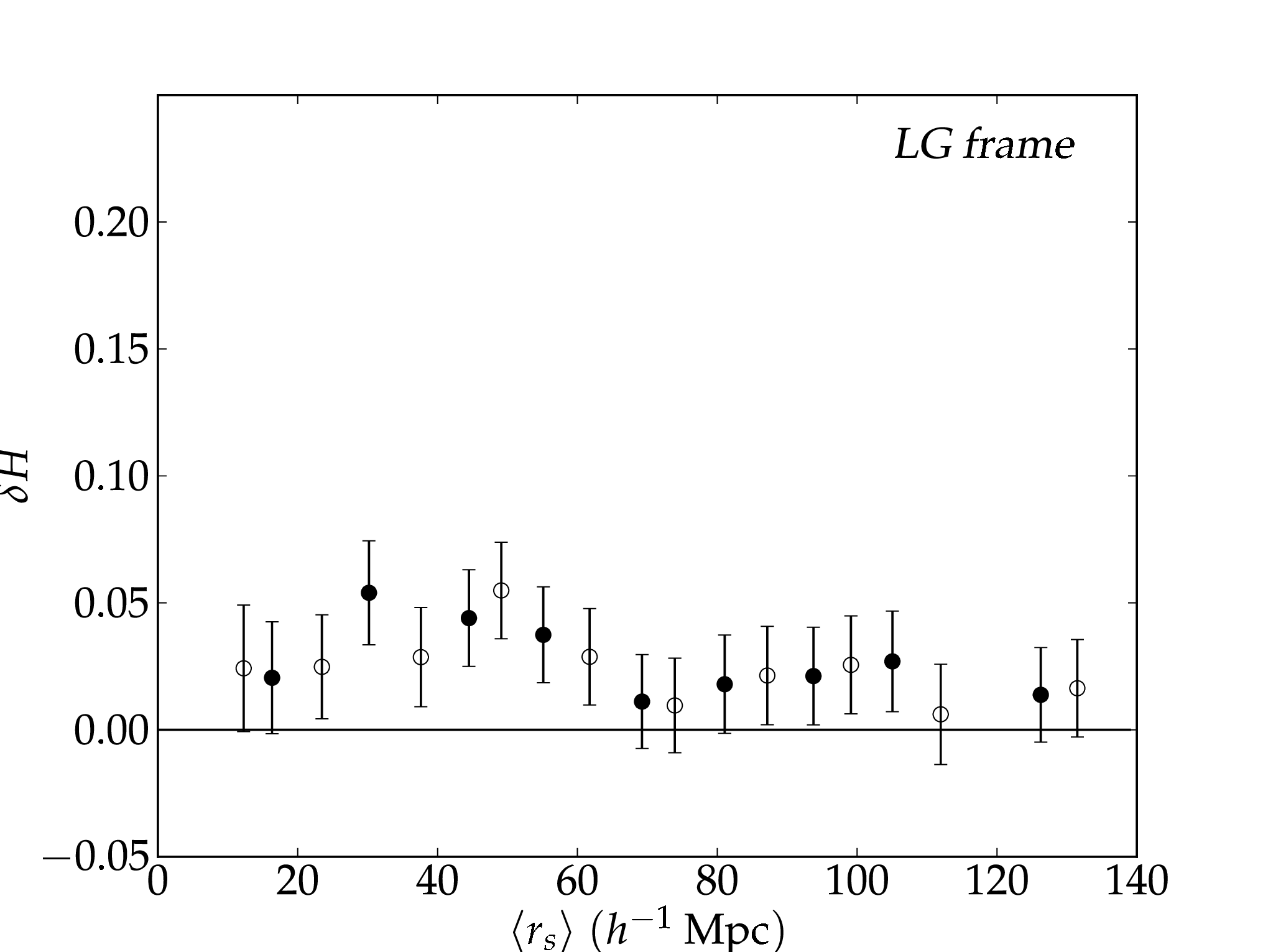}}\cr
\noalign{\vskip-5pt}\qquad{\bf(b)}\cr}}
}
\centerline{\vbox{\halign{#\hfil\cr
\rotatebox{270}{\scalebox{0.4}{\includegraphics{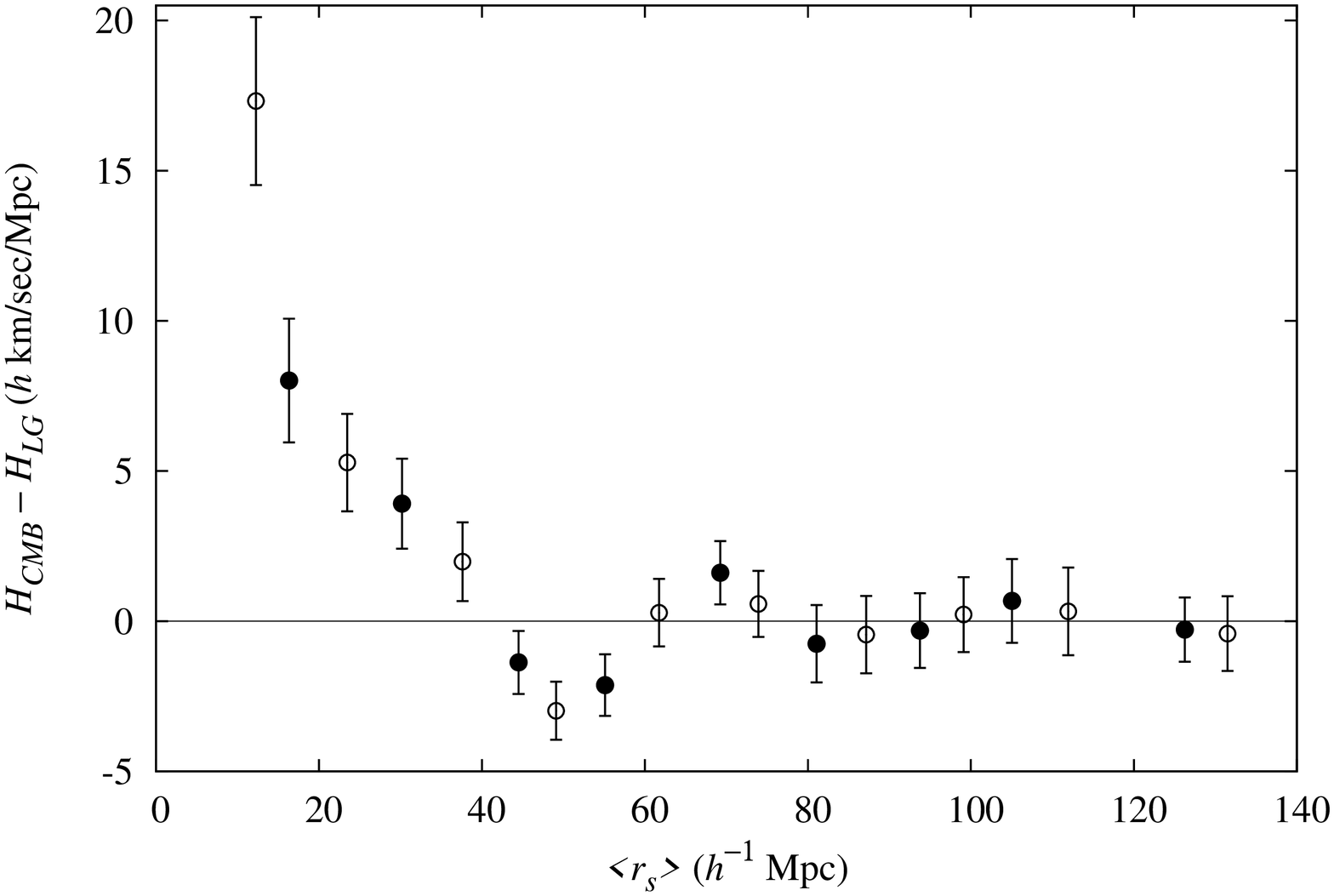}}}\cr
\noalign{\vskip-15pt}\qquad{\bf(c)}\cr}}}
\caption{%
Variation in the Hubble flow $\de H_s=\left(H_s-\bH\Z0\right)/\bH\Z0$ in
spherical shells as a function of weighted mean shell distance:\break
(a) CMB frame; (b) LG frame. In panel (c) the difference $\left(H_s\right)\Ns
{CMB}-\left(H_s\right)\Ns{LG}$ is plotted. In each case the filled data points
represent the
first choice of shells in Table~\ref{shell}, and the unfilled circles the
alternative second choice of shells. We have omitted the first shell from the
plots since $\de H$ is so large in the CMB frame that it is off-scale -- for
the first shell: with a mean weighted distance of $\langle r_s\rangle=5.43\hm$
we have $\de H\Ns{CMB}=0.737\pm0.029$, $\de H\Ns{LG}=0.168\pm0.007$ and
$H\Ns{CMB}-H\Ns{LG}=56.0\pm8.2\kmsMpc$.
\label{dHs}}}
\end{figure*}

\subsection{Results}\label{res}
The results of our analysis are shown in Fig.~\ref{dHs}, where we plot
$\de H_s$ for both the CMB and LG frames, for the two independent choices
of shells given in Table~\ref{shell}, along with the difference
$(H_s)_{\rm CMB}-(H_s)_{\rm LG}$. We computed the result for the LS also;
it is essentially indistinguishable from the LG frame.

The values of $\de H_s$ are positive, consistent with the results of LS08,
and consistent with the fact that in a universe whose volume is dominated
by voids a spherical average will inevitably include more voids than the
denser filaments and walls if one averages on scales comparable to the
diameters of the largest typical voids, leading to a higher than average
$H_s$ as compared to the asymptotic value $\bH\Z0$.

It is clear that the variance of the spherically averaged LG or LS frame
Hubble flow is less than that of the CMB frame. In both frames the Hubble
flow averaged in spherical shells gives $\de H_s$ within $2\si$ of uniform in
almost all shells\footnote{The one small exception is that $\de H\Ns{CMB}$ is
$2.1\si$ from uniform for shell 9 with $100\h<r\le 112.5\hm$ and $\mr\Z9=
105.0\hm$. In general, the LG frame flow is still somewhat closer to uniform
than the CMB frame flow in the outer regions. For all shells with $\mr_s\ge
69.2\hm$ the LG frame flow is within 1.36$\si$ of uniform.} for $\mr_s\ge55.1
\hm$. However, particularly for values $\mr_s<37.6\hm$, the LG/LS frame has
$\de H_s$ much closer to uniform than the CMB frame, and the average LG/LS
frame flow is even within 1.2$\si$ of uniform in the range $12.3\h<\mr_s\le
23.5\hm$, whereas the average CMB frame flow is 4.0$\si$ -- 7.0$\si$
from uniform in the same range. Since the Local Sheet itself is defined
within $r<6.25\hm$, this is not a result that would be readily
expected with the standard interpretation of peculiar velocities.

The statistical significance of the relative uniformity of the averaged
flow in the two frames can now be determined by summing the mean square
differences from a uniform $\de H=0$ expectation,
\beq\chi^2(r_s)=\sum_{j=s}^{12}{\bH\Z0^{\;4}\,\de H_j^2\over\bH\Z0^{\;2}\,
\bs\Z{H_j}^2+H_j^{\;2}\,\bs\Z{\bar H_0}^2},
\eeq
for each choice of rest frame in all shells outside an inner cutoff shell,
$r_s$, as the inner cutoff is varied. An inner cutoff is commonly applied to
eliminate the contribution of large peculiar velocities near the origin, and
given the reinterpretation we follow in this paper, the effect of varying the
cutoff is particularly interesting.
The probability, $P\Ns{CMB}(r_s)$ or $P\Ns{LG}(r_s)$, of a uniform Hubble
flow for each choice of rest frame and cutoff is then calculated directly from
the complementary incomplete gamma function for the chi square distribution
with the relevant number of degrees of freedom. A Bayes factor $B(r_s)=P\Ns{LG}
/P\Ns{CMB}$ for each choice of inner cutoff is determined for the two
independent choices of shells in Table~\ref{shell}. The resulting values of
$\ln B$ are tabulated in Table~\ref{shell}, and plotted as a function of $r_s$
in Fig.~\ref{lnB}. We also determined $P\Ns{LS}/P\Ns{CMB}$ for the
LS relative to the CMB; however, the values obtained gave Bayes factors
which were essentially indistinguishable from those tabulated for the LG
relative to the CMB.

\begin{figure}[htb]
\vbox{
\centerline{\scalebox{0.45
}{\includegraphics{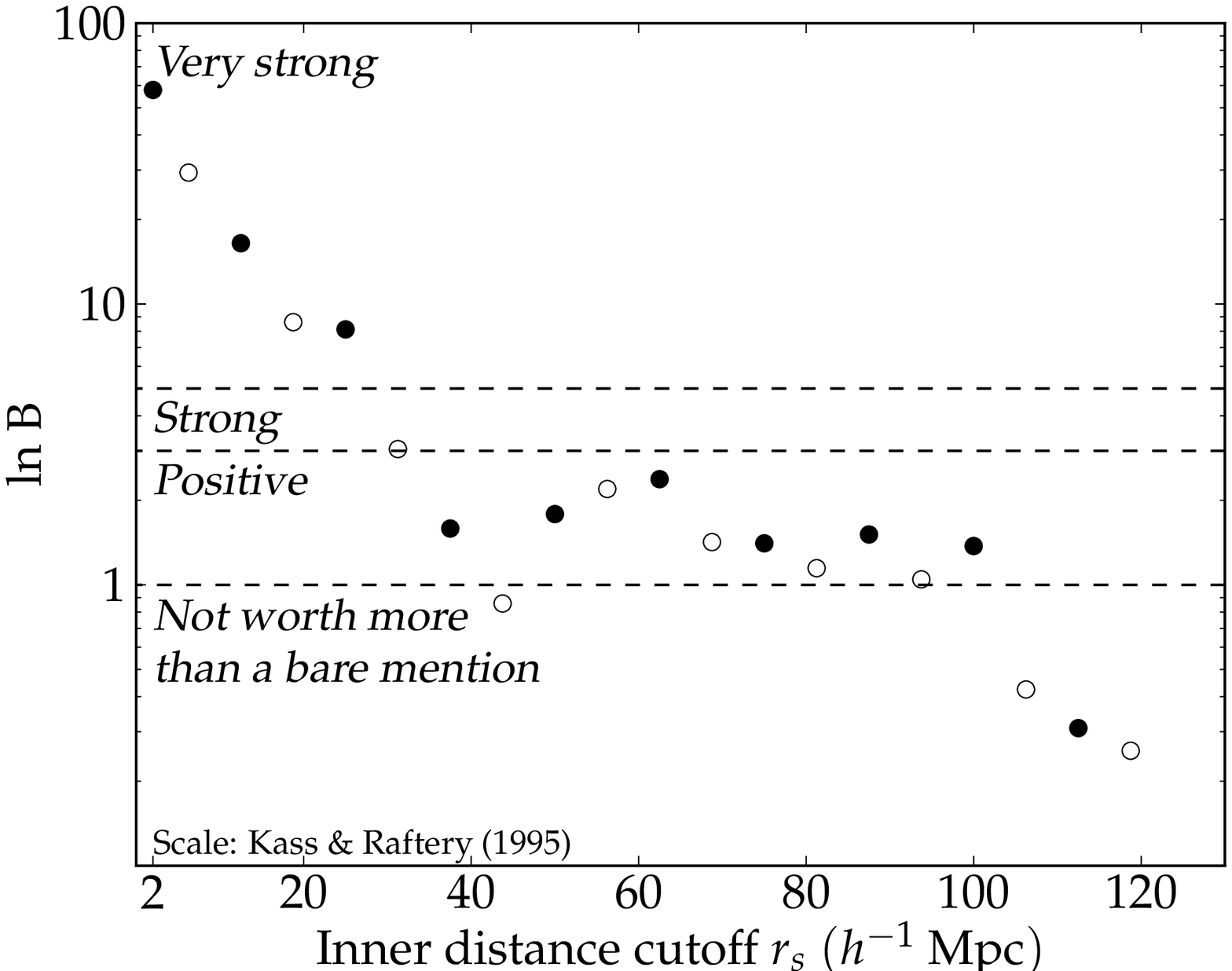}}}
\caption{%
The Bayes factor $\ln B$, where $B$ is the Bayes factor for the ratio
of probability that the LG frame Hubble flow is uniform in the region
$r>r_s$ outside the shell with inner radius $r_s$, divided by the equivalent
probability for the CMB frame.\label{lnB}}}
\end{figure}

Fig.~\ref{lnB} reveals a number of interesting features. The fact that
overall the LG frame is more uniform is demonstrated by $\ln B$ being
everywhere positive. If we consider a large inner cutoff, $r_s\ge106.25\hm$
then the difference in uniformity of the two frames has $\ln B\le1$, which is
not statistically significant. With cutoffs in the range $37.5\h\le r_s\le100
\hm$ we find $1<\ln B<3$ with positive evidence in favour of the LG frame
being the more uniform. Bringing the cutoff down to $r_s=37.5\hm$ gives
$\ln B=3.6$ increasing the Bayesian evidence to strong. For cutoffs $r_s\le
25\hm$ the Bayesian evidence becomes very strong, $\ln B>5$.
Different adjectives are used to describe the strongest Bayesian
evidence \cite{kr95,t07}; since $\ln B>10$ for any inner cutoff with
$r_s<14.5\hm$ Jeffreys' terminology of ``decisive evidence'' in favour of the
relative uniformity of the Hubble flow in the LG frame seems to be appropriate.

We must be careful, however, in the determination of statistical
confidence, since there are also statistically significant departures from
uniformity in the LG and LS frames also, as is consistent with the presence of
foreground structures.

The nonlinear effects of the foreground structures can be seen by computing
the goodness of fit probability, $Q_s$, given by the complementary incomplete
gamma function for $\chi_s^2$ in shell $s$ with $\nu=N_s-1$ degrees of freedom.
In Table~\ref{shell} a bad linear fit is indicated in both the CMB and LG
frames for shells $s\le4$ (unprimed) or $s\le3'$ (primed) since $Q_s$ is less
than 0.1 and equivalently the reduced $\chi^2$ per degree of freedom
is significantly in excess of one.

We have investigated the extent to which the relative magnitude of the Bayes
factor is driven by the greater scatter relative to a linear law, rather than
by the difference of the linear fit of the Hubble constant from its asymptotic
value. The results of this investigation are presented in Appendix~\ref{nonl}.
We find that when the data points which contribute the greatest scatter
relative to a linear law are excluded, leading to a high goodness of the fit,
then the Bayes factors are somewhat reduced. However, the Bayesian evidence
for the greater uniformity of the LG frame Hubble flow, relative to that
of the CMB frame, remains very strong.

The outer shells with $s\ge 5$ (unprimed) or $s\ge 4'$ (primed) all have
a strong goodness of fit in the full data set of Table~\ref{shell}. This
is also true in the outermost, $r>156.25\hm$ shell, although it only contains
91 points. This outer shell, which is used to anchor the
asymptotic value of the Hubble constant and its uncertainty, has an almost
perfect goodness of fit $Q_s=0.999$ and a reduced $\chi^2$ of $0.59$ per
degree of freedom in both frames.

Some hints about the nature of the effects which contribute to the deviations
from a uniform linear Hubble law can be discerned by comparing $\de H_s$ in
the shells where the deviations from uniformity become statistically
significant. Perhaps the most interesting feature is that counter to the
overall trend, the individual CMB frame shells $\{4,4',5\}$ with mean
distances in the range\footnote{Each bound is the average of the mean
distances of the shell where the CMB frame is more uniform with the mean
distance of the neighbouring shell where the LG frame is more uniform.}
$41.0\h\le\mr_s\le58.4\hm$ have a closer to uniform $H_s$ than the
corresponding LG frame shells. In the cumulative Bayes factor this adds a
negative contribution, and reduces the overall $\ln B$ to $0.92$ at $\mr_s=
49.1\hm$, even though adjacent points have $\ln B>2$.

\subsection{Systematic offsets from choice of reference frame}\label{vsys}
Another important point is to consider how the nonlinear dependence of $H_s$
on the individual $c z_i$ in the regression formula (\ref{Hs}) can lead
to systematic offsets when applying boosts. Suppose we are in a frame in
which the spherically averaged variance in the Hubble flow is minimized,
which of course can be a frame other than the LG or LS one. Now change
reference frame by applying a uniform boost to all data points, so that
\beq
cz_i\to cz_i'=cz_i+v\cos\ph_i
\eeq
where $\ph_i$ is the angle on the sky between the data point and the
boost direction. Then $(c z_i)^2\to(cz_i')^2=cz_i^2+2cz_i v\cos\ph_i+v^2\cos^2
\ph_i$ in the numerator of (\ref{Hs}), and $c z_i r_i\to cz_i'r_i=cz_i r_i+r_i
\cos(\ph_i)$ in the denominator.

If we perform a spherically symmetric average (\ref{Hs}) on data which is
reasonably uniformly distributed over the celestial sphere\footnote{The absence
of data in the ZoA does not affect this argument, since the
gaps in the data set are symmetrically distributed on opposite sides of the
sky. The argument would fail if there was a significant lack of data
on one side of the sky only.} then on average each positive contribution of
the linear terms $v\cos\ph_i$ in the sums in the numerator and denominator
of the boosted frame $H_s'$ will be counterbalanced by a negative contribution
from a $v\cos\ph_j$ on the opposite side of the sky. The linear contributions
are therefore roughly self-canceling, leaving a dominant contribution to
the difference
\bea
H'_s-H_s&\goesas&\left(\sum_{i=1}^{N_s}{(v\cos\ph_i)^2\over\si_i^2}\right)
\left(\sum_{i=1}^{N_s}{cz_i r_i\over\si_i^2}\right)^{-1}\nonumber\\
&=&{\ave{(v\cos\ph_i)^2}\over\ave{cz_ir_i}}\,.
\label{Hsd}
\eea
where $\ave{f_i}\equiv(\sum_{i=1}^{N_s}f_i\si_i^{-2})(\sum_{i=1}^{N_s}
\si_i^{-2})^{-1}$ is a weighted average.
If we now consider (\ref{Hsd}) applied to successive shells with different
values of $s$, then given a uniformly symmetrical distribution of data
the weighted average $\ave{(v\cos\ph_i)^2}\goesas\frac12v^2$ in the numerator
will be roughly constant from shell to shell, while putting the leading order
approximation $cz_i\simeq\Hm r_i$ in the denominator we find
\beq
H'_s-H_s\goesas{v^2\over2\Hm\ave{r_i^2}}\,.
\eeq

Consequently, for symmetrically distributed data the effect of incorporating
a boost in the redshift data is an additional contribution to the spherically
averaged Hubble constant which is inversely proportional to the averaged
square distance. The difference between the CMB and LG frames
in Fig.~\ref{dHs} does indeed show hints of such a dependence. Of course,
the LG frame itself may incorporate such a dependence with respect to
whatever frame has the minimum variance in $H_s$, only to a lesser extent.

We stress that by our method of analysis the effect of a spurious boost
is to add a spherically symmetric, or monopole, ``Hubble bubble'' type
variation to the Hubble relation.
This feature makes the present analysis very different to the standard
peculiar velocity approach, where the focus is on dipole or higher multipole
variations.

We summarize the results of this section as follows. Although there are
significant foreground structures which distort the spherically averaged
Hubble flow in a statistically significant manner, the LG frame has a much
smaller monopole Hubble flow variance than the CMB frame, counter to standard
expectations. Nonetheless, there is a particular range of distances at roughly
$40\h\lsim r\lsim60\hm$ for which the boost to the standard CMB frame produces
an apparently more uniform spherically symmetric average flow. This
is the first evidence for the hypothesis we will present in
Sec.~\ref{origin}, namely that rather than being a transformation
which puts us in the frame in which the Hubble flow is most uniform
at our own point, the boost to the CMB frame is actually compensating for
the effect of foreground structures largely associated with distance
scales of order $40\h$ to $60\hm$. To better understand these
structures we now consider angular averages.

\begin{figure*}[htb]
\vskip10pt
\centerline{\vbox to 170pt{\vfill\halign{#\hfil\cr\scalebox{0.52}
{\includegraphics{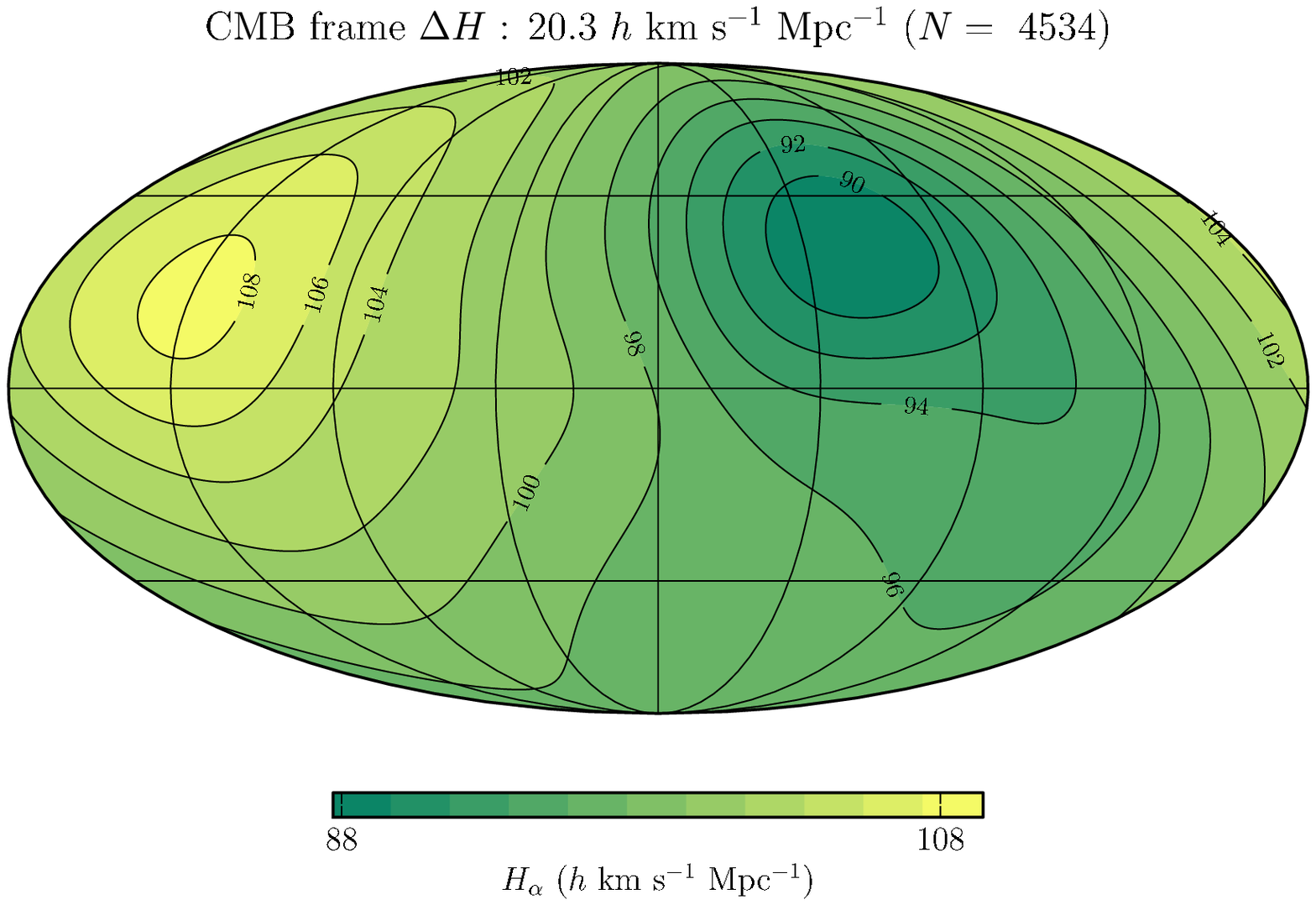}}\cr
\noalign{\vskip-30pt}\qquad{\bf(a)}\cr}}\hskip31pt
\vbox to170pt{\vfill\halign{#\hfil\cr\scalebox{0.52}
{\includegraphics{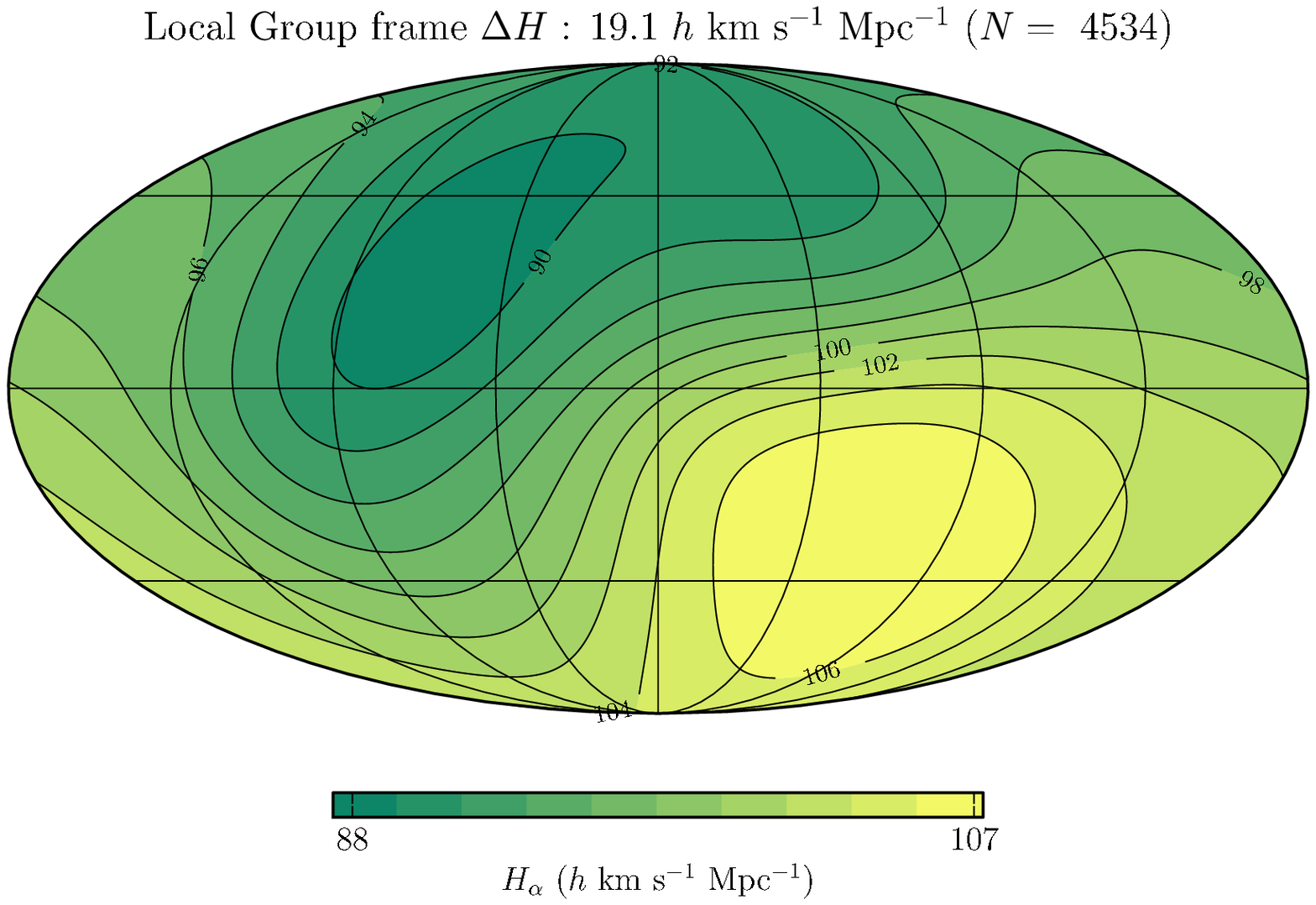}}\cr
\noalign{\vskip-30pt}\qquad{\bf(b)}\cr}}}\vskip22pt
\caption{%
Angular variance of the Hubble flow in the full \CS sample: (a)
CMB rest frame; (b) LG rest frame. \latitudes
\label{wholesky}}
\end{figure*}

\section{Angular averages}\label{ang}
In order to associate variance in the Hubble flow with particular foreground
structures angular information is also required. The angular variance of the
Hubble flow in the same HST Key Data \cite{freedman01} investigated by
\Rls{ls08} has been studied by \Rmd{md07} (henceforth MD07) in the CMB
reference frame. \Rmd{md07} used all 76 points in the Key Data set, and
concluded that a 13\% variation in $\Hm$ existed in the data. Once again,
although individual distances in the \CS sample are noisier, the 60-fold
increase in the size of data set enables a more detailed analysis.

Angular averages will in general have a complex multipole structure. The higher
the order of the multipole probed, the greater the amount of data required to
achieve sufficient angular resolution. Ideally one would split the sample
into radial shells, as in Sec.~\ref{rad}, and perform a separate multipole
analysis in each shell. Given limited data a trade-off can be made by making
the width of the radial shells larger.

We find that the \CS sample has sufficient data to establish
the existence of a simple dipole Hubble law in several of the shells of
Sec.~\ref{rad} with 99.99\% confidence. While our main results in the present
paper relate to the monopole and dipole variations, it is also necessary to
perform a full multipole analysis in order to:\begin{enumerate}[(i)]
\item check that multipoles higher than a dipole are sufficiently small that
the dipole amplitude will not be significantly affected by leakage into higher
multipoles if only a dipole Hubble law is considered; \item establish a means
for determining the degree of correlation between the map of angular Hubble
flow variation, with all its multipoles, to the residual pure temperature
dipole of the CMB in the rest frame of the LG or LS, as discussed in
Sec.~\ref{corr}.
\end{enumerate}

\subsection{Results of Gaussian window averages}\label{avar}
We have adapted the technique of MD07 to produce a series of maps of an angular
average value of $cz/r$ over the sky. At each grid point on the sky, a mean
$\Hth$ is calculated in which the value of $cz_i/r_i$ for each data point is
weighted according to its angular separation from the grid point, with a
Gaussian window function whose standard deviation, $\sith$, determines the
smoothing scale. The technical details are described in Appendix~\ref{mdtech}.
We applied the averages (\ref{Hal})--(\ref{sial}) both without an inverse
variance (IV) weighting using (\ref{Wi}), and alternatively with an IV
weighting using (\ref{Wi2}). The small differences between these two methods
did not affect our conclusions.

In Fig.~\ref{wholesky} we plot the contour maps of angular Hubble flow
variance produced using the whole \CS data set in a single sphere, in
both the CMB and LG rest frames. This allows a direct comparison to MD07
who considered only 76 points in a single sphere in the CMB rest frame.

\begin{figure*}[htb]
\vbox{\centerline{\scalebox{0.57}
{\includegraphics{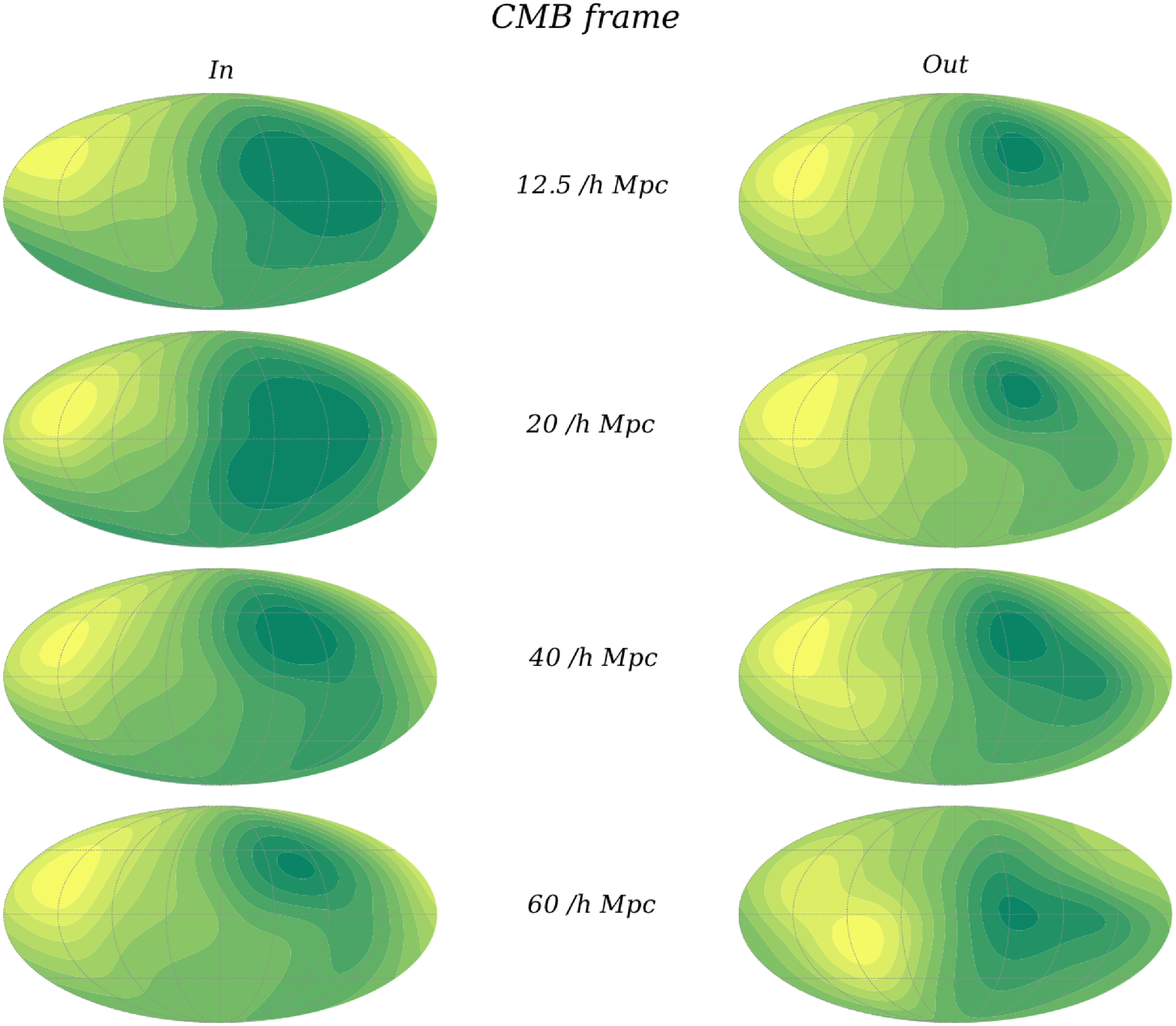}}}
\caption{%
Angular variance in the Hubble flow in the CMB rest frame for inner
($r<r_o$ left panel) and outer ($r>r_o$ right panel) spheres as $r_o$
is varied over the values 
$12.5$, $20$, $40$ and $60$ $\hm$. \latitudes
\label{cmbcutsky}}}
\end{figure*}

\begin{figure*}[htb]
\vbox{
\centerline{\scalebox{0.57}
{\includegraphics{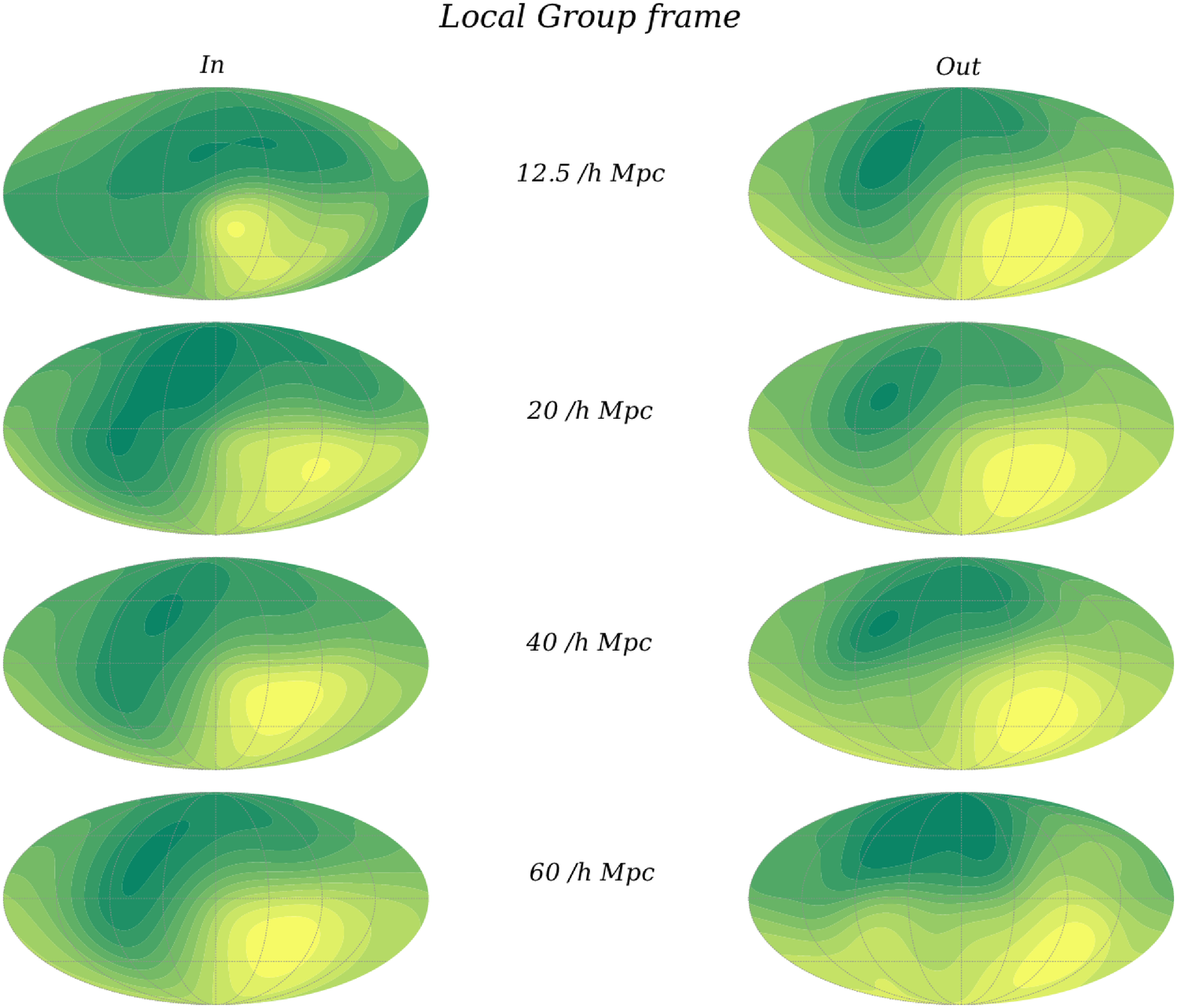}}}
\caption{%
Angular variance in the Hubble flow in the LG rest frame for inner
($r<r_o$ left panel) and outer ($r>r_o$ right panel) spheres as $r_o$
is varied over the values 
$12.5$, $20$, $40$ and $60$ $\hm$. \latitudes
\label{lgcutsky}}}
\end{figure*}

\bigtable{*}{lrrrrrrrrrrrrr}
$r<r_o$ ($\hm$)&&$<12.5$&$<15$&$<20$&$<30$&$<40$&$<50$&$<60$&$<70$&$<80$&$<90$&$<100$\\
\noalign{\smallskip}\tableline\noalign{\smallskip}
CMB $C_2/C_1$&&0.123&0.061&0.044&0.098&0.136&0.191&0.187&0.167&0.141&0.134&0.120\\
CMB $C_3/C_1$&&0.010&0.011&0.007&0.003&0.005&0.010&0.009&0.011&0.011&0.012&0.012\\
LG $C_2/C_1$&&0.653&0.179&0.123&0.135&0.116&0.103&0.104&0.103&0.092&0.089&0.085\\
LG $C_3/C_1$&&0.067&0.018&0.008&0.005&0.008&0.006&0.007&0.009&0.011&0.011&0.011\\
LS $C_2/C_1$&&0.861&0.197&0.133&0.146&0.124&0.112&0.113&0.112&0.101&0.097&0.093\\
LS $C_3/C_1$&&0.068&0.015&0.006&0.005&0.007&0.006&0.007&0.009&0.011&0.012&0.011\\
\noalign{\smallskip}\tableline\noalign{\smallskip}
$r>r_o$ ($\hm$)&$>2$&$>12.5$&$>15$&$>20$&$>30$&$>40$&$>50$&$>60$&$>70$&$>80$&$>90$&$>100$\\
\noalign{\smallskip}\tableline\noalign{\smallskip}
CMB $C_2/C_1$&0.102&0.096&0.115&0.124&0.073&0.038&0.023&0.041&0.093&0.093&0.090&0.327\\
CMB $C_3/C_1$&0.010&0.009&0.013&0.015&0.017&0.009&0.007&0.011&0.018&0.078&0.069&0.076\\
LG $C_2/C_1$&0.072&0.061&0.064&0.064&0.053&0.042&0.032&0.045&0.068&0.077&0.066&0.151\\
LG $C_3/C_1$&0.010&0.009&0.012&0.014&0.019&0.013&0.013&0.014&0.010&0.048&0.051&0.016\\
LS $C_2/C_1$&0.079&0.065&0.068&0.068&0.054&0.044&0.033&0.048&0.074&0.079&0.070&0.162\\
LS $C_3/C_1$&0.010&0.009&0.012&0.014&0.019&0.013&0.013&0.014&0.011&0.053&0.062&0.020
\enddata
\caption{Ratios $C_2/C_1$, $C_3/C_1$ of quadrupole/dipole and
octupole/dipole for the multipoles of the angular Hubble variance maps in the
CMB, LG and LS frames, using (\ref{Htp}) with no IV variance weighting.
In each case the multipole ratios are computed inside ($r<r_o$) and
outside ($r>r_o$) a bounding shell.}
\label{cratios}\end{table*}\endgroup

To study radial variations one can simultaneously break the sample into
independent shells as in Sec.~\ref{rad}. There is enough data in the
\CS sample to reliably establish a quadrupole and perhaps higher order
multipoles in many of the shells of Table~\ref{shell}. However, our first
aim is to determine the gross features of the relative angular
variation. We will therefore perform the most simple of radial separations:
we divide the data into an inner ($r<r_o$) and an outer ($r>r_o$) sphere,
with a boundary $r_o$ which we vary, and reperform the Gaussian window
averages.

We show a subset of the resulting sky maps in
Figs.~\ref{cmbcutsky} and \ref{lgcutsky} for the CMB and LG rest frames,
with the boundary between the inner and outer spheres taking the values
$12.5\h$, $20\h$, $40\h$
and $60\h$ Mpc.

The maps are of course not entirely independent, as there is overlap of data
between the outer shells for small $r_o$ and the inner shells of maps with
larger $r_o$. The extent of overlap of sources, and their angular distribution,
can be determined roughly from the numbers given in Table~\ref{shell} and
in Fig.~\ref{skycover}, where points within individual shells are shown.
Working with maps which are not independent shows how power in the dipole is
transferred from the outer to inner sphere as $r_o$ is varied.

The first observation we make is that although both frames reveal a dipole
structure, the nature of the dipole has important differences between the two
frames. In the CMB frame the difference between the inner and outer spheres is
not very strong. In the outer sphere the two poles migrate from being both
in the northern hemisphere in the $r>12.5\hm$ map to both being close to the
galactic equator in the $r>60\hm$ map, while the poles in the corresponding
interior spheres become localized to the northern hemisphere. However, the
strength of the dipole feature does not vary significantly between the inner
and outer spheres, nor with the variation of the boundary $r_o$ between the
inner and outer spheres. The fact that both poles are in the northern
hemisphere in most of the CMB frame plots also means of course that the
dipole is less strong than for example in the $r>12.5\hm$ LG frame map, for
which the poles are closer to $180\deg$ apart.

By contrast to the CMB frame, in the LG frame there is a significant radial
dependence to the Hubble variance dipole evident in Fig.~\ref{lgcutsky}. With
the division set at $r_o=12.5\hm$ there is very strong dipole feature in the
outer $r>12.5\hm$ sphere, which is stronger than in the full sample map of
Fig.~\ref{wholesky}. By contrast, within the inner $12.5\hm$ sphere any
dipole signature is masked by other multipoles which appear equally as strong.

As the division scale $r_o$ is increased the relative power in the dipole in
the inner sphere in the LG frame maps increases substantially, so that by the
time we reach $r_o=60\hm$ the inner sphere shows a dipole almost as distinct
as the outer sphere of the $r_o=12.5\hm$ map. At the same time the dipole in
the outer $r>60\hm$ map becomes less distinct. This is consistent with our
finding in the previous section that the structures principally responsible
for the Hubble flow variance lie within $r<65\hm$. This conclusion will be
confirmed by an independent analysis of the data in Sec.~\ref{dip} below.

We remark that the dipole feature in the LG frame can
be seen by eye in the colour coded peculiar velocities relative to $H_s$ in
each shell, as shown in Fig.~\ref{skycover}: in shells 2 and 3,
which cover the range $12.5\h<r<37.5\hm$ there is a clear concentration of
negative peculiar velocities (blue) in the upper left quadrant and
positive peculiar velocities (red) in the lower right quadrant, which
correlate with the dipole structure in Fig.~\ref{lgcutsky}. These
concentrations of peculiar velocities become more and more diluted by
the contributions of peculiar velocities of the opposite sign in shells
4 and 5, where $37.5\h<r<62.5\hm$. In shells with $r>62.5\hm$ the areas
previously associated with the dipole feature contain similar numbers of
positive and negative peculiar velocities.

The fact that the CMB frame dipole shows far less variation than
the LG frame dipole as $r_o$ is varied is consistent with the hypothesis
that it is not directly associated with the structures defining the Hubble
flow variance but is rather due to an overall systematic, namely the relative
boost to the CMB frame, as discussed in Sec.~\ref{vsys}.

\begin{figure}[htb]
\vbox{\centerline{\rotatebox{270}{\scalebox{0.35
}{\includegraphics{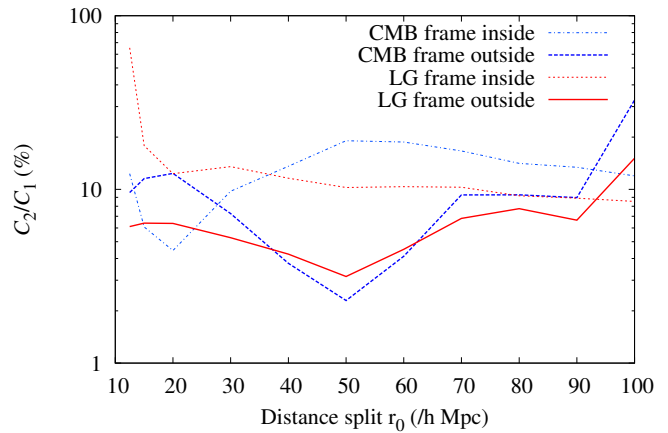}}}}
\caption{%
Ratios $C_2/C_1$, of quadrupole to dipole, on inner ($r<r_o$) and outer
($r>r_o$) spheres, as $r_o$ is varied in the CMB and LG rest frames.
\label{qd}}}
\end{figure}

The above statements are of course made from a simple inspection of
the sky maps by eye. However, the statements can be quantified by performing
a spectral analysis on the sky maps. To this end we digitized the contour
maps into 1 square degree regions and performed a multipole analysis using
HEALPIX\footnote{http://healpix.jpl.nasa.gov/ \cite{gorski05}} to determine
the relative power in the $C_\ell$ coefficients. On account of the Gaussian
window averaging there is aliasing at the $25\deg$ scale, and information for
the high multipoles is not reliable. However, since multipoles with $\ell\ge4$
are very much suppressed a good measure of the significance of the dipole can
be estimated by determining the quadrupole to dipole ratio, $C_2/C_1$, and
octupole to dipole ratio, $C_3/C_1$, as listed in Table~\ref{cratios} in the
inner and outer spheres as the boundary, $r_o$, is varied in the CMB, LG and
LS frames. The inner and outer $C_2/C_1$ ratios are also illustrated
graphically for the CMB and LG frames in Fig.~\ref{qd}.

In the LG frame $C_2/C_1=0.061$ in the outer $r>12.5\hm$ sphere,
representing a small quadrupole relative to dipole while $C_2/C_1=0.653$ in
the corresponding inner sphere representing a quadrupole roughly comparable to
the dipole. By contrast in the CMB frame with $r_o=12.5\hm$, the respective
ratios are $C_2/C_1=0.096$ ($0.123$) in the outer (inner) sphere, indicating a
dipole which is similar in both spheres, and less clearly defined than in the
outer LG frame.

In the inner sphere the ratio $C_2/C_1$ in the LG frame drops
substantially for $r_o\ge30\hm$, and maintains a value in the range $0.09$ --
$0.12$ when $40\h\le r_o\le90\hm$. This is of course higher than the same
ratio in the outer sphere; but the inner sphere value includes in every
case a contribution from the innermost shell in which the dipole and
quadrupole are comparable.

The outer dipole is stronger in the LG frame than the CMB frame except for the
values $40\h\lsim r\lsim60\hm$, for which the situation is reversed. The outer
CMB dipole is strongest for $r>50\hm$, when the quadrupole/dipole
ratio drops to a minimum $C_2/C_1=0.023$ as compared to
$C_2/C_1=0.032$ in the LG frame. However, when $r_o\ge70\hm$ the outer CMB
dipole becomes less distinct again. As we saw earlier
the variance in the spherically averaged Hubble flow was less in the CMB frame
in the range $35\h\lsim r\lsim60$ $\hm$. It appears that the boost to the CMB
frame is also having the effect of making the angular variance of the CMB
frame Hubble flow more dipole-like over this particular radial range.

Table~\ref{cratios} shows that the quadrupole and higher order multipoles
are at least an order of magnitude smaller than the dipole in the range
$15\lsim r\lsim65\hm$, and therefore a simple dipole law can be reliably
used in this range. However, for $r_o>90\hm$ the ratio $C_2/C_1$
increases substantially, so caution should be exercised about fitting
a simple dipole in the outermost shells.

\subsection{Dipole law averages in radial shells}\label{dip}
It is difficult to provide statistical bounds on the angular orientation
and magnitude of the Hubble flow variance dipole with Gaussian window
averaging. However, a completely independent analysis can be made by fitting
the raw data to a simple linear dipole law
\beq\frac{cz}r=H_d+\be\cos\ph\,,\label{di}\eeq
for the LG and CMB rest frames, where in each case $\ph$ is the angle on the
sky between each galaxy and the direction $(\ld,\bd)$ which defines the
best fit dipole axis. This method is similar to that used in Fig.~9 of
Ref.\ \cite{sand86} or Fig.~8 of Ref.\ \cite{kash09}.

In each case we determine the four parameters $H_d$, $\be$, $\ld$ and $\bd$ by
a least squares fit of $H_d$ and the linear parameters $\be_x=\be\cos\ld\cos
\bd$, $\be_y=\be\cos\ld\sin\bd$, and $\be_z=\be\sin b_d$. Details are given in
Appendix~\ref{dipest}. The results of the analysis for the same independent
shells chosen in Table~\ref{shell} are tabulated in Table~\ref{dshell}, and
the corresponding dipole amplitudes are plotted in Fig.~\ref{sdipfit}. Here
statistical 1$\si$ uncertainties are shown.

\begin{figure}[htb]
\medskip
\vbox{\centerline{\scalebox{0.42
}{{\includegraphics{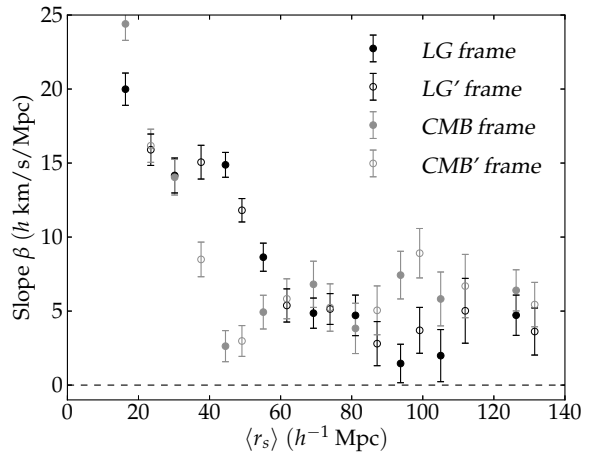}}}}
\caption{%
The slope $\be$ of the linear dipole relation $cz/r=H_d+\be\cos\ph$, as
given in Table~\ref{dshell}, is plotted by shell in the CMB and LG rest
frames. The filled (unfilled) circles correspond to the unprimed (primed)
shells.
\label{sdipfit}}}
\end{figure}

Monte Carlo simulations, discussed in Appendix \ref{mc}, establish that
random reshuffles of the angular data in each shell are consistent with zero
dipole. We also determine a probability in excess of random of the dipole law
providing a better fit, and consequently a statistical confidence that the
dipole in each shell is indeed non-zero.

We see from Fig.~\ref{sdipfit} that the range of radial shells over which the
CMB frame provides a better fit to the monopole Hubble law than the LG frame
also coincides with a dramatic difference in changes to the dipole in the two
frames. The magnitudes of the dipoles in the two frames coincide in shell $3$
with $\mr\Z3=30.2\hm$, taking the values $\be\Ns{CMB}=(14.0\pm1.2)h\kmsMpc$
and $\be \Ns{LG}=(14.2\pm1.2)h\kmsMpc$ respectively. They also coincide in
shell $5'$, with $\mr\Z{5'}=61.7\hm$, where they take the reduced values $\be
\Ns{CMB}=(5.8\pm1.3)h\kmsMpc$ and $\be\Ns{LG}=(5.4\pm1.2)h\kmsMpc$. However,
the dipoles exhibit very different behaviour for the shells in between. In
particular, the CMB dipole magnitude reaches a minimum of $\be=(2.6\pm0.6)h
\kmsMpc$ (close to zero dipole) in shell 4, for which $\mr\Z4=44.5\hm$, whereas
for the LG frame $\be=(14.9\pm0.8)h\kmsMpc$ in the same shell. The CMB frame
dipole then increases while the LG frame dipole decreases so the two take
similar values at $\mr\Z{5'}=61.7\hm$. The dipole directions in each frame
are strongly consistent in shells 4 to 6 in the range $37.5\h\le r\le62.5\hm$.

The analysis of Appendix~\ref{mc} shows that in shells 3 to 5' our principal
conclusions above concerning the relative magnitudes of the dipoles are
supported to the level of at least 99.9\% confidence. We therefore have
a statistically robust justification for the conclusion that {\em the boost
from the LG to CMB frame is compensating for structures in the
range} $30\h\lsim r\lsim62\hm$.

There are some further changes to the dipoles in the outer shells. A small
residual dipole of amplitude $\goesas5\kmsMpc$ and roughly consistent direction
is maintained in the LG frame in shells 5' to 7, at the 90\% confidence
level. Beyond this scale there is no significant LG frame dipole, with the
exception of shell 10. By contrast a dipole is found with at least 90\%
confidence level in all shells 5' to 10 the CMB frame, and with more than 95\%
confidence the shells 8 and 9 which lack a significant LG frame dipole.

Although there appears to be a significant dipole in both frames in shell 10,
we note from Fig.~\ref{qd} that the quadrupole to dipole ratio in the Gaussian
window averages begins to increase significantly at $r_o\goesas100\hm$. Given
a strong quadrupole, and the fact that the LG dipole axis is roughly orthogonal
to the largest mass concentration in shell 10 -- the Shapley Concentration --
we should be careful not to draw strong conclusions from the fit of a simple
dipole law (\ref{di}) in the outer shell, as the magnitude may change once
higher multipoles are included. The \CS sample does not have enough data
in shell 10 to constrain the quadrupole; considerably more data is required.

\subsection{Smoothed dipole law variation}
The analysis of the previous subsection provides the strongest direct evidence
that there is a correlation between the structures responsible for both the
monopole and dipole variations of the Hubble law in the range $30\h\lsim r
\lsim62\hm$. However, to make contact with the result of analyses in the
peculiar velocity formalism, and with the Gaussian window averages, it is also
useful to consider the results of a dipole law fit on all data
outside (inside) a sphere $r>r_o$ ($r\le r_o$).

\begin{figure}[htb]
\vbox{\vskip6pt\centerline{\scalebox{0.42
}{{\includegraphics{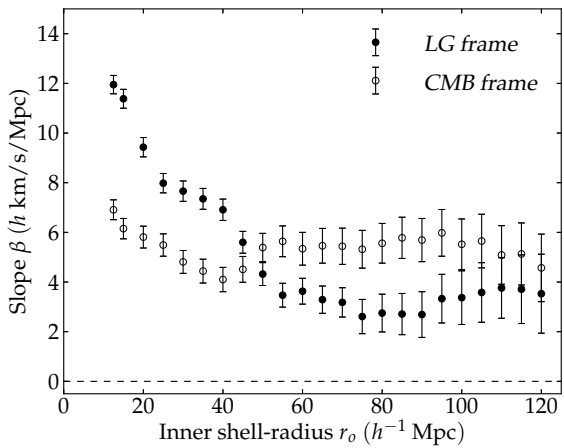}}}}
\caption{%
All data in a sphere $r>r_o$ is fit to the linear dipole relation
(\ref{di}) in the CMB and LG rest frames. We plot the slope, $\be$, of this
relation as a function of the radius $r_o$ outside which the data is included.
Open data points represent fits for the CMB rest frame and solid data points
fits for the LG rest frame.
\label{dipfit}}}
\end{figure}

The results averaged in an outer sphere are tabulated in Table~\ref{oshell}
of Appendix \ref{dipest}, and the corresponding dipole magnitudes are
plotted in Fig.~\ref{dipfit}. In the LG frame there
is a very strong dipole with magnitude $\be=11.4\pm0.4\kmsMpc$ for $r_o=15\hm$,
which decreases to a typical value $\be\goesas3.5\kmsMpc$ for the largest
values of $r_o$ in Table~\ref{oshell}, which is of order $3\si$ different from
zero. If we consider only those
cases with goodness of fit $Q>0.1$ then the most abrupt decrease in $\be$
occurs in the range $40\h\le r_o\le55\hm$ where $\be$ decreases from $6.9\pm
0.4\kmsMpc$ to $3.5\pm0.5\kmsMpc$.

The diminishing of the LG frame dipole at is consistent with the results
of the previous two subsections. Furthermore the angular position of the
dipole for $20\h\lsim r_o\lsim45\hm$ is consistently in the range $(\ld,\bd)=
(83\deg\pm6\deg,-39\deg\pm3\deg)$ while the dipole is strong, but the
angular position then wanders once its magnitude is reduced to residual levels.
For $r_o\gsim80\hm$ the typical position of the residual dipole differs from
that of the inner dipole by $80\deg$ -- $100\deg$ in galactic longitude. The
direction coincides with that of the data in shell 10, which contains the
only significant dipole in the outer regions.

By contrast, the magnitude of $\be$ is initially smaller in the CMB frame for
small values of $r_o$, with a value $\be=6.1\pm0.4\kmsMpc$ at $15\hm$ which
decreases somewhat to $\be=4.1\pm0.5$ at $r_o=40\hm$. However, $\be$ then
increases to $5.6\pm0.6\kmsMpc$ at $r_o=55\hm$, and for larger $r_o$ its value
remains roughly constant at this level, which is of order 4$\si$ -- 7$\si$
different from zero. Furthermore, over the entire range $15\h\le r\le120\hm$
the direction $(\ld,\bd)$ remains within 1$\si$ of the ``dark flow'' direction
$(\ell,b)=(296\deg,14\deg)\pm13\deg$ found by \Rkash{kash09} for X--ray
clusters in the range\footnote{When restricted to the larger scales, $0.12\le
z\le0.3$, the dark flow direction is $(267\deg,34\deg)\pm15\deg$, which
coincides with the direction of the residual CMB dipole in the LG frame.}
$0.05\le z\le0.3$. It also within 1$\si$ of the bulk flow direction $(\ell,b)=
(287\deg\pm9\deg,8\deg\pm6\deg)$ found by \Rwfh{wfh09} for all values $20\h\le
r_o\le115\hm$. For the largest values of $r_o$ in Table~\ref{oshell} the
direction\footnote{In the outermost shell $11$ of Table~\ref{dshell} with $r>
156.25\hm$, the uncertainty in the dipole position in the CMB frame is
essentially the whole sky, meaning that there is not enough data to constrain
the dipole in this range, as is confirmed by the analysis of section \ref{mc}.
In the LG frame the magnitude of the dipole in this shell is just 1.2$\si$ from
$\be=0$.} remains consistent with the bulk flow direction $(\ell,b)=(319\deg\pm
18\deg,7\deg\pm14\deg)$ of \Rturnbull{turnbull11}. This suggests that as far as
the dipole direction is concerned the slight difference between the results of
WFH09, FWH10 and Ref.~\cite{turnbull11} is accounted for by the latter study
having a greater mean depth.

For the largest values of $r_o$ in Table~\ref{oshell} the CMB frame dipole
direction is also consistent with some other cosmic dipoles that have been
observed: a dipole in the fine structure constant \cite{fine1}--\cite{fine3}
and the maximum temperature asymmetry \cite{mta}.

A comparison of Fig.~\ref{sdipfit} and Fig.~\ref{dipfit} shows that analysing
the data only in terms of smoothed dipoles on large scales -- as is implicit
in the peculiar velocities approach -- can hide much
information. In particular, the residual LG frame smoothed dipole at $r\gsim
80\hm$ is accounted for by a feature in shell 10, whereas there is no
significant LG frame dipole in shells 8 and 9. By contrast there is a CMB
frame dipole in the shell by shell analysis; thus its nature is different,
consistent with our hypothesis of Sec.~\ref{vsys}.

\subsection{Identification of Hubble variance with particular structures}
\label{align}
A large dipole structure in the Hubble flow across the sky is consistent with
a foreground density gradient leading to concentrations of more rapidly
expanding void regions in one sector of the sky, and less rapidly expanding
wall regions in the opposite sector. A detailed understanding of the structures
within $30\hm$ may be gained from the work of Ref.\ \cite{tsk08}, and a
viewing of the associated video\footnote{%
http://ifa.hawaii.edu/$\goesas$tully/pecv\_12min\_sound\_qt.mov}
is particularly instructive. Sky maps of structures on larger scales are
given by \Rel{el06} using the 2 Micron All-Sky Redshift Survey (2MASS).

The distinctive feature of the location of our galaxy is that it is in a
thin filamentary sheet, the LS, which defines the supergalactic $(X,Y)$ plane,
on the edge of a Local Void of at least $30\hm$ diameter. While large void
regions dominate one side of the sky, wall regions dominate the other side of
the sky with the superclusters of Centaurus, Hydra and Norma being
particularly prominent. Our Local Sheet and nearby filamentary sheets
such as the Leo Spur are of modest density. The Virgo Cluster appears
to be the closest region of the thick section of a dense nearby wall;
however, it lies almost in the supergalactic plane of the LS, rather than
along the axis which defines the greatest density contrast between the nearby
voids and walls.

In this particular situation our galaxy is neither in the middle of one of the
largest typical voids of $30\hm$ diameter, nor is it in the middle of one of
the thick wall regions. Rather it is in the transition zone between void and
modest wall structures, close to the edge of both. In this circumstance the
observed dipole pattern of Hubble flow variance might be expected to be the
dominant one. Since the spatial width of typical walls is generally smaller
than the diameter of the largest typical voids, observers
located in the middle of a thick wall region with extent in their $(X,Y)$
plane, with larger typical voids some way off and equidistant along their
$\pm Z$ axes, might in contrast to our situation expect to see a more dominant
quadrupole Hubble variance.

The angular extent of various structures must also be important in determining
how close the pattern of Hubble flow variance is to a dipole. A pure step
function contains many higher multipoles, so a simple division of the whole
sky into two hemispheres of uniform faster and slower expansion would contain
many higher multipoles. Since voids have a purer ellipsoidal geometry
than walls, in terms of defining the relevant angular scales it is the voids
which will more clearly delineate the dipole density gradient.

Some estimate of the angular scales of the nearest voids can be obtained from
the work of \Rtully{tsk08}. However, since the distances of many galaxies in
their survey are not known \Rtully{tsk08} present their diagrams in redshift
space, which are subject to redshift space distortions as large as the Hubble
flow variance that we are endeavouring to measure.

Our own Local Void comprises three separate smaller volumes: the Inner Local
Void, and the Local Voids North and South, which are separated by filamentary
thin wall structures \cite{tsk08}. Here ``north'' and ``south'' refer to
directions orthogonal to the LG in supergalactic coordinates; and since the
plane of our galaxy is roughly orthogonal to that of the LS, this means that
supergalactic ``north'' and ``south'' indicate directions principally along
the galactic longitude axis relative to the supergalactic north pole at
$\ell=47.37\deg$, $b=6.32\deg$.

The Inner Local Void, which is ellipsoidal with its major axis roughly
parallel to the Local Sheet, is the structure that covers the largest fraction
of the sky in the Local Void complex. From Fig.~10 of \cite{tsk08} we
estimate that it covers at least 40--60\% of one hemisphere, given the
uncertainties of redshift space distortions. In any case we expect it to be
too large a fraction of the sky to give a pure dipole. This is confirmed by
splitting the inner and outer shells at $r_o=12.5\hm$, since the inner shell
should just exclude the Inner Local Void while retaining the Local Voids North
and South. As shown in the first panel of Fig.~\ref{lgcutsky} and in
Table~\ref{cratios}, the inner sphere in these cases has similar power in the
quadrupole.

The dipole axis appears to be principally defined by structures within
the range $30\h$ -- $62\hm$, which lies beyond the scales explicitly identified
by \Rtully{tsk08}. However, using the 2MASS survey \Rel{el06} have
reconstructed the density field in shells every $20\hm$ out to $160\hm$. To
define a dipole, rather than simply locating the largest overdensity or
underdensity, one must find an axis where the {\em integrated} density
gradient, including foregrounds, is maximized. If we compare\footnote{Note
that Ref.\ \cite{el06} places galactic longitude $\ell=0\deg$ in the centre
of their skymaps, rather than to the right edge as we do.} the results of
Sec.~\ref{dip} and \ref{winu} to Fig.~3 of Ref.\
\cite{el06} we see that the minimum Hubble variance pole coincides with the
near side of the Centaurus--Hydra Wall on one side of the sky and the maximum
Hubble variance pole coincides with the Andromeda Void on the opposite side of
the sky. Our axis to the Andromeda Void passes through the Inner Local Void and
the edge of the Local Void North\footnote{In the terminology of Ref.\
\cite{tsk08} the ``Local Void North'' comprises the region
denoted ``Delphinus'' in Fig.~3 of Ref.\ \cite{el06} together with
an adjacent large $\de<0$ area extending to just above the galactic plane,
$b\goesas 6$, with $47\deg<\ell<90\deg$. The ``Local Void South'' of
Ref.\ \cite{tsk08} is similarly much larger than the area marked ``LV'' in
Fig.~3 of Ref.\ \cite{el06} and extends to adjacent $\de<0$ regions above the
galactic plane, with $\ell<47\deg$.}.

\begin{figure*}[htb]
\vbox{\vskip10pt
\centerline{\vbox to 170pt{\halign{#\hfil\cr\scalebox{0.52
}{\includegraphics{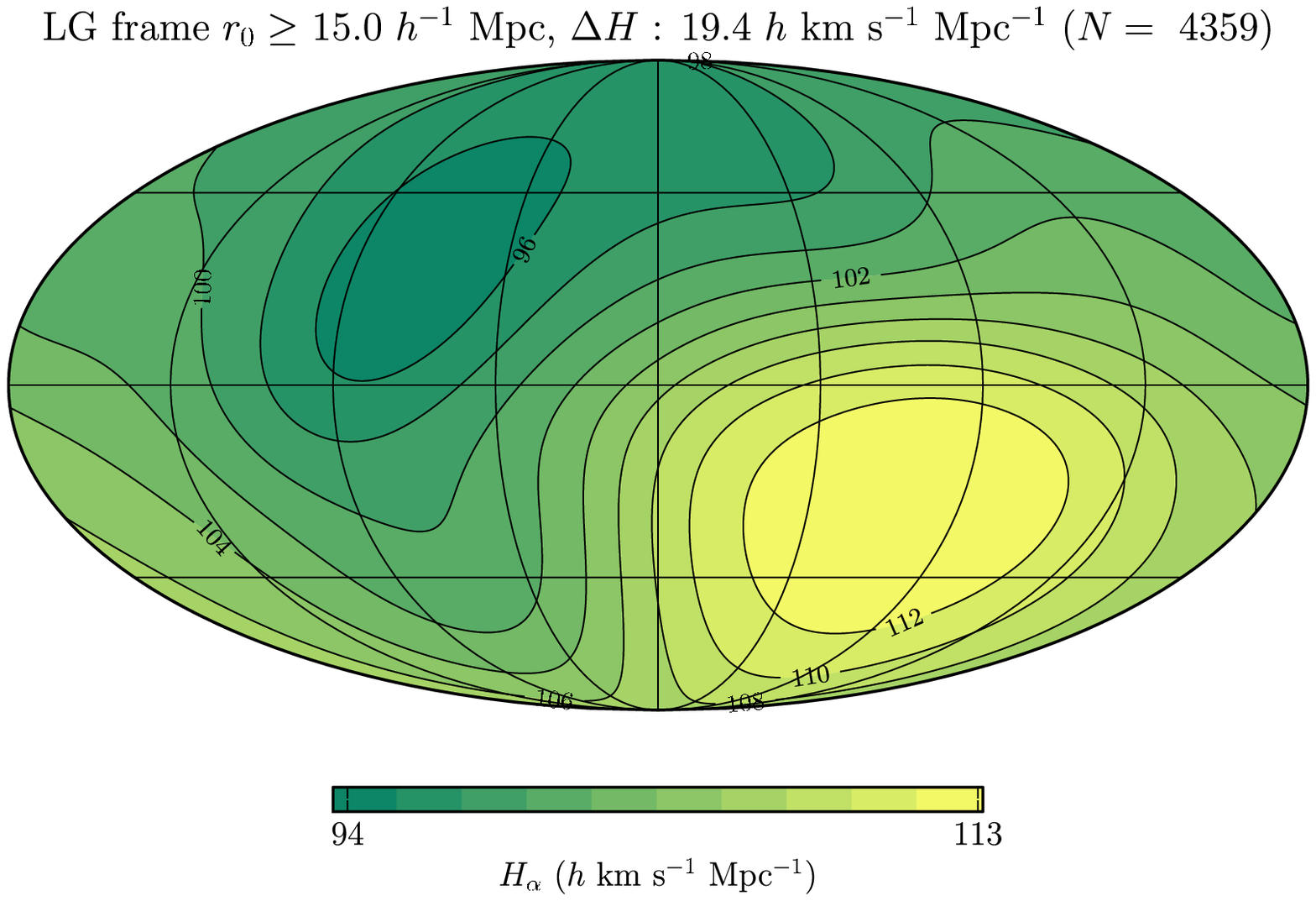}}\cr
\noalign{\vskip-30pt}\qquad{\bf(a)}\cr}\vfil}\hskip20pt
\vbox to 170pt{\vskip-3pt\halign{#\hfil\cr\scalebox{0.33
}{\includegraphics{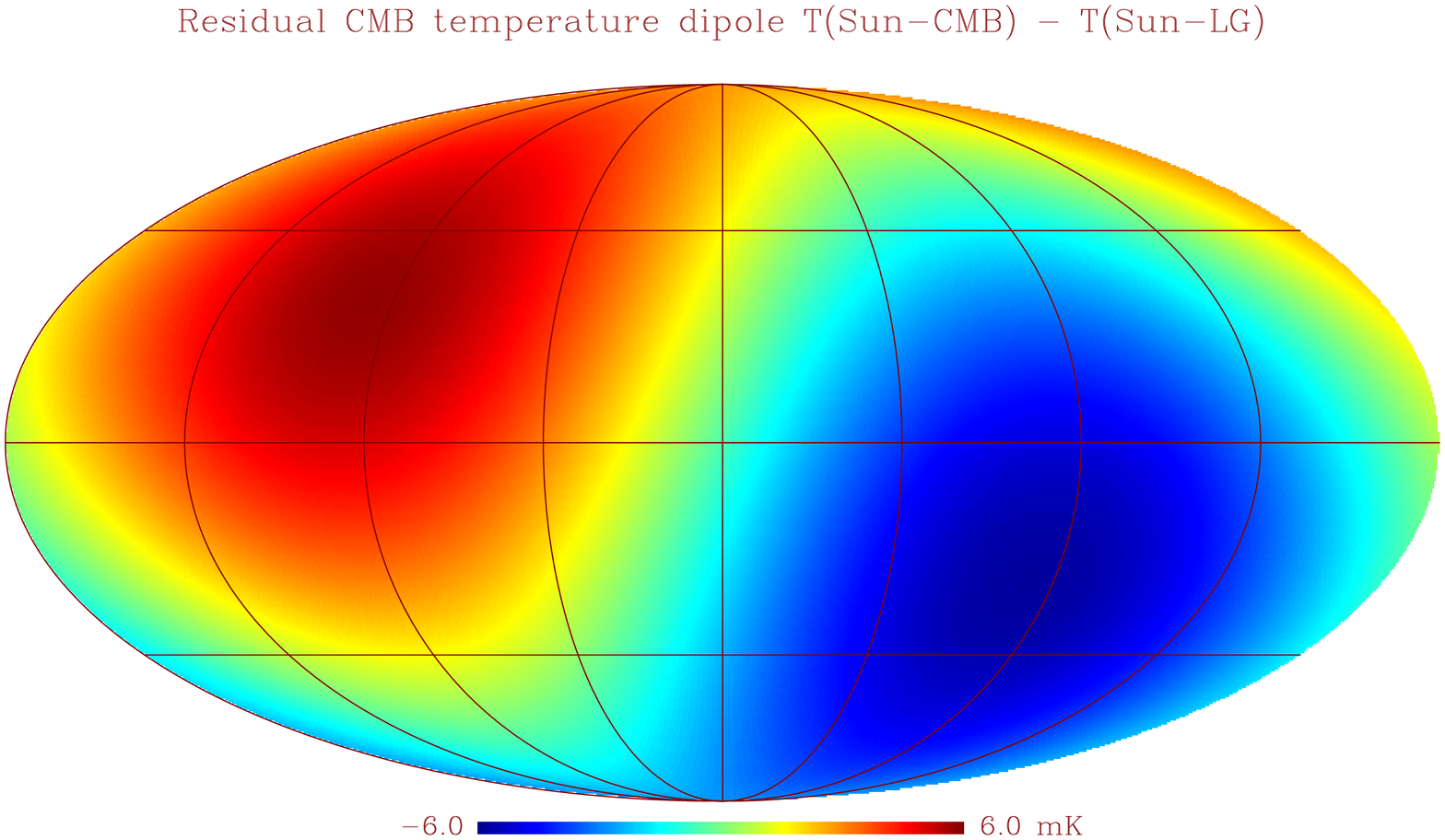}}\cr
\noalign{\vskip-3pt}\qquad{\bf(b)}\cr}}}}
\caption{LG frame Hubble flow variance map for $r>15\hm$ with IV weightings
(panel(a)) compared to residual CMB temperature dipole in the LG rest frame
(panel (b)). \latitudes\label{2dipoles}}
\end{figure*}

The centre of the Hydra Supercluster is at $r=(34.9\pm2.5)\hm$, $(\ell,b)=
(269.6\deg,26.5\deg)$, and the centre of the Centaurus Supercluster is at
$(31.5\pm2.6)\hm$, $(\ell,b)=(302.4\deg,21.6\deg)$. These radial distances are
close to the radial scale at which the CMB frame Hubble flow begins to be more
uniform than the LG one. In the next sky map plotted in Ref.~\cite{el06}, at
$r=40\hm$, the Hydra Supercluster remains very dense near the Hubble variance
angular minimum while on the opposite side of the sky the Andromeda void has
begun to close up, but an adjacent void complex Cygnus--Aquarius has opened up,
maintaining the dipole density gradient. By Fig.~5 of Ref.\ \cite{el06}
at $r=60\hm$, on the other hand, there are now large overdensities, Pegasus
and Pisces, in the angular patch that previously contained the Hubble
variance maximum, while on the opposite side of the sky in Hydra
underdensities have emerged. These opposing influences will now even
out the density gradients along the dipole axis. We can therefore
understand why the dipole diminishes beyond $r\goesas60\hm$.

As remarked above there is no significant LG frame dipole in shells 8 and
9 there is some evidence for such dipole in shell 10.
Its direction, $(\ell,b)=(348\deg\pm24\deg,-38\deg\pm14\deg)$ is at
$\goesas83\deg$, roughly orthogonal to the Shapley Concentration (SC), which
is centred in shell 10 and extends into parts of shells 9 and 11. Thus the
small LG frame dipole in shell 10 (which points from an underdense region to
Abell 576) is not correlated to the SC. Indeed in the region, $300\deg<\ell
<330\deg$, $15\deg<b<45\deg$, bounding the SC the numbers of positive and
negative peculiar velocities with respect to the shell
mean Hubble constant are equal in both shells 10 and 11, and almost equal in
shell 9. The fact that the SC does not participate in a strong dipole may be
due to significant mass concentrations on roughly the opposite side of the sky:
see Ref.\ \cite{el06}, Figs.~8, 9.

Since the quadrupole/dipole ratio is strong in shell 10, extra data is required
to isolate the quadrupole before drawing strong conclusions about the magnitude
of the dipole in this shell. However, we remark that effects on the Hubble
flow at this scale might indeed be expected if the wall--to--wall
distance--redshift is modified at the BAO scale: we are near one wall (defined
by Virgo--Centaurus--Hydra) which is separated from more distant structures
such as the SC by the $100\hm$ BAO distance. Since the BAO enhancement is
treated in the linear regime of perturbation theory, we might naturally expect
the magnitude of nearby Hubble flow variations driven by a BAO enhancement to
be significantly smaller than the ``nonlinear regime'' dipole amplitude
observed at $r\lsim55\hm$. This would also suggest that much high quality data
in the range $100\h\lsim r\lsim150\hm$ is needed to fully constrain any
potential variations.

\section{Correlation of Hubble variance and CMB dipoles}\label{corr}
Having demonstrated that Hubble flow is more uniform in the LG and LS frames
as compared to the CMB frame, and that there is a strong dipole in these
frames with an amplitude correlated to the residual monopole variations, the
natural question to ask is: to what degree is the Hubble flow variance dipole
correlated with the component of the CMB dipole that is usually attributed
to the motion of the LG?

To answer this question we must compensate for our heliocentric motion
with respect to the rest frame of the LG or LS by performing a boost to
the relevant rest frame and examine the residual CMB temperature dipole
in the rest frame in question. We create an artificial residual CMB dipole
temperature map by subtracting a boosted CMB sky with temperature
\beq T'=\frac{T\Z0}{\ga(1-(v/c)\cos\th')}\label{Tdip}\eeq
from the corresponding observed pure temperature monopole plus dipole
maps using the values of Ref.~\cite{fixsen96} assumed in Sec.~\ref{rad}. Here
$v=v\Ns{LG}$ or $v=v\Ns{LS}$ as appropriate, and $\gamma\simeq1$ since
velocities are nonrelativistic. This leaves us with a residual temperature
dipole with poles $\pm5.77\,$mK at $(\ell,b)=\left\{(96.4,-29.3),(276.4,29.3)
\right\}$ in the LG frame, and $\pm5.73\,$mK at $(\ell,b)=\left\{(90.3,-26.7),
(270.3,26.7)\right\}$ in the LS frame. The dipole amplitudes have a 6.3\%
uncertainty arising principally from the uncertainty in the heliocentric
velocity of the LG and LS frames. The LG residual temperature dipole is
shown in Fig.~\ref{2dipoles}.

We compute a correlation function directly by using HEALPIX to digitize both
the residual temperature dipole map, and also the corresponding Hubble flow
variance maps for the LG or LS frame as relevant. We quantify the correlation
between the variance of that Hubble expansion and the residual CMB temperature
dipole by the Pearson correlation coefficient
\bea
\rho\Z{HT}=&\nonumber\\ &\dsp\hskip-7mm
\frac{\sqrt{N_p}\,\sum_\al \bsa^{-2}(H_\al-\bar{H})(T_\al-\bar{T})}
{\sqrt{\left[\sum_\al \bsa^{-2}\right]\left[\sum_\al \bsa^{-2}(H_\al-\bar{H})^2
\right]\left[\sum_\al(T_\al-\bar{T})^2\right]}}\,,\nonumber\\
\eea
where $T_\al$ is the temperature in the pixel with angular coordinates $\al$,
$\bar T$ is the mean temperature,
\beq \bar{H}=\frac{\sum_{\al}^{N_p}\bsa^{-2}H_\al}
{\sum_{\al}^{N_p}\bsa^{-2}}\,\eeq
$H_\al$ is given by (\ref{Hal}),
$\bsa$ by (\ref{sdal}), and $N_p$ denotes the total
number of pixels distributed over the sky. As we are considering a pure
residual CMB temperature dipole there are no uncertainties in $T_\al$.
Since HEALPIX partitions the celestial sphere into pixels of equal area, and
since the CMB temperature dipole is assumed to be ideal, the only weighting
in the sum comes from the measurement uncertainties of the Hubble flow.

With $\sith=25\deg$ we performed a correlation analysis between the Hubble
variance dipole and the residual CMB temperature dipole in both the LG and
LS frames, as the division radius, $r_o$, between the inner and outer spheres
was varied. The results are shown in Fig.~\ref{cc}. We observe firstly
that the correlation coefficient is negative since the maximum value of
the Hubble parameter coincides with the minimum residual CMB temperature.
The strongest anticorrelation is therefore represented by those values
which are closest to $-1$.

\begin{figure}[htb]
\vbox{\centerline{\rotatebox{270}{\scalebox{0.36
}{{\includegraphics{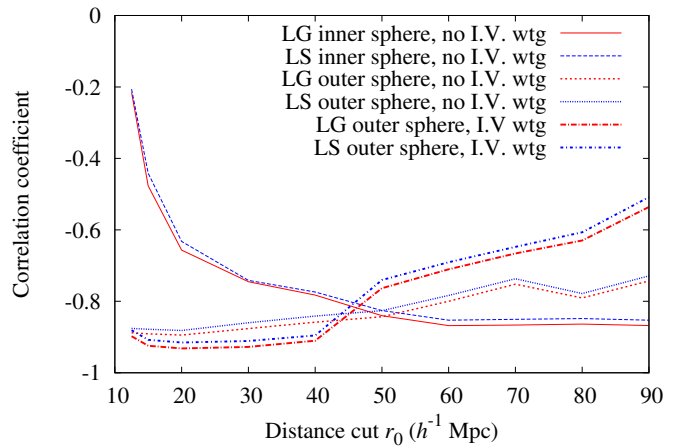}}}}}
\caption{%
Pearson correlation coefficient for the correlation of the residual CMB
temperature dipole in the LG and LS rest frames, as compared to the Hubble
variance sky map in inner ($r<r_o$) and outer ($r>r_o$) spheres, as $r_o$ is
varied in the LG and LS rest frames. The correlation is computed both with
and without the IV weighting.
\label{cc}}}
\end{figure}

\begin{table}
\begin{center}
\begin{tabular}{lrrrr}
\tableline\noalign{\smallskip}
$\sith$&LG u&LG w&LS u&LS w\\
$15\deg$&-0.8909&-0.9056&-0.8695&-0.8750\\
$20\deg$&-0.8945&-0.9197&-0.8774&-0.8965\\
$25\deg$&-0.8905&-0.9240&-0.8782&-0.9077\\
$30\deg$&-0.8847&-0.9237&-0.8769&-0.9133\\
$40\deg$&-0.8752&-0.9187&-0.8739&-0.9160\\
\tableline
\end{tabular}
\caption{Correlation coefficient between the residual CMB temperature
dipole sky map in the LG and LS rest frames, as compared to the Hubble flow
variance sky map in the $r>15\hm$ sphere, for different choices of Gaussian
smoothing angle $\sith$. Unweighted (u) and IV weighted (w) angular averaging
is considered in each case.}\end{center}
\end{table}

In all cases the result for the LS frame does not differ greatly from
that of the LG frame. However, the anticorrelation is generally a bit
stronger in the LG frame. The anticorrelation is stronger for the IV
weighted sky maps in both frames. The anticorrelation is strongest in
the outer sphere for $r_o=15\hm$. As might be expected from Sec.~\ref{avar},
the anticorrelation remains strong in the outer sky maps for
values of $r_o$ up to $40\hm$. By contrast, the anticorrelation in the inner
sphere is not at all strong for $r\le r_o$ with small values of $r_o$.
However, the anticorrelation in the inner sphere improves dramatically
as $r_o$ is increased, and by the stage that we reach $r_o=50\hm$ the
anticorrelation is comparable in both spheres for the unweighted case,
and stronger in the inner sphere than in the outer sphere for the IV
weighted case.
There is no further improvement in the anticorrelation in the inner
sphere for $r_o>60\hm$, which is again consistent with the earlier
indications that the structures responsible for the Hubble variance
dipole are within $65\hm$.

One final question is the extent to which the correlation depends on
the Gaussian smoothing width, $\sith$. We have checked this is two ways.
Firstly, we have recomputed the correlation coefficient for a range of values
of $\sith$ for the $r>15\hm$ map, the case which shows the strongest
anticorrelation. The results are shown in Table~III. We find that the
anticorrelation in the IV weighted map is stronger than the unweighted map for
different choices of $\sith$. Moreover, as well as the correlation coefficient
in the LS frame being slightly weaker, it also varies a little more
with smoothing angle. In the LG frame the correlation coefficient varies
the least for varying $\sith$ in the IV weighted case. Indeed to two
significant figures the correlation coefficient of $-0.92$ is unchanged
for $\sith=25\deg\pm5\deg$.

A second check on the relation between the CMB temperature dipole and Hubble
flow variance, that is completely independent of the Gaussian window averaging
procedure, is given by evaluating a correlation coefficient between the
residual temperature dipole and the raw \CS peculiar velocity data in
the LG frame. In this case the relevant correlation coefficient is given by
\bea \rho\Z{vT}=&\nonumber\\
&\dsp\hskip-7mm \frac{\sqrt{N}\,\sum_i\si_i^{-2}(v_i-\bar{v})
(T_i-\bar{T})}{\sqrt{\left[\sum_i\si_i^{-2}\right]\left[\sum_i\si_i^{-2}
(v_i-\bar{v})^2\right]\left[\sum_i(T_i-\bar{T})^2\right]}}~.\nonumber\\
\eea
where $v_i$ denote the peculiar velocities and $N$ is the number of data
points. The weighted average peculiar velocity should approach zero for a
large number of data points: here $\bar{v}=-64.9\kms$ with a standard
deviation of $722.4\kms$. For the \CS LG-frame velocities from the data
with $r\geq15\hm$, $N=4359$ and we obtain $\rho\Z{vT}=-0.35$. The magnitude
of this correlation is naturally lower than it is for the weighted grid data
which we calculated above due to the scatter in these data, but the well
defined number of points implies we have better statistical tools to quantify
our confidence that the correlation is indeed nonzero. We test this by
evaluating the variable
\beq t= \rho\Z{vT} \frac{\sqrt{\nu}}{\sqrt{1-\rho\Z{vT}^2}}~,
\label{ttest}\eeq
where $\nu=N-2$ is the number of degrees of freedom. If there is no
correlation, the test variable $t$ in (\ref{ttest}) should follow the
standard normal distribution $N(0,1)$. For $\nu=4357$ and $\rho\Z{vT}=-0.346$,
we obtain the value $t=-24.35$, i.e., a deviation of more than $24\sigma$.
This is extremely strong statistical evidence for a nonzero (anti)correlation.

\begin{figure*}[htb]
\vbox{\vskip10pt
\centerline{\vbox to 170pt{\vfill\halign{#\hfil\cr\scalebox{0.52
}{\includegraphics{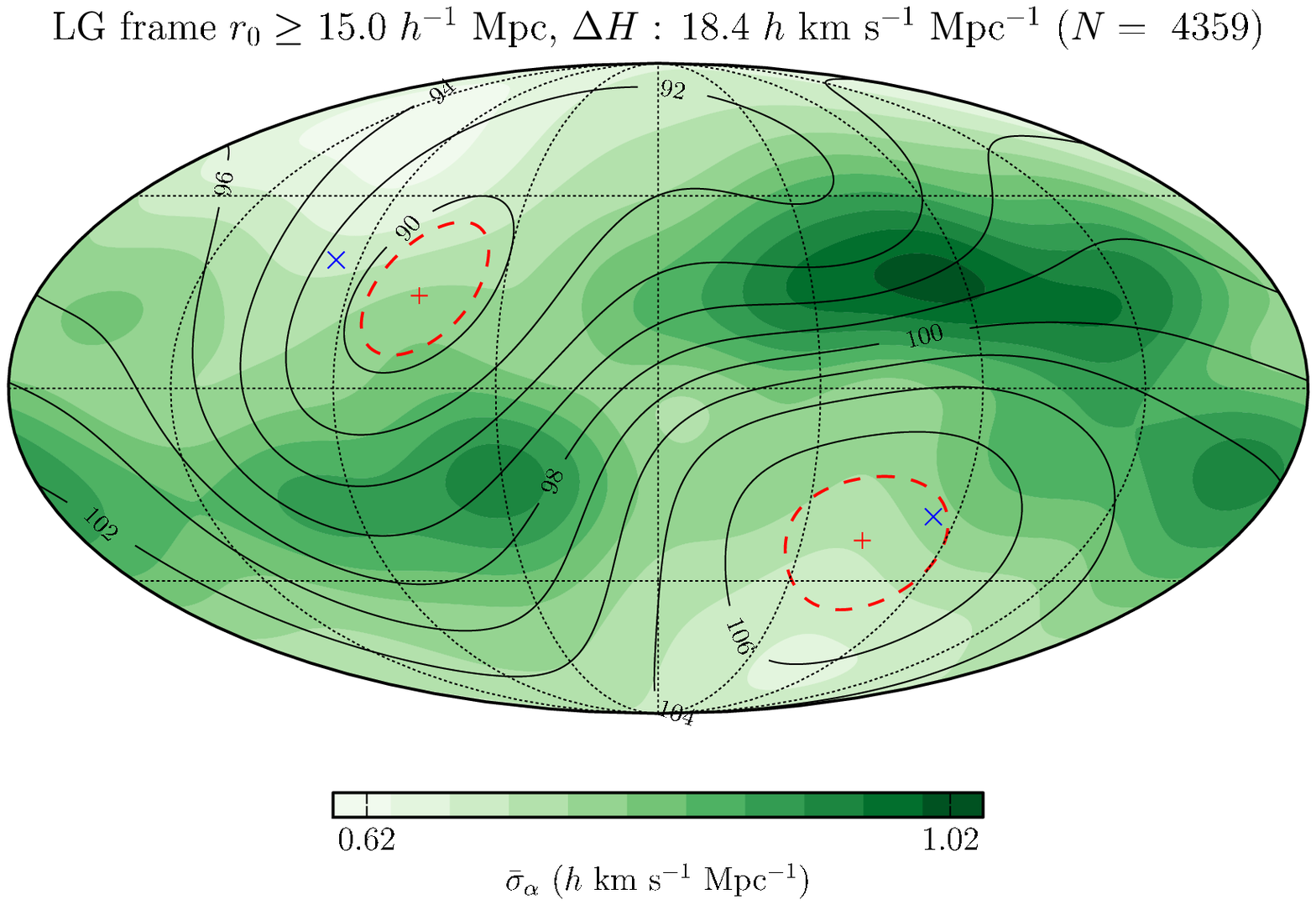}}\cr
\noalign{\vskip-30pt}\qquad{\bf(a)}\cr}}\hskip31pt
\vbox to170pt{\vfill\halign{#\hfil\cr\scalebox{0.52
}{\includegraphics{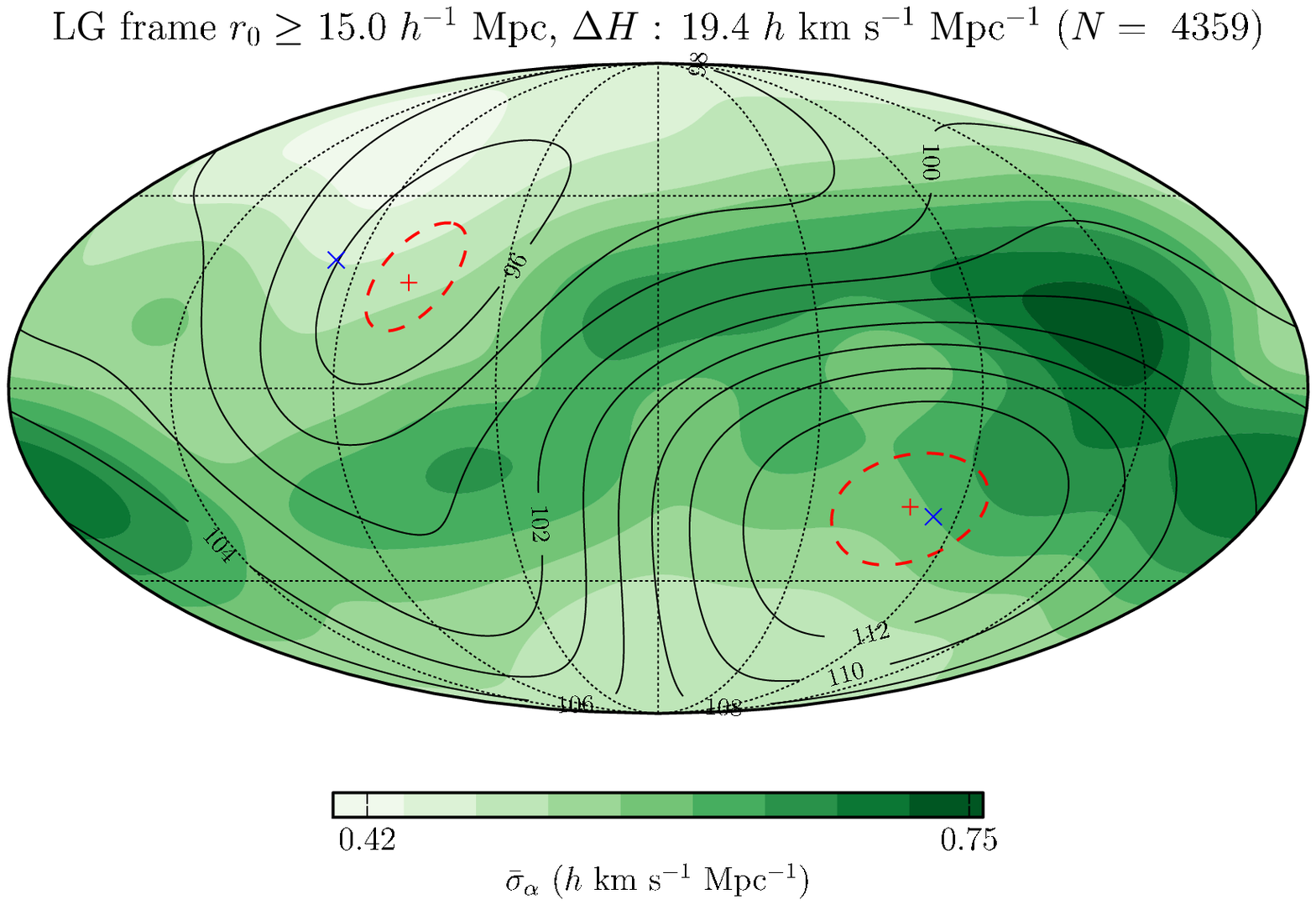}}\cr
\noalign{\vskip-30pt}\qquad{\bf(b)}\cr}}}}\vskip22pt
\caption{%
Angular variance of the LG rest frame Hubble flow given by $\Hth$ in the
$r>15\hm$ range (solid lines), overlaid with angular uncertainties $\bsa$
(colour map contours): (a) with no IV weights; (b) with IV weights (\ref{Wi2}).
The red dashed circle indicates a 1$\si$ region around the maximum/minimum.
Blue crosses indicate the residual CMB temperature dipole poles. \latitudes
\label{errsky}}
\end{figure*}

We therefore have strong evidence that the dipole features of the two maps
in Fig.~\ref{2dipoles} are related. From
Fig.~\ref{errsky} we see that in the IV weighted sky map the cooler
residual CMB temperature pole (marked by a cross) lies just $7.4\deg$ from
the maximum of the Hubble flow variance, well within the 1$\si$ contour.
The hotter residual CMB temperature pole is separated by $22\deg$ from the
minimum of the Hubble flow variance, however, and lies $10\deg$ outside
the 1$\si$ contour but is within $3\si$. It remains to be seen whether
the uncertainty estimates in this case are significantly affected by
the choice of weighting scheme. In particular, the left hand panel of
Fig.~1 of FWH10 shows that with a maximum likelihood estimate based
on traditional IV weightings there are a substantial number of very strongly
weighted data points in the \CS sample in the region which coincides
with that of minimum $\bsa$ to the north of residual CMB temperature pole in
Fig.~\ref{errsky}(b). With the alternative minimum variance weightings
shown in the right hand panel of Fig.~1 of FWH10 the same data
points are not strongly weighted.

\section{Origin of the CMB dipole}\label{origin}
Ever since the first bounds were placed on the anisotropy of the CMB in the
mid 1960s \cite{pw67} it has been assumed that the dipole anisotropy
represents a measurement of our motion with respect to our surface of average
homogeneity \cite{ss67,pw68}. Our results clearly show, however, that for
spherical (monopole) averages the Hubble flow is closer to being uniform in
the frame of the LG or LS, rather than the frame indicated by treating the
CMB dipole as being entirely due to a boost. This is completely unexpected
in the standard framework, since the cosmic rest frame and the frame of
minimum Hubble flow variance should be one and the same. Moreover, we
find a dipole in the LG frame whose amplitude diminishes in
correlation to the diminishing of the monopole variations. This dipole is
strongly correlated with the residual CMB temperature dipole.

\subsection{Puzzles of the bulk flow formalism}
If we set aside the question of the monopole variation and its relative
uniformity in different rest frames, then a dipole is of course expected in
the standard peculiar velocity framework. Indeed, in the standard framework the
dipole pattern we observe was detected by \Rjrk{jrk} in a sample of 69
MLCS2k2-reduced SneIa distances in the range $0.005\le z\le0.025$, and
interpreted by them as a positive detection of the motion of the LG with
magnitude $541\pm75\kms$ towards $(\ell,b)=(258\deg\pm18\deg,51\deg\pm12\deg)$,
which is $2\si$ consistent with the amplitude and magnitude of the
boost of the LG with respect to the CMB frame. The problem was re-examined by
\Rgls{gls} using a sample of 61 SALT-reduced SneIa in the range $0.0076\le z
\le0.124$, who found $v\Ns{LG}=697\pm137$ in the direction $(\ell,b)=(220\deg
\pm14\deg,32\deg\pm11\deg)$. When the linear theory was corrected to account
for correlated peculiar velocities, the uncertainties on these values
were increased giving $v\Ns{LG}=690\pm201\kmsMpc$ towards $(\ell,b)=(257\deg
\pm24\deg,29\deg\pm16\deg)$ \cite{gls}, which is 1$\si$ consistent
with the LG boost with respect to the CMB frame.

Phenomenologically, the above results in the linear theory show agreement with
the expected LG boost, and the SneIa sample of Ref.~\cite{jrk} appears to show
convergence to the expected result within the same radial scale we find in the
\CS sample: in Sec.~\ref{rad} we saw that the LG frame is within
1.36$\si$ of uniform in spherical shells with $\mr_s\ge69\hm$, and in
Sec.~\ref{dip} we saw that the large dipole feature in the LG frame
diminished to its residual value by $r_o=55\hm$. Thus the structures
responsible for both the monopole and the dipole variation in the Hubble
flow appear to be foregrounds within roughly\footnote{There is some
ambiguity in defining this transition scale. Conservatively, we take
the scale to be the average of the mean distances of shells $5'$ and $6$,
where $\de H_s$ first drops below 1.5$\si$ difference from uniform.}
$65\hm$, or $z\simeq0.022$.

In the standard peculiar velocity framework, however, demonstrating the
convergence of bulk flows on this scale has proved challenging. In this
framework peculiar velocities are estimated from linearly perturbed FLRW model
according to
\cite{peebles93}
\beq {\mathbf v}({\mathbf r})={\Hm\OmMn^{0.55}\over4\pi}\int\dd^3{\mathbf r'}\,
\delta_m({\mathbf r'})\,{({\mathbf r'}-{\mathbf r})\over|{\mathbf r'}-
{\mathbf r}|^3}\label{vlin}\eeq
where $\delta_m({\mathbf r})=(\rho-\bar\rho)/\bar\rho$ is the density contrast,
and the power $0.55$ of the matter density parameter $\OmMn$, rather than
$0.6$, gives improved accuracy for models with dark energy \cite{linder05}.

The search for convergence of bulk flows within this framework has a three
decade history, which is summarized by \Rltmc{ltmc} and by \Rbcmj{bcmj}.
Contrary to earlier investigations \cite{el06} \Rltmc{ltmc} failed to find
convergence in the 2MASS survey on scales up to
120$\hm$: less than half the amplitude was generated on scales $40\hm$, and
whereas most of the amplitude was generated within 120$\hm$ the direction did
not agree. \Rbcmj{bcmj} analysed a larger sample in the 2MASS survey using a
different methodology and failed to find convergence within 150$\hm$.

In the \CS sample \Rwfh{wfh09} failed to see convergence of peculiar
velocities to the CMB dipole on scales of $50\hm$. Even more puzzling in their
results is the suggestion that the bulk flow actually increases with increasing
scale above $20\hm$ (see their Fig.~5). In the peculiar velocity framework,
the only way to understand how a larger volume can have a larger bulk flow
than a smaller volume contained within it is to posit that the inner volume
has an additional compensating motion in the opposite direction. While
possible, this arrangement seems unexpected at best. Furthermore, it offers
no explanation for the surprising degree of uniformity of the spherically
averaged Hubble law in the LG frame found in Sec.~\ref{rad}.

\subsection{Differential expansion of space as a foreground anisotropy}
The problems and puzzles of the standard linear theory alone might not
justify a radical revision of the standard formalism. However, our new results
concerning the monopole Hubble flow variations, their radial correlation with
change of amplitude of the dipole variations, and the relative uniformity of
the Hubble flow in the two frames, all defy explanation in the standard
framework. Theoretically therefore the natural course of action is to go back
to first principles.

Another potential source of the dipole anisotropy has been previously
considered \cite{mm96}, namely the Rees--Sciama
effect \cite{rs68} (the nonlinear version of the integrated Sachs--Wolfe
effect). However, to generate a dipole anisotropy of magnitude $\de T/T
\goesas10^{-3}$ in this fashion would require a void of radius $300\hm$
-- an order of magnitude larger than than that of the observed voids --
and furthermore it would also induce a quadrupole of the same magnitude as
the dipole \cite{mm96}. This effect therefore fails as a potential
explanation.

The Rees-Sciama effect and other late-epoch sources of CMB anisotropies
such as the Sunyaev--Zel'dovich effect are conventionally considered as the
effects of inhomogeneities embedded in a background FLRW cosmology that
applies at all scales. However, from first principles, differences in the
observed mean CMB temperature will be also generated if the universe has grown
sufficiently inhomogeneous by the present epoch that the distance to the
surface of photon decoupling is slightly different in different directions.
Over large distances photon paths see an average of all the structures,
but the last section of a photon journey below the scale of statistical
homogeneity will be influenced by the peculiar foregrounds.

Our universe is in fact sufficiently inhomogeneous on scales $\lsim100\hm$
that the differential expansion of void and wall regions can be expected
to produce differences in the distance to the surface of last scattering,
if we are abandon the assumption that an exact FLRW description must apply at
all scales. Our proposal is potentially consistent with a variety of
non-FLRW models, such as the timescape scenario \cite{clocks,sol}, an
alternative to the standard cosmology which has remained viable in tests to
date \cite{obs,lnw,sw,grb,dnw}. We do not wish to focus on any particular
model-dependent estimates here, but rather to point out that the global FLRW
geometry is a very special one which encourages conceptual simplifications
that are not demanded by full general relativity. We will determine
the magnitude of the effect required to produce the observed residual
temperature dipole using the \LCDM\ model phenomenologically for
distance estimates.

As long as the average evolution of the universe can be described by an
average cosmic scale factor\footnote{It is not necessary for the average
evolution to obey the Friedmann equation for this statement to be true.
In particular, it also applies to the timescape cosmology \cite{clocks,sol,%
obs} which describes the average evolution of an ensemble of void and wall
regions in the Buchert averaging scheme \cite{buch00,buch08}, with additional
physical assumptions about the interpretation of physical observables.}
which is related to the observed cosmological redshift by $a\Z0/a=1+z$, and
which is inversely proportional to the mean CMB temperature, $T\propto1/a$,
then a small change, $\de z$, in the redshift of
the surface of photon decoupling -- due to foreground structures -- will
induce a CMB temperature increment $T=T\Z0+\de T$, where $T\Z0=2.725\,$K
is the present epoch mean temperature and
\beq {\de T\over T\Z0}={-\de z\over1+z\ns{dec}}\,,\label{delT}\eeq
$z\ns{dec}=1089$ being the mean redshift of photon decoupling. For the LG
residual dipole the increment $\de T=\pm(5.77\pm0.36)\,$mK represents a
redshift increment $\de z=\mp(2.31\pm0.15)$.

The comoving distance of the surface of photon decoupling is given by
\beq D={c\over\Hm}\int_1^{1+z\ns{dec}}{\dd x\over\sqrt{\OmLn+\OmMn x^3+\OmRn
x^4}}\eeq
in the standard spatially flat \LCDM\ model, where $\OmLn=1-\OmMn-\OmRn$ and
$\OmRn=4.15h^{-2}\times10^{-5}$. If we take $\OmMn=0.25$ and $h=0.72$ we
find that the comoving distance increment of $\de D=\mp(0.33\pm0.02)\hm$ is
what is required to generate the residual CMB dipole in the LG frame. For
$\OmMn=0.30$ the value is slightly reduced to $\de D=\mp(0.32\pm0.02)\hm$. For
the timescape model \cite{obs} the value is similar, with possible small
differences depending on parameter values.

The results of the previous sections suggest that the structures responsible
for the Hubble flow variance dipole lie at most within 65$\hm$. Since the
differences in the distance to the surface of last scattering occur
effectively at $z=0$, a 0.35$\hm$ difference in distance therefore would amount
to a maximum 0.5\% difference on these scales. Even if the whole difference was
taken up within an average distance scale of $30\hm$, leading to a 1\%
effect, this is still within the regime of plausibility given the degree of
Hubble flow variance we observe in the \CS sample.

Our picture then, as in Fig.~\ref{Hshells}, is one of differential expansion of
void and wall regions at late epochs leading to distance differences of up to
the order of 1\% between walls and voids on $30\h$ -- $70\hm$ average distance
scales, a scale determined by the size of the largest typical nonlinear
structures \cite{hv02,hv04} and their random correlations. While such
differences are not isolated to our own immediate vicinity, when light travels
over scales larger than the scale of statistical homogeneity the differences
generally average out on any typical line of sight. It is on the last stretch
of the journey, when the typical nonlinear structures subtend a large angle on
the sky, that the particular foreground inhomogeneities peculiar to our own
location give a strong net anisotropic contribution\footnote{The BAO scale is
close to the statistical homogeneity scale, near $\goesas110\hm$, and
therefore we may also expect a small anisotropic enhancement to the Hubble flow
near the BAO scale. Since this is a linear perturbation the amplitude of the
anisotropy is likely to be smaller. Furthermore, once one considers larger
shells then the angles subtended by typical nonlinear structures are smaller
relative to the centre (see Fig.~\ref{Hshells}), so multipoles higher than
the dipole are likely to come into play.}. In our case the largest foreground
density gradient defines an axis with the void direction yielding a slightly
larger distance than average and a net CMB temperature decrement, and the
opposite wall direction a slightly smaller distance than average and a CMB
temperature increment.

Our proposal for an alternative origin for part of the CMB dipole should only
have a very significant effect on the angular power spectrum on scales larger
than $15\deg$, on which anomalies are seen. High precision measurements of
the acoustic peaks \cite{keisler11,Pparm} on scales of less than $1\deg$
will be influenced in subtle ways, as discussed in Sec.~\ref{kin} below.

\subsection{Ray-tracing estimates of dipole and quadrupole strengths}
One important question remains: why is the CMB dipole so large
compared to other multipoles if a substantial contribution is due to a
foreground anisotropy in the distance--redshift relation? In the case of the
Rees--Sciama effect, for example, the quadrupole is comparable to the dipole
\cite{mm96}. The Rees--Sciama effect for structures on scales $\lsim100\hm$
is estimated to be $\de T/T\goesas10^{-7}$--$10^{-6}$ \cite{mdwse,brs},
and is a secondary effect compared to the one we propose. Nonetheless, the
relative strength of the observed CMB multipoles must be understood if our
proposal is to be a viable explanation for the unexpected results concerning
the uniformity of the Hubble flow.

The problem of describing the propagation of light through a realistic
inhomogeneous structure is a complicated one. However, the
plausibility of our proposal is readily demonstrated by adapting results
\cite{aa06} found in previous studies of ray tracing of the CMB sky as seen
by an off-centre observer in a Lema\^{\i}tre--Tolman--Bondi (LTB) void.
Although past investigations focused on large voids as toy models for
explaining luminosity distances of supernovae without dark energy
\cite{aa06,gp11}, the ray-tracing results can also be applied to voids of any
scale, and in particular to those of the size of observed local structures
\cite{el06}.

\Raa{aa06} showed that a Newtonian approximation, with an effective peculiar
velocity
\beq{v\ns p\over c}={(h\ns{in}-h\ns{out})d\ns{off}\over
2998\w{Mpc}}\label{aa1}\eeq
where $H\ns{in$\,$0}=100\,h\ns{in}\kmsMpc$ and $H\ns{out$\,$0}=100\,h\ns{out}
\kmsMpc$ are the Hubble constants inside and outside the void, and $d\ns{off}$
is the distance of the observer from the centre in Mpc, yields results
for CMB multipoles which are numerically close to the ray-traced result.
In this approximation the ratio of the quadrupole coefficient, $a\Z{20}$, to
dipole coefficient, $a\Z{10}$, is
\beq
{a\Z{20}\over a\Z{10}}=\sqrt{4\over15}{(h\ns{in}-h\ns{out})d\ns{off}\over
2998\w{Mpc}}\,.\label{aa2}
\eeq
Our position in the thin filamentary Local Sheet on the edge of the
Local Void \cite{tsk08} complex, at a distance $\gsim12\hm$ from the
nearest wall region (Virgo Cluster), means that our location is sufficiently
void-like for the approximation (\ref{aa1}), (\ref{aa2}) to be reasonable.

A value of $H\ns{in$\,$0}-H\ns{out$\,$0}$ can be estimated from the magnitude
of the dipole in the LG frame from the first 4 shells in Table~\ref{dshell},
which correspond to the range over which the LG frame dipole is strong before
rapidly decreasing. The weighted mean is $\be=16.1\,h\kmsMpc$ in the unprimed
shells 1--4 , and $\be=14.1\,h\kmsMpc$ in the primed shells 1'--4', so that
$H\ns{in$\,$0}-H\ns{out$\,$0}\simeq\be=(15.1\pm1.0)h\kmsMpc$. To reproduce the
effective peculiar velocity $v\ns p=635\pm38\kms$ inferred for Local
Group\footnote{Numerical values in Ref.\ \cite{aa06} differ by roughly a factor
two, as they neglected to subtract the contributions of the known motions
of the sun and galaxy to the CMB dipole.} \cite{tsk08}, eq.~(\ref{aa1})
therefore requires us to be $d\ns{off}=(42\pm3)\hm$ from the centre of a void.
As seen in Fig.~\ref{sdipfit} this does indeed match to the scale at which the
dipole magnitude $\be$ decreases rapidly in the LG frame, demonstrating the
consistency of the approximation (\ref{aa1}).

Using (\ref{aa1}), (\ref{aa2}) we find that for the values of $\be$ and $v_p$
we find $a\Z{20}/a\Z{10}=0.0011\pm0.0002$. A quadrupole/dipole ratio $\goesas
0.1$\% for relative CMB anisotropies is certainly within observational bounds.
In fact, the Newtonian approximation (\ref{aa2}) is known to overestimate the
quadrupole by a factor of 2 \cite{aa06}. This is confirmed in explicit ray
tracing simulations in an LTB model with structures of the scales above. In
particular, using the parametrization of Ref.~\cite{bw09} (model 1), with
parameters adjusted so that $\be=15.1h\kmsMpc$ and $d\ns{off}=42\hm$ within a
void of radius $54\hm$ embedded in a background FLRW model with $\OmMn=0.3=1-
\OmLn$, one finds $a\Z{20}/a\Z{10}\approx0.0005$ \cite{KB}.

The actual matter distribution is of course more complicated than that of a
single LTB void in a homogeneous background, and the problem of determining the
average propagation of CMB photons through the foregrounds is closely related
to how to realistically average foreground density fields. Work is in progress
to trace rays through exact solutions of Einstein's equations which more
closely emulate our peculiar density foregrounds \cite{bnw}. The nonsymmetric
Szekeres models \cite{Sz} enable one to model the density gradient provided by
a void and adjacent cluster, with the parametrization of Sec.~III$\,$C of
Ref.~\cite{bc10}. For example, by placing an observer $12\hm$ from the centre
of a $27\hm$ radius void and at a distance of $15\hm$ from an overdensity, one
finds $a\Z{20}/a\Z{10}\approx0.01$ \cite{KB}, which is larger than in the LTB
case but still observationally acceptable. (Reducing this configuration to
spherical symmetry \cite{b09,sb12}, decreases the ratio to $0.0001$,
which is within a factor of 2 agreement with Eq.\ (\ref{aa2}).)
More detailed results of this ray-tracing analysis will be
presented in a forthcoming article \cite{bnw}.

\section{Discussion}\label{dis}
In summary, we have shown with decisive Bayesian evidence that when averaged
in spherical shells on scales $\lsim150\hm$ the Hubble flow is more uniform
in the rest frame of the LG or LS than in the standard ``rest frame'' of the
CMB. An exception occurs for shells in the range $40\h\lsim r\lsim60\hm$,
where the boost to the CMB frame ``almost works''. This is echoed in
the dipole variation: the CMB frame dipole in shell 4 at mean distance $\mr\Z4
=44.5\hm$ is not statistically significant, but subsequently increases at
larger radial distances. By contrast the LG frame dipole diminishes greatly,
becoming consistent with zero in shells 8 and 9. A significant CMB frame dipole
remains in all outer shells, its direction being consistent with previous
peculiar velocity studies. While a smaller magnitude LG frame
dipole reappears in shell 10 in the range $110\h\lsim r\lsim150\hm$, its
direction is different to the inner dipole, and unlike the inner dipole is
not correlated with the residual CMB temperature dipole in the LG frame.

These results are difficult to reconcile with the standard kinematic
interpretation of the Local Group moving in response to the gravitational
attraction of the clustering dipole. However, they are consistent with a
foreground different expansion of space of order $0.5$\% due to the density
gradients of nonlinear structures on scales $30\h\lsim r\lsim 62\hm$, which
we have tentatively identified in Sec.~\ref{align}. Such a
foreground would also affect measurements on large scales
$\gg100\hm$ as the typical distance for a given redshift would vary from the
average distance by up to $0.35\hm$, with a roughly dipolar distribution
on the sky.

This suggestion challenges a basic assumption of observational cosmology,
and if upheld by future investigations, will clearly have important
consequences. From general relativity there is no {\em a priori} reason for
assuming that space is flat with a simple Doppler law expansion on scales
$\lsim150\hm$.
Nonetheless this assumption is so firmly embedded in much of the practice of
observational cosmology that it is nontrivial to disentangle the consequences
of revisiting this assumption. Here we will sketch just some of the directions
which should be pursued in more detailed investigations.

\subsection{From bulk flows to Hubble flow variance}\label{formal}
A differential expansion of space arising from the differing histories of
regions of different density may simply lead to an alternative interpretation
of many of the phenomenological results of the peculiar velocity
framework. For example, redshift space distortions are well understood in
terms of the Kaiser effect \cite{kaiser87}. Before tackling the broader
treatment of redshift space distortions, we need to begin by understanding how
convergence of the nearby Hubble flow to the CMB dipole should be defined.

The fractional dipole anisotropy, typically of order $\be/H_d\goesas9$\% for
shells in the range $30\h\lsim\mr_s\lsim62\hm$ (c.f.~Fig.~\ref{sdipfit}), is
much larger than the $0.5$\% differential expansion required on these scales
to produce the CMB dipole. However, these values of $\be/H_d$ also include
the monopole variation (c.f.~Fig.~\ref{dHs}), which first needs to be factored
out before additional angular variation is considered.

A starting point for a multipole expansion of the Hubble flow variation would
be to define the average comoving distance, $D$, to a redshift, $z<\zhom$,
within the scale of statistical homogeneity, $\zhom$, according to
\beq D(z)=c\int_0^z{\dd z_s\over H_s(z_s)}\,.
\label{D0}\eeq
Here $H_s(z_s)$ is computed exactly as in (\ref{Hs}) except that the
shells on which the linear regression is performed are chosen by
redshift ranges, $z_s<z\le z_s+\si_z$, where $\si_z$ represents the
width of the radial shells in redshift. For example, in Sec.~\ref{rad} we
chose shells of radial width $12.5\hm$, which corresponds to taking
$c\si_z=1250\kms$, or $\si_z=0.0042$. The average luminosity and angular
diameter distances are related to $D(z)$ by $D\Z L=(1+z)D=(1+z)^2 D\Z A$.

The radial shell width, $\si_z$, is analogous to the angular smoothing width,
$\sith$, of Sec.~\ref{avar}. The minimum shell width possible is set by
the largest bound structures that exist, since a regression
(\ref{Hs}) can only be calculated on scales over which space is expanding
and a Hubble law is defined. Thus $c\si_z/\Hm$ should be larger than the
diameter of the largest rich clusters of galaxies, which justifies the
choice made in Sec.~\ref{rad}. There will be similar restrictions on
the choice of smoothing angle, $\sith$, depending on the details of
angular averaging procedure.

For each shell redshift, $z_s$,
the angular corrections $H(z_s,\th,\ph)-H(z_s)$ will lead to corrections
to the mean comoving distance (\ref{D0}) which might be
expanded as multipoles. Eq.~(\ref{D0}) defines
the monopole contribution to the distance--redshift relation. Convergence
of the Hubble flow variance in a large data set would then be obtained
if the dipole anisotropy converges to a fixed value for $z>z\ns{conv}$, where
our preliminary investigations suggest that the convergence scale is at least
of order $z\ns{conv}\simeq0.022$. To consistently account for the residual
CMB temperature dipole the residual anisotropy in $D(z)$ would be up to the
order of $\pm0.35\hm$, with the exact value depending on the cosmological
model.

A further question to be resolved by future surveys is the split between the
linear and nonlinear regimes in cosmology. In particular, the BAO enhancement
is assumed to be in the linear
regime of perturbation theory about a FLRW model in the standard model.
Our results are tentatively consistent with the possibility that convergence
of Hubble flow variance by $z\ns{conv}\simeq0.022$, apart from very small
features at $r\ns{bao}\pm r\ns{off}$, where $r\ns{bao}$ is the effective
comoving BAO scale and $r\ns{off}$ is the distance by which we are offset from
the centre of the nearest (Virgo--Centaurus--Hydra) wall. Much more data
is required in the range $100\h\lsim r\lsim150\hm$ to confirm this. If
confirmed, it would be consistent with the notion that scales $z\lsim z\ns
{conv}$ are in the ``nonlinear regime'' while the BAO scale is in the ``linear
regime''\footnote{Such an interpretation just relies on there
existing a scale of statistical homogeneity above which average cosmological
evolution can be described. It is not necessary for the average evolution to
be exactly that of a homogeneous isotropic FLRW model.}.

\subsection{The minimum Hubble variance rest frame}
The multipole expansion of the Hubble flow variance proposed in
Sec.~\ref{formal} should ideally be performed in the rest frame in which
the radial variance in the Hubble flow with respect to the asymptotic
global average $\bar\Hm$ is minimized. We have shown that this frame is closer
to the LG rest frame than the CMB rest frame. The flow in LS frame is very
close to that of the LG frame, but very slightly more variable.

We should also consider the possibility that the LG has an additional
peculiar velocity with respect to the frame in which variance in the Hubble
flow is minimized. Such a task is feasible, even if computationally
intensive, and we will report on this in future work.
A further comparison is to independently determine the frame with the
greatest anticorrelation between the residual CMB temperature dipole
and the Hubble flow variance dipole. Does such a frame agree with the
minimum Hubble variance frame, within uncertainties?

\subsection{The Hubble bubble and type Ia supernova systematics}
Type Ia supernovae (SneIa) provide the standardizable candles which are the
cornerstone of many current cosmological tests. The use of SneIa is
currently limited by systematic uncertainties, and differences in
cosmological parameter estimations can be obtained when different light
curve reduction methods are used. In the SALT/SALT--II method \cite{guy05,%
guy07} empirical light curve parameters are marginalized together with
cosmological parameters over the whole data set, whereas in the
MLCS2k2 method~\cite{jrk} template light curves are determined by
minimizing the distance modulus residuals of a training set of nearby SneIa,
which lie within the range in which the Hubble flow is linear, yet are
sufficiently distant for peculiar velocities to be negligible compared
to the Hubble-flow $cz$.

If the cosmic rest frame is taken to be that of minimum Hubble flow variance
on $\lsim100\hm$ scales, and if such a frame is close to the LG frame, then an
interesting systematic issue arises. In both light curve calibration methods
one seeks to minimize the distance modulus residuals with respect to a nearby
global linear Hubble law, and by convention such a minimization
in the rest frame of the CMB rather than the LG frame. The Union
\cite{union}, Constitution \cite{hicken09} and Union2 samples \cite{union2}
contain a significant number of data points in the range\footnote{By contrast
\Rkessler{kessler09}, for their full MLCS2k2 Nearby+SDSS+SNLS+ESSENCE+HST
sample, took a minimum redshift of $z=0.0218$. There are differences in
cosmological parameters estimated from the SDSS sample \cite{kessler09} and
the Union, Constitution and Union2 samples.}
$0.015\lsim z\lsim0.02$ which is below the scale $z\ns{conv}$
but is still conventionally deemed to be ``within the Hubble flow''.

Interestingly, the redshift range $0.012\lsim z\lsim0.02$ corresponds to the
range $36\h\lsim r\lsim60\hm$ over which the boost to the CMB rest frame was
found to produce a smaller deviation from a uniform Hubble flow than in the
LG frame (c.f.~Fig.~\ref{dHs}), even though the Hubble flow is
significantly more uniform in the LG frame overall. Thus the
fact that the boost to the CMB rest frame appears to best compensate for
structures in the range $30\h\lsim r\lsim62\hm$ may have led to a
misidentification of the minimum redshift, $z\ns{conv}$, at which a single
global linear Hubble law can be safely assumed.

Fig.~\ref{dHs} and Table~\ref{shell} indicate that in the LG frame
convergence to an almost uniform Hubble flow is achieved by $\mr\simeq65\hm$
or $z\ns{conv}\simeq0.022$. This scale coincides roughly with the cutoff
scale of the Hubble bubble identified in the supernovae data by \Rzrkd{zrkd98}
at $z=0.24$, and confirmed by \Rjrk{jrk}, using a MLCS2k2 sample with a
reddening parameter $\RV=3.1$.
We note that over the range $60\h\lsim r\lsim70\hm$ the Hubble flow is somewhat
closer to uniform in the LG frame as opposed to the CMB frame, and the variance
in these shells in either frame is less than the $6.5\pm2.2$\% found in Ref.\
\cite{zrkd98}. However, Refs.~\cite{jrk,zrkd98} worked with a far
simpler model of Hubble flow variance in which the sample was divided into
inner and outer spheres.

We have checked the result of \Rzrkd{zrkd98} by repeating their analysis
on the \CS sample, fitting a simple linear Hubble law
to the 2222 data points in the interval $30\h<r\le 70\hm$ chosen for the
inner shell in Ref.\ \cite{zrkd98}. We find $\Hm=(104.5\pm0.6)
\,h\kmsMpc$ in the CMB frame, and $\Hm=(105.1\pm0.6)\,h\kmsMpc$ in the LG
frame. These values are respectively $4.40\pm0.08$\% and $4.06\pm0.07$\% larger
than the global asymptotic values of $\bH\Z0$ determined in Sec.~\ref{rad}.
They are somewhat lower but consistent with the $6.5\pm2.2$\% effect found
in Ref.~\cite{zrkd98}.

The $30\h<r\le 70\hm$ range chosen in Ref.\ \cite{zrkd98} is equally divided
between regimes in which the LG frame Hubble flow is closer to uniform in
the \CS data, and alternatively in which the CMB frame is closer to
uniform, as seen in Fig.~\ref{dHs}. This explains why the average values
of $\Hm$ in this range are closer to each other than those determined
in Sec.~\ref{rad} by fitting a simple linear Hubble law to the whole
sample. The latter values, which amounted to $8.8\pm0.2$\% in the CMB frame
and $3.37\pm0.07$\% in the LG frame, might be taken as a sharper estimate
of the Hubble bubble effect.

The existence of a Hubble bubble has been controversial since as far as SneIa
data analysis is concerned the presence of the effect is dependent on the
details of the treatment of extinction and reddening by dust \cite{conley07}.
A Hubble bubble is found if dust in other galaxies has the same reddening
properties as dust in the Milky Way but not if the reddening parameter is
significantly reduced. \Rhicken{hicken09} find no evidence for a Hubble bubble
at $z=0.024$ if the reddening parameter is set to $\RV=1.7$.

Our results suggest that the combination of the boost to the rest frame
of the CMB compensating for structures in the range $30\h\lsim r\lsim62\hm$,
together with the treatment of parameters such as $\RV$ as adjustable in
light curve reduction, may contribute significantly to the systematic
uncertainties associated with SneIa. Reddening by dust in other galaxies is
after all a physical quantity which should be determined independently of
SneIa. Ideally it should not be treated as a parameter which one can freely
adjust to obtain the best fit of Hubble residuals.

This issue has been studied independently by \Rfinkelman{fi08,fi10}
who investigated dust lanes in 15 E/S0 galaxies and determined extinction
properties by fitting model galaxies to the unextinguished parts of the
images in each of six spectral bands, and then subtracting these from the
actual images. They found an average value $\RV=2.82\pm0.38$ for 8 galaxies
in their first study \cite{fi08}, and $\RV=2.71\pm0.43$ for 7 galaxies
in their second investigation \cite{fi10}. For the combined sample
$\RV=2.77\pm0.41$. This value is a little
lower than the Milky Way value $\RV=3.1$ but consistent with it within the
uncertainty.

Our results suggest that the convergence scale $z\ns{conv}\simeq0.022$
is close to that of the Hubble bubble originally proposed by \Rzrkd{zrkd98},
but the magnitude of the Hubble bubble effect is smaller when viewed in
the LG frame. For consistency MLCS2k2 SneIa data should be reduced using
$\RV$ values consistent with independent determinations, e.g., $\RV=2.77\pm
0.41$ as suggested by the work of \Rfinkelman{fi08,fi10}. As discussed
in Ref.\ \cite{sw} this is also important for cosmological model comparison.

Since the First Amendment SneIa data of Ref.\ \cite{turnbull11} was reduced
with $\RV=1.7$, one should check to what extent the difference
in the amplitude of the bulk flow velocity from that of the \CS sample
of \cite{wfh09} is due to the choice of the $\RV$ parameter. We suggest that
the data set of Ref.\ \cite{turnbull11} should be reanalysed with $\RV=2.77\pm
0.41$, and by the method of Sec.~\ref{rad} in the LG frame.

\subsection{The asymptotic global Hubble constant}
The \CS sample enables us to determine the relative Hubble flow,
but does not constrain the overall normalization of the distance scale and
consequently the precise value of the global asymptotic Hubble constant.

The Hubble constant has recently been determined to high accuracy by
the SH0ES survey as $\Hm=73.8\pm2.4\kmsMpc$ \cite{shoes}. Independent
estimates of the Hubble constant using BAO data at a variety of redshifts
\cite{percival10,beutler11,blake11}, have yielded values $\Hm=\{68.2\pm2.2,%
67\pm3.2,68.1\pm1.7\}\kmsMpc$ respectively in the (possibly curved) \LCDM\
model. The Planck fit of the CMB anisotropies to a spatially flat \LCDM\ model
yields \cite{Pparm} $\Hm=67.4\pm1.4\kmsMpc$. While these $\Hm$ values
are consistent with Ref.\ \cite{shoes} at the 2$\si$ level, a further increase
in precision could lead to tension. The BAO and Planck analyses rely on fits
to the \LCDM\ model at large redshifts, whereas the SH0ES survey is
less model dependent but relies on an empirical
ladder of cosmic distance indicators on very nearby scales.

If we identify the cosmic rest frame with that of minimum Hubble flow
variance, then the impact of performing all cosmological tests in such
a frame rather than in the CMB frame needs to be carefully considered.
The impact is likely to be most significant on those tests which directly
use measurements on $z\lsim0.022$ scales. Whether this has an impact on
measurements that establish the cosmic distance ladder is an intriguing
question which should be investigated once the minimum Hubble flow variance
rest frame is positively identified.

\subsection{Large angle CMB anomalies}\label{ano}
There are several observations concerning the large angle multipoles
of the CMB anisotropy spectrum, which may be considered anomalous to varying
degrees of statistical significance. These include: (i) the power asymmetry
between the northern and southern hemispheres \cite{toh03,ehb04,ebg07,heb09};
(ii) the low quadrupole power \cite{toh03,dOC04}; (iii)
the alignment of the quadrupole and octupole \cite{dOC04,sshc04,lm05,%
chss06}; and (iv) the parity asymmetry \cite{kn10}. The significance of some
of these problems has increased with the recent release of Planck satellite
data \cite{Piso}.

It is beyond the scope of the present paper to investigate all these anomalies.
However, it is clear that our proposal to revisit a significant feature of the
CMB anisotropy analysis, namely the nature of the dipole, will introduce
systematics which would necessitate a reassessment of all of these issues.

There are two obvious potential lines of inquiry:
\begin{itemize}
\item The propagation of photons through the foregrounds contributing to
the Hubble flow variance may produce a multipole signature which differs subtly
from the pure dipole signature (\ref{Tdip}) associated with a Lorentz boost;
\item Since the dipole subtraction is an integral part of the map-making
procedure, differences in dipole subtraction may lead to subtle differences
in the cleaning of galactic foregrounds \cite{toh03}.
\end{itemize}
A study by \Rfreeman{freeman06} found that of several possible systematic
errors, a 1--2\% error in the CMB dipole subtraction stood out as being an
effect which could potentially resolve the power asymmetry anomaly.

We note that while a 1--2\% change in the dipole would not affect the
power on small angles, its effect on the large angle multipoles would
require a redrawing of the CMB sky maps. Any such redrawing may potentially
alter other large angle features, such as the Cold Spot.

\subsection{The dark flow}
The determination of peculiar velocities via the kinematic Sunyaev-Zel'dovich
(kSZ) effect \cite{kash08,kash09,kash10,hand12,lah12} is particularly
interesting, since it is a method which uses the CMB anisotropy spectrum,
rather than directly testing the distance--redshift relation. Furthermore, the
very large bulk flow that has been claimed by \Rkash{kash08,kash10} has been
controversial. The effect was not reproduced in the subsequent studies of
\Rkeisler{keisler09} and \Rosborne{osborne11}. However, \Rakeke{akeke10,akek12}
showed that these differences could be attributed to flaws in the filtering
methodology of their critics. Following the first release of Planck data,
the debate has continued \cite{Ppec,minrep}.

A significant feature of the measurements of \Rkash{kash08,kash09,kash10}
is the claim that the result is independent of many systematics, since
both foreground and cosmological dipoles and quadrupoles have been subtracted
in a consistent way. It is claimed that the peculiar velocities inferred
are those of the galaxy clusters with respect to the CMB in their own
rest frames. Nonetheless, the direction of the reported bulk ``dark
flow'' on scales $0.12\le z\le0.3$, $(\ell,b)=(267\deg,34\deg)\pm15\deg$,
coincides closely with that of the residual CMB dipole in the LG rest frame,
while its amplitude is consistent with the boost of $635\pm38\kms$ that the LG
would be required to have with respect to the CMB frame to account for the
residual dipole in the standard framework. Given these coincidences all
potential systematics need to be considered.

Our results suggest in particular, that if the CMB rest frame is to
be identified as that of minimum Hubble flow variance then such
a frame is closer to that of the LG than to the conventionally assumed
rest frame. We are unable to identify any obvious systematic
errors in the highly technical analysis of Ref.\ \cite{kash09}. However,
they do of course implicitly assume that the conventional CMB frame is our
actual relevant cosmic rest frame, an assumption which is not questioned
in the standard framework but which our analysis leads us to question.

This issue might be resolved by carefully reperforming the analysis of
Refs.\ \cite{kash08,kash09,kash10} with each step in the pipeline that implicitly
or explicitly assumes a normalization to the CMB frame redone as if the
observation was made by observers in the LG frame. In particular, in
the case of the WMAP data this should be done before applying the Wiener
filter, and in the case of the X--ray clusters this should be done before
any calibration of any quantity which depends on redshift. If the frame
transformation has any consequences for the cluster redshifts this would be
likely to affect the nearer sample more; however, an overall boost (\ref{Tdip})
of the CMB sky might affect the whole analysis. One should begin with raw
skymaps in which only the dipole and all higher order multipoles which
correspond to a boost from the heliocentric to LG frame have been removed. This
will of course leave a global dipole that will be removed in the data analysis
pipeline. The more pertinent issue is whether starting in the LG frame
changes the way in which the SZ components are treated by the Wiener filter.
The intrinsic optical depths, $\tau$, of the clusters are convolved with the
filter in the process of estimating the final effective $\tau$, which is a
crucial physical quantity in the kSZ determination. A transformation to the
LG frame may therefore potentially affect the result.

The intrinsic kSZ effect is of course due to the local CMB dipole
at the cluster location. With our interpretation, this temperature dipole
will include contributions from both a peculiar velocity and from the
differential expansion of space due to inhomogeneities in the vicinity of the
galaxy cluster. Estimates of the maximum possible Hubble flow variance
based on void/wall statistics should therefore put bounds on the magnitude
of what is assigned to individual ``peculiar velocities'' in the standard
framework.

\subsection{Direct tests of a nonkinematic dipole}\label{kin}
It is possible to directly test the extent to which the CMB dipole is
kinematic by considering the combined effects of the frequency shift
plus aberration that arise from the local boost of an observer who views
the spectrum of an otherwise isotropic background of sources \cite{eb84}.

The Planck satellite team has recently examined the effects of Doppler
boosting on the primary CMB anisotropies, and claims evidence in favour of the
standard boost interpretation \cite{Pboost}. It is notable, however,
their results agree with the conventional boost direction, $(\ell,b)=(264\deg,
48\deg)$ only if small angle multipoles are included in the analysis, with
a definitive measurement for multipoles in the range $l\ns{min}=500<l<l\ns{max}
=2000$. If large angle multipoles are included and $l\ns{max}$ is reduced to
$l\ns{max}<100$, then the inferred boost direction moves across the sky to
coincide with the modulation dipole anomaly direction \cite{heb09},
$(\ell,b)=(224\deg,-22\deg)\pm24\deg$.

This provides evidence that a nonkinematic boost component and the large
angle anomalies may be related, as pointed out in Sec.~\ref{ano}. In
Sec.~\ref{dip} we found there is no statistically significant CMB frame
dipole in shell 4 with $\mr=44.5\hm$, within the range $40\lsim r\lsim60\hm$
for which the CMB frame also has a smaller monopole variation in $\Hm$ than
the LG frame. Thus the observed dipole has features very close those of a
Lorentz boost dipole, which smooth the Hubble flow at a particular radial
scale, but not exactly. Work needs to be done to establish whether the angular
scale dependence of Ref.\ \cite{Pboost} can be explained in terms of a
component of the dipole, with properties close to but not exactly those
produced by a boost.

The effects of aberration and frequency modulation can also be readily tested
in the radio spectrum.
In recent work \Rrs{rs13} have found that the assumption of kinematic origin
for the cosmic radio dipole is inconsistent at the 99.5\% confidence level,
using the NRAO VLA Sky Survey (NVSS). Furthermore, through a careful analysis
of the biases introduced by different statistical estimators \Rrs{rs13}
have been able to resolve apparently discrepant results \cite{bw02,s11,gh12}
from previous studies.

The direction of the radio dipole in NVSS is found to be $(\hbox{RA},\hbox{dec}
)=(154\deg\pm21\deg,-2\deg\pm21\deg)$ \cite{rs13} in equatorial coordinates
(epoch J2000). By comparison, in Sec.~\ref{dip} the smoothed
dipole in the LG frame was found to have a direction $(180+\ld,-\bd)=(263\deg
\pm6\deg,39\deg\pm3\deg)$ for the spheres $r>r_o$ with $20\h\lsim r_o\lsim45
\hm$ which produced a strong amplitude dipole. In equatorial coordinates this
corresponds to $(\hbox{RA},\hbox{dec})=(162\deg\pm4\deg,-14\deg\pm3\deg)$,
which lies well within the error circle of the NVSS radio dipole. Since the
directions are consistent, the nonkinematic nature of the dipole found by
\Rrs{rs13} provides independent evidence for our hypothesis of
Sec.~\ref{origin}.

The NVSS data is of course sampled over much larger distances than the
\CS sample, with a mean redshift $z\goesas1$, and even if the radio
dipole is nonkinematic it could in principle be generated by a completely
different effect to the foregrounds we have identified. While there are
numerous systematic uncertainties which complicate the analysis, future
surveys such as the Square Kilometre Array should have the power to potentially
falsify the hypothesis of Sec.~\ref{origin}. In particular, it should be
possible to eventually determine just what fraction of the radio dipole is
kinematic, along with the magnitude and direction of the associated boost.

\subsection{Conclusion}
While much work remains to be done, our results suggest that a fundamental
revision of the treatment of peculiar velocities may shed new light on many
of the puzzles raised by bulk flows, and perhaps even resolve some of the
associated anomalies. If a large fraction of the CMB dipole is due to a
residual anisotropy in the distance--redshift relation, as our results seem
to suggest, then this may also have important consequences for many
aspects of theoretical and observational cosmology.

Peculiar velocities will always play a role in observational
cosmology -- galaxies in clusters exhibit peculiar velocities with respect
to the cluster barycentres, and this is directly observed in the ``fingers
of god'' redshift space distortions. However, the most natural conclusion
of our work is that on scales larger than gravitationally bound systems the
variance of the Hubble flow should be treated as the differential expansion
of regions of varying density, which have decelerated by different amounts
from the initial uniform distribution at the epoch of last scattering. One
should only talk about ``infall'' if the physical distance between objects
is actually decreasing with time, rather than applying it to the case of
denser regions which are expanding slower than the average. While directional
forces are the basis of Newtonian mechanics in Euclidean space, there is
nothing in general relativity which demands that such notions should apply
to scales of tens of megaparsecs over which space is expanding.

In recent years there has been some discussion about whether it is
conceptually more correct to think of space as expanding, or whether the
treatment of the expansion by a simple Doppler law on a fixed background is
sufficient \cite{ablm,chod07,ge07,fbjl,lfbj,bh09,roukema09,faraoni10}. In
particular, \Rablm{ablm} showed that the expansion of space can in principle
be observationally determined. Some other authors, e.g., \Rbh{bh09},
maintained that the Doppler law picture is still useful. The debate involved
thought experiments conducted within homogeneous isotropic cosmological models.
Our results suggest that, as far as actual observations are concerned, variance
in the Hubble law over scales of tens of megaparsecs cannot be simply reduced
to a boost at a point; space really is expanding, and by differential amounts.

\bigskip\noindent{\em Acknowledgements} This work was supported by the
Marsden Fund of the Royal Society of New Zealand. We are especially grateful
to Krzysztof Bolejko for assistance with ray tracing studies. We also thank
Anthony Brown, Chris Gordon, Loretta Dunne, Steve Maddox and Maria Mattsson for
helpful discussions, and Fernando Atrio-Barandela and Alexander Kashlinsky for
correspondence.

\def\R{\\\noalign{\vskip-.2pt}}\bigtable{*}{lrrrrrrrrrrr}
Shell $s$&1&2&3&4&5&6&7&8&9&10&11\\
$N_s$&78&457&494&713&799&555&413&303&221&274&91\R
$r_s$ ($h^{-1}$Mpc)&2.00&12.50&25.00&37.50&50.00&62.50&75.00&87.50&100.00&112.50&156.25\R
$\mr_s$ ($h^{-1}$Mpc)&7.23&16.22&30.12&44.52&55.11&69.25&81.06&93.68&105.10&126.51&182.59\R
$(H_s)\Ns{CMB}$&152.7&109.5&108.2&103.5&101.4&103.0&102.0&103.1&104.1&102.1&100.1\R
$(\bs_s)\Ns{CMB}$&4.9&1.5&1.1&0.7&0.7&0.7&0.9&0.9&1.0&0.8&1.7\R
$(Q_s)\Ns{CMB}$&0.000&0.000&0.024&1.000&1.000&1.000&1.000&1.000&1.000&1.000&0.999\R
$(\chi^2/\nu)\Ns{CMB}$&9.092&1.818&1.130&0.739&0.606&0.664&0.580&0.577&0.603&0.685&0.581\R
$(H_s)\Ns{LG}$&107.8&98.2&103.7&105.4&103.6&101.4&102.7&103.5&103.4&102.4&101.0\R
$(\bs_s)\Ns{LG}$&3.4&1.4&1.0&0.8&0.7&0.7&0.9&0.9&1.0&0.8&1.7\R
$(Q_s)\Ns{LG}$&0.000&0.000&0.000&0.980&1.000&1.000&1.000&1.000&1.000&1.000&0.999\R
$(\chi^2/\nu)\Ns{LG}$&4.048&2.340&1.431&0.894&0.583&0.723&0.578&0.513&0.595&0.667&0.590\R
$\ln B$ ($r\ge r_s$)&50.84&11.42&5.97&1.04&1.58&2.16&1.53&1.67&1.52&0.44&\\
\noalign{\smallskip}\tableline\noalign{\smallskip}
Shell $s$&1'&2'&3'&4'&5'&6'&7'&8'&9'&10'&11\\
$N_s$&284&488&532&869&669&481&343&271&160&204&91\R
$r_s$ ($h^{-1}$Mpc)&6.25&18.75&31.25&43.75&56.25&68.75&81.25&93.75&106.25&118.75&156.25\R
$\mr_s$ ($h^{-1}$Mpc)&12.25&24.05&37.47&49.17&61.75&73.92&87.15&99.13&111.81&131.52&182.59\R
$(H_s)\Ns{CMB}$&119.5&107.3&105.3&102.4&103.0&102.2&102.7&103.9&101.7&102.3&100.1\R
$(\bs_s)\Ns{CMB}$&2.2&1.2&1.0&0.7&0.8&0.8&0.9&0.9&1.1&0.9&1.7\R
$(Q_s)\Ns{CMB}$&0.000&0.000&0.649&1.000&1.000&1.000&1.000&1.000&1.000&1.000&0.999\R
$(\chi^2/\nu)\Ns{CMB}$&4.042&1.313&0.975&0.645&0.638&0.587&0.624&0.604&0.561&0.690&0.581\R
$(H_s)\Ns{LG}$&99.7&101.1&103.6&105.7&102.7&101.6&103.2&103.6&101.5&102.7&101.0\R
$(\bs_s)\Ns{LG}$&1.8&1.1&0.9&0.7&0.8&0.8&0.9&0.9&1.0&0.9&1.7\R
$(Q_s)\Ns{LG}$&0.000&0.000&0.001&1.000&1.000&1.000&1.000&1.000&1.000&1.000&0.999\R
$(\chi^2/\nu)\Ns{LG}$&3.581&1.496&1.197&0.699&0.653&0.650&0.592&0.552&0.563&0.655&0.590\R
$\ln B$ ($r\ge r_s$)&23.97&6.29&2.31&0.59&1.86&1.53&1.32&1.24&0.57&0.42&
\enddata
\caption{The analysis of Table~\ref{shell} is repeated removing all
points which contribute a value $\chi^2>5$ in both the CMB and LG frames in
both the primed and unprimed shells. A total 4398 points remain.}
\label{shellAA}\end{table*}\endgroup

\bigtable{*}{lrrrrrrrrrrr}
Shell $s$&1&2&3&4&5&6&7&8&9&10&11\\
$N_s$&56&385&454&692&788&548&410&301&217&270&91\R
$r_s$ ($h^{-1}$Mpc)&2.00&12.50&25.00&37.50&50.00&62.50&75.00&87.50&100.00&112.50&156.25\R
$\mr_s$ ($h^{-1}$Mpc)&9.04&17.87&30.33&44.58&55.10&68.74&80.95&93.74&105.08&126.20&182.59\R
$(H_s)\Ns{CMB}$&155.4&109.3&107.6&103.0&101.4&103.9&102.0&103.4&104.6&102.6&100.1\R
$(\bs_s)\Ns{CMB}$&5.0&1.6&1.2&0.8&0.7&0.8&0.9&0.9&1.0&0.8&1.7\R
$(Q_s)\Ns{CMB}$&0.000&0.228&0.991&1.000&1.000&1.000&1.000&1.000&1.000&1.000&0.999\R
$(\chi^2/\nu)\Ns{CMB}$&2.895&1.053&0.850&0.632&0.546&0.616&0.550&0.538&0.498&0.613&0.581\R
$(H_s)\Ns{LG}$&114.3&101.3&104.7&105.0&103.5&102.8&102.9&103.6&103.7&102.8&101.0\R
$(\bs_s)\Ns{LG}$&3.6&1.5&1.1&0.8&0.7&0.8&0.9&0.9&1.0&0.8&1.7\R
$(Q_s)\Ns{LG}$&0.000&0.529&0.778&1.000&1.000&1.000&1.000&1.000&1.000&1.000&0.999\R
$(\chi^2/\nu)\Ns{LG}$&2.275&0.993&0.949&0.745&0.523&0.647&0.549&0.482&0.519&0.606&0.590\R
$\ln B$ ($r\ge r_s$)&50.81&10.83&5.31&1.78&2.40&3.02&1.99&2.19&1.98&0.59&\\
\noalign{\smallskip}\tableline\noalign{\smallskip}
Shell $s$&1'&2'&3'&4'&5'&6'&7'&8'&9'&10'&11\\
$N_s$&223&436&501&856&657&477&339&268&158&201&91\R
$r_s$ ($h^{-1}$Mpc)&6.25&18.75&31.25&43.75&56.25&68.75&81.25&93.75&106.25&118.75&156.25\R
$\mr_s$ ($h^{-1}$Mpc)&12.92&24.39&37.80&49.23&61.51&73.84&87.08&99.03&111.89&131.38&182.59\R
$(H_s)\Ns{CMB}$&124.5&106.4&105.7&102.3&102.9&103.0&103.0&104.0&101.9&103.1&100.1\R
$(\bs_s)\Ns{CMB}$&2.4&1.3&1.0&0.7&0.8&0.8&0.9&0.9&1.1&0.9&1.7\R
$(Q_s)\Ns{CMB}$&0.000&0.994&1.000&1.000&1.000&1.000&1.000&1.000&1.000&1.000&0.999\R
$(\chi^2/\nu)\Ns{CMB}$&2.414&0.839&0.794&0.588&0.568&0.556&0.568&0.554&0.486&0.610&0.581\R
$(H_s)\Ns{LG}$&105.2&101.3&104.4&105.3&102.9&102.9&103.5&103.6&101.6&103.4&101.0\R
$(\bs_s)\Ns{LG}$&2.0&1.2&1.0&0.7&0.8&0.8&0.9&0.9&1.1&0.9&1.7\R
$(Q_s)\Ns{LG}$&0.000&0.863&0.972&1.000&1.000&1.000&1.000&1.000&1.000&1.000&0.999\R
$(\chi^2/\nu)\Ns{LG}$&1.444&0.926&0.883&0.616&0.566&0.588&0.539&0.506&0.510&0.591&0.590\R
$\ln B$ ($r\ge r_s$)&29.83&5.66&2.75&1.08&2.29&2.01&1.66&1.59&0.83&0.62&
\enddata
\caption{The analysis of Table~\ref{shell} is repeated removing all
points which contribute a value $\chi^2>5$ in either the CMB or LG frames in
both the primed and unprimed shells. A total 4212 points remain.}
\label{shellOA}\end{table*}\endgroup

\appendix\section{Nonlinearity from foreground structures and statistical
issues}\label{nonl}

We have investigated the extent to which the relative magnitude of the Bayes
factors in Fig.~\ref{lnB} is driven by the greater scatter relative to a
linear law, rather than by the difference of the linear fit of the Hubble
constant from its asymptotic value. To investigate this in
Tables~\ref{shellAA}, \ref{shellOA} and \ref{shellOO}, we have repeated the
analysis of Secs.~\ref{meth}, \ref{res} that led to Table~\ref{shell} by
successively removing points which contribute the greatest scatter:
\begin{enumerate}[(1)] \item Firstly, in
Table~\ref{shellAA} we remove points which contribute an individual value
$\chis\equiv(r_i-cz_i/H_s)^2/\si_i^2$ with $\chis>5$ in both the
CMB and LG frames for both the primed and unprimed choices of shell boundaries.
\item Secondly, in Table~\ref{shellOA} we remove points with $\chis>
5$ in either the CMB or LG frame for both choices of shell boundaries.
\item Finally, in Table~\ref{shellOO} we remove points with $\chis>
5$ in either the CMB or LG frame for either choice of shell boundaries.
\end{enumerate}

Since the underlying Hubble relation is not linear, it is clear that we are
rejecting some of the points with the strongest discriminating power in
such a procedure. Indeed, only the points excluded in Table~\ref{shellAA}
could be said to be outliers in any sense\footnote{The points excluded
in Table~\ref{shellAA} are still only outliers relative to the CMB and LG
frames. Since we have not yet established a ``minimum variance Hubble frame'',
it is perfectly possible that some of the points so excluded in fact
have little scatter relative to a linear Hubble law in some other minimum
variance frame.} since all other points have $\chis<5$ in at least one
frame and shell slicing.

Nonetheless, although this procedure is not a perfect one it does illustrate
that as the linear goodness of fit is improved the relative Bayes factor
is somewhat reduced but still remains significant. In Table~\ref{shellAA}
only shell 4 has a significantly improved goodness of fit in both frames.
However, in Table~\ref{shellOA} the shells $\{2,2',3,3'\}$ now all have an
acceptable goodness of fit and a reduced $\chi^2$ close to unity. Although
$\ln B$ is somewhat reduced, a very strong value $\ln B>5$
is still found in shells $\{2,2',3\}$ and indeed in shell 2 we still have
$\ln B>10$. Our statistical conclusions are thus robust.

In the final Table~\ref{shellOO} even shells $1$ and $1'$ have an acceptable
goodness of fit. However, for shell 1 this comes at the expense of having
removed two thirds of the 92 points originally present. The $12.5\hm$ radius
of shell 1 is simply so small relative to the foreground structures that a
notion of a spherically averaged linear Hubble law is not really applicable.
However, our statistical conclusions do not rely on the innermost shell.

A robust optimization procedure could also be used to estimate $H_s$ in the
inner shells. However, that would also require
modeling the nonlinearity of the inner shells, and in the present paper
we aim to be model-independent, without any particular theoretical biases.
To this end, we believe the very strong evidence for the relative uniformity
of the LG and LS frames as compared to the CMB frame, has been clearly
demonstrated.

\bigtable{*}{lrrrrrrrrrrr}
Shell $s$&1&2&3&4&5&6&7&8&9&10&11\\
$N_s$&30&366&447&691&786&547&410&301&217&270&91\R
$r_s$ ($h^{-1}$Mpc)&2.00&12.50&25.00&37.50&50.00&62.50&75.00&87.50&100.00&112.50&156.25\R
$\mr_s$ ($h^{-1}$Mpc)&8.44&18.32&30.20&44.58&55.10&68.74&80.95&93.74&105.08&126.20&182.59\R
$(H_s)\Ns{CMB}$&138.4&107.2&107.1&103.0&101.2&103.8&102.0&103.4&104.6&102.6&100.1\R
$(\bs_s)\Ns{CMB}$&6.3&1.6&1.2&0.8&0.7&0.8&0.9&0.9&1.0&0.8&1.7\R
$(Q_s)\Ns{CMB}$&0.320&0.992&0.997&1.000&1.000&1.000&1.000&1.000&1.000&1.000&0.999\R
$(\chi^2/\nu)\Ns{CMB}$&1.103&0.829&0.823&0.627&0.531&0.611&0.550&0.538&0.498&0.613&0.581\R
$(H_s)\Ns{LG}$&106.9&101.7&104.2&104.9&103.4&102.7&102.9&103.6&103.7&102.8&101.0\R
$(\bs_s)\Ns{LG}$&4.9&1.5&1.1&0.8&0.7&0.8&0.9&0.9&1.0&0.8&1.7\R
$(Q_s)\Ns{LG}$&0.464&0.530&0.907&1.000&1.000&1.000&1.000&1.000&1.000&1.000&0.999\R
$(\chi^2/\nu)\Ns{LG}$&1.001&0.993&0.912&0.738&0.519&0.638&0.549&0.482&0.519&0.606&0.590\R
$\ln B$ ($r\ge r_s$)&21.00&8.00&4.88&1.74&2.33&2.96&1.99&2.19&1.98&0.59&\\
\noalign{\smallskip}\tableline\noalign{\smallskip}
Shell $s$&1'&2'&3'&4'&5'&6'&7'&8'&9'&10'&11\\
$N_s$&181&429&498&854&655&477&339&268&158&201&91\R
$r_s$ ($h^{-1}$Mpc)&6.25&18.75&31.25&43.75&56.25&68.75&81.25&93.75&106.25&118.75&156.25\R
$\mr_s$ ($h^{-1}$Mpc)&14.27&24.55&37.83&49.22&61.49&73.84&87.08&99.03&111.89&131.38&182.59\R
$(H_s)\Ns{CMB}$&112.5&106.3&104.9&102.2&102.7&103.0&103.0&104.0&101.9&103.1&100.1\R
$(\bs_s)\Ns{CMB}$&2.3&1.3&1.0&0.7&0.8&0.8&0.9&0.9&1.1&0.9&1.7\R
$(Q_s)\Ns{CMB}$&0.504&0.999&1.000&1.000&1.000&1.000&1.000&1.000&1.000&1.000&0.999\R
$(\chi^2/\nu)\Ns{CMB}$&0.995&0.803&0.770&0.578&0.555&0.556&0.568&0.554&0.486&0.610&0.581\R
$(H_s)\Ns{LG}$&103.0&100.9&104.3&105.2&102.7&102.9&103.5&103.6&101.6&103.4&101.0\R
$(\bs_s)\Ns{LG}$&2.1&1.2&1.0&0.7&0.8&0.8&0.9&0.9&1.1&0.9&1.7\R
$(Q_s)\Ns{LG}$&0.243&0.966&0.980&1.000&1.000&1.000&1.000&1.000&1.000&1.000&0.999\R
$(\chi^2/\nu)\Ns{LG}$&1.071&0.879&0.875&0.610&0.555&0.588&0.539&0.506&0.510&0.591&0.590\R
$\ln B$ ($r\ge r_s$)&11.34&4.80&2.07&1.03&2.22&2.01&1.66&1.59&0.83&0.62&
\enddata
\caption{The analysis of Table~\ref{shell} is repeated removing all
points which contribute a value $\chi^2>5$ in either the CMB or LG frames in
either the primed or unprimed shells. A total 4156 points remain.}
\label{shellOO}\end{table*}\endgroup

\section{Angular Gaussian window averages}\label{mdtech}
In Sec.~\ref{ang} we follow McClure and Dyer \cite{md07} (MD07) to produce
contour maps of the angular variation in the Hubble flow.
At each grid point on the sky, a mean $\Hth$ is calculated in which
the value of $cz_i/r_i$ for each data point is weighted according to its
angular separation from the grid point. The $i$th data point is weighted by
\beq
\Wi=\frac{1}{\sqrt{2\pi}\sith}\exp\Big(\frac{-\th_i^2}{2\sith^2}\Big),
\label{Wi}\eeq
where $\cos\th_i=\vec{r}_{\mathrm{grid}}\cdot\vec{r_i}$, $\sith=25\deg$
is the smoothing scale and the Greek subscript $\al$ is used to represent
the angular dependence on the spherical polar coordinates, $(\ell,b)$ encoded
in $\th_i$.

The choice of smoothing scale, $\sith$, is constrained by the fact that
the diameter $2\sith$ of the Gaussian window function should be wider
than the angular width of the Zone of Avoidance (ZoA). If it were smaller
then windows centred on gridpoints close to the galactic plane would have
insufficient data to give reliable results in those regions. The ZoA is
typically $30\deg$ wide for the \CS sample, which means that the
smoothing scale must be greater than $15\deg$. On the other hand we cannot
make the smoothing scale so large that we lose all angular resolution. This
determines the choice $\sith=25\deg$, which matches that made by MD07.
This angle subtends an area 0.59 steradians, 4.8\% of the full sky. We have
checked that varying the smoothing scale in the range $15\deg<\sith<40\deg$
does not significantly change our results.

Since (\ref{Wi}) determines a mean value of $\Hth$ at each grid point on the
sky in which each data point is weighted by its distance from the grid point,
there will be larger uncertainties for grid points near the ZoA, as can be
seen in Fig.~\ref{errsky}.

In the method adopted by MD07, the weighted mean $\Hth$ is
calculated at each spherical polar grid point by
\beq\Hth={\sum_{i=1}^N\Wi\,cz_i\,r_i^{-1}\over\sum_{j=1}^N\Wj}\,,
\label{Htp}\eeq
with the weight (\ref{Wi}). The variance of this sample mean at each grid point
is given by
\beq
\bsa^2=\frac{\sum_{i=1}^N\Wi^2\si\Z{H_i}^2}
{(\sum_{j=1}^N\Wj)^2}\,,\eeq
where
\beq \sHi={cz_i\si_i\over r_i^2} \label{sih}\eeq
is the standard uncertainty from error propagation of the uncertainty $\si_i$
in the $i$th distance $r_i$ in (\ref{Htp})--(\ref{sih}) produced values.

No additional uncertainty weighting of the weight $\Wi$ was used by MD07
in the determination of $\Hth$, since they claimed that the impact of the
errors in the individual data points is averaged out by the Gaussian weighting
procedure. The individual uncertainties in the \CS sample are larger,
and therefore the question of the treatment of the uncertainty weightings in
the determination of the mean (\ref{Htp}) is an important one.

In order to manage the uncertainty weightings, rather than using
Eqs.~(\ref{Htp})--(\ref{sih}), we will instead
determine the weighted mean $\Hth$ at each spherical polar grid point
$(\ell,b)$, by first evaluating its inverse
\beq\Hth^{-1}={\sum_{i=1}^N\Wi\,r_i\,(cz_i)^{-1}\over\sum_{j=1}^N\Wj}\,,
\label{Hal}\eeq
with the weight (\ref{Wi}). The
variance of $\Hth^{-1}$ at each grid point is then given by
\beq
\bs\Z{H^{-1}_{\al}}^2=\frac{\sum_{i=1}^N\Wi^2\si\Z{H^{-1}_i}^2}
{(\sum_{j=1}^N\Wj)^2}\,,\eeq
where
\beq \si\Z{H^{-1}_i}={\si_i\over cz_i} \label{sial}\eeq
is the standard uncertainty in $H_i^{-1}=r_i/(cz_i)$. Then
\beq \bsa=\bs\Z{H^{-1}_{\al}}\Hth^2\label{sdal}\eeq
is the standard uncertainty in $\Hth$. If, following MD07, no
additional uncertainty weightings are used then in (\ref{Hal}) $\Wi$ is given
by (\ref{Wi}). Alternatively, if inverse variance (IV) uncertainty
weightings are used then we replace (\ref{Wi}) by
\beq
\Wi=\frac{1}{\si\Z{H^{-1}_i}^2\sqrt{2\pi}\sith}\exp\Big(\frac{-\th_i^2}
{2\sith^2}\Big).
\label{Wi2}\eeq

The reason that it is preferable to work with $\Hth^{-1}$ is a consequence of
the dominant uncertainties in the \CS sample being those associated with
the distance measure, $\si_i$. In the case of the radial shells we chose to
minimize $\sum_i\left[\si_i^{-1}(r_i-cz_i/H)\right]^2$ with respect to $H$ for
the same reason. The Gaussian window averaging adds a nonlinear weighting to
what is otherwise a linear regression. Using (\ref{Wi2}) ensures that the
nonlinear weighting is added to IV weightings determined from uncertainties
(\ref{sial}) which are themselves linear in the measurement uncertainties. The
alternative procedure of (\ref{Htp})--(\ref{sih}) introduces a different
distance weighting of each point in (\ref{sih}) which leads to different
results\footnote{We found that using Eqs.~(\ref{Htp})--(\ref{sih}) in
place of (\ref{Hal})--(\ref{sial}) gives results which differ very little
from each other if IV weightings are not used. However, once IV weightings
are included using (\ref{Htp})--(\ref{sih}) gives values of $\Hth$ with a
mean which is 10\% lower than the mean values determined from the averages
in spherical shells.} when combined with the Gaussian window averaging in an
equation analogous to (\ref{Wi2}).

Use of a standard IV weighting in (\ref{Wi2}) may not be the
most robust method for uncertainty estimates for this data set. In particular,
as has been discussed in WFH09 and FWH10 in the standard peculiar velocity
framework the nonideal geometry of typical surveys can lead to an aliasing
of small scale power. Where the data is sparse biases can be introduced
relative to the bulk flow of a regular volume that one is ideally interested
in. To deal with these issues \Rwfh{wfh09,fwh10} have developed a minimum
variance weighting method with respect to the leading peculiar velocity
moment amplitudes (dipole, quadrupole and octupole).
To treat these issues in our framework requires the development of a
similar methodology, as is discussed in Sec.~\ref{formal}, and is left
for future work.

\Rmd{md07} performed Monte Carlo simulations to assess the significance of
the variation that they found. While a similar analysis has been undertaken
by one of us \cite{s12}, in the present paper we will use the more direct
method of plotting the contours of the uncertainty $\bsa$ -- as given by
(\ref{sdal}) using either (\ref{Wi}) or (\ref{Wi2}) -- on the same map as
the Hubble flow contours. Such a plot of the angular variation of the
uncertainty contains much detailed information, uncertainties being larger
in some angular regions rather than others.

\subsection{Uncertainties in Gaussian window averages}\label{winu}
Dealing with the uncertainties in the Gaussian window averaged sky maps is
complicated if we wish to constrain individual large angle multipoles.
Since we have not done a Fourier analysis on the raw data, we have avoided
the direct problems of nonorthogonality of Fourier modes that arise from an
analysis on a cut sky determined by the Zone of Avoidance. However, the finite
area of the Gaussian window function may introduce its own systematic issues.
This area is 0.59 steradians for $\sith=25\deg$, giving effectively
21.4 independent patches on the sky to constrain the 5
independent degrees of freedom that define the quadrupole, and the 7
that define the octupole.

Putting aside the question of the multipole decomposition, the overall
uncertainties (\ref{sdal}) in the Gaussian window averages are readily
represented as a function of angular position. In Fig.~\ref{errsky} we show
the example of the outer $r>15\hm$ LG frame sky map, with the contours of the
Hubble flow variance plotted as solid lines overlaid with colour map contours
showing the angular uncertainties as a function of angular position on the sky.
In each case we have used $\sith=25\deg$.

The angular uncertainties are somewhat greater in a curved band near the
galactic plane. This effect is due to the absence of data in the ZoA as
well as the propagation of measurement uncertainties through the Gaussian
window averaging procedure. We have checked that the same band of greater
uncertainties is obtained in the CMB frame map, even though the positions of
the extrema are quite different in that case.

In Fig.~\ref{errsky} we identify the poles corresponding to the
maximum and minimum Hubble variance in the LG frame, and plot a
$1\si$ contour around each pole, where the $1\si$ value is taken as the
maximum on the map, i.e., $1.02h\kmsMpc$ in the unweighted case and $0.75\,h
\kmsMpc$ in the weighted case. In the unweighted case, we find $(\ell,b)=
(116\deg,-35\deg)$ and $(\ell,b)=(249\deg,21\deg)$ for the maximum and minimum
respectively. Similarly, $(\ell,b)=(105\deg,-27\deg)$ and $(\ell,b)=(253\deg,
24\deg)$ for the maximum and minimum in the IV weighted case. The poles
are somewhat squeezed in galactic longitude as compared to a pure
dipole\footnote{The latitude of the unweighted maximum matches that of the
corresponding pure dipole law fit, $(\ld,\bd)
=(68\deg\pm3,-38\deg\pm2\deg)$ from Table~\ref{oshell} within 1$\si$, while
the longitude of the minimum matches that of $(\ld+180\deg,-\bd)$.}.

The mean value of $\Hm$ obtained by Gaussian window averaging differs in
general from the values obtained by spherical averages.
For the $r>15\hm$ LG
frame Gaussian window average, for example, $\Hm$ is $97.3\,h\kmsMpc$ in the
unweighted case and $102.6\,h\kmsMpc$ in the IV weighted case. In the
unweighted case the maximum and minimum values of $H_\al$ are +9.3\% and
-9.4\% from the mean respectively.
With IV weightings the differences are
+11.0\% and -7.9\% respectively\footnote{The maximum variation is comparable to
the ratio $\be/H_d$ in Table~\ref{oshell} below, which is $12.0\pm0.5$\% for
$r>15\hm$.}. In each case these differences are considerably larger than the
standard uncertainty at any angle, which is of order 1\% of the mean $\Hm$.

\section{Dipole estimation}\label{dipest}
In Sec.~\ref{dip} we perform a $\chi^2$ minimization of the simple dipole
Hubble law (\ref{di}).
For each radial shell we minimize $\chi^2=\sum_i\left[
\sHi^{-1}\left(H_i-H_d-\Be\cdot{\mathbf n}_i\right)\right]^2$, where
$\Be\equiv$ ($\be_x$, $\be_y$, $\be_z)=(\be\cos\ld\cos\bd$, $\be\cos\ld\sin\bd
$, $\be\sin b_d)$, ${\mathbf n}_i\equiv(n_{xi}$, $n_{yi}$, $n_{zi})=(\cos\ell_i
\cos b_i$, $\cos\ell_i\sin b_i$, $\sin b_i)$, $H_i=cz_i/r_i$ and its
uncertainty $\sHi$ is given by (\ref{sih}). Minimization with respect to the
four independent parameters $X^a\equiv(H_d,\be_x,\be_y,\be_z)$ yields the
linear system
\def\obox{\hbox to63pt{\hfil}}\def\N{\hskip-15pt}
\bea&\N&H_d\SumH{1}+\be_x\SumH{n_{xi}}+\be_y\SumH{n_{yi}}\nonumber\\
&\N&\obox+\be_z\SumH{n_{zi}}=\SumH{H_i},\\
&\N&H_d\SumH{n_{xi}}+\be_x\SumH{n_{xi}^2}+\be_y\SumH{n_{xi}\,n_{yi}}\nonumber\\
&\N&\obox+\be_z\SumH{n_{xi}\,n_{zi}}=\SumH{H_i\,n_{xi}},\\
&\N&H_d\SumH{n_{yi}}+\be_x\SumH{n_{yi}\,n_{xi}}+\be_y\SumH{n_{yi}^2}\nonumber\\
&\N&\obox+\be_z\SumH{n_{yi}\,n_{zi}}=\SumH{H_i\,n_{yi}},\\
&\N&H_d\SumH{n_{zi}}+\be_x\SumH{n_{zi}\,n_{xi}}+\be_y\SumH{n_{zi}\,n_{yi}}
\nonumber\\&\N&\obox+\be_z\SumH{n_{zi}^2}=\SumH{H_i\,n_{zi}},
\eea
which is readily solved. The covariance matrix for the original variables
$Y^a\equiv(H_d,\be,\ld,\bd)$ is obtained straightforwardly from the
covariance matrix Cov$(X^c,X^d)$ by the standard relation
\bea
\hbox{Cov}(Y^a,Y^b)={\pt Y^a\over\pt X^c}\,\hbox{Cov}(X^c,X^d)\,
{\pt Y^b\over\pt X^d}
\eea
where $\pt Y^a/\pt X^c$ is the relevant Jacobian.

\bigtable{*}{lrrrrrrrrrrr}
Shell $s$&1&2&3&4&5&6&7&8&9&10&11\\
$N_s$&92&505&514&731&819&562&414&304&222&280&91\R
$r_s$ ($h^{-1}$Mpc)&2.00&12.50&25.00&37.50&50.00&62.50&75.00&87.50&100.00&112.50&156.25\R
$\mr_s$ ($h^{-1}$Mpc)&5.43&16.33&30.18&44.48&55.12&69.24&81.06&93.75&105.04&126.27&182.59\R
$(H_{d\,s})\Ns{CMB}$&95.2&87.2&94.7&97.5&96.0&97.9&97.0&99.7&100.3&98.4&95.8\R
$(\si_{H_{d\,s}})\Ns{CMB}$&1.3&0.6&0.7&0.6&0.6&0.7&0.9&0.8&1.0&0.7&1.9\R
$(\be_s)\Ns{CMB}$&95.8&24.4&14.0&2.6&4.9&6.8&3.8&7.4&5.8&6.4&0.9\R
$(\si_{\be\,s})\Ns{CMB}$&1.5&1.1&1.2&1.1&1.1&1.6&1.7&1.6&1.8&1.4&2.9\R
$(\ell_{d\,s})\Ns{CMB}$&308.7&300.1&290.2&309.6&285.5&279.3&264.6&274.7&282.0&309.0&147.2\R
$(\si_{\ell_{d\,s}})\Ns{CMB}$&1.9&3.0&6.0&33.9&13.6&12.7&25.8&14.7&21.4&14.5&389.1\R
$(b_{d\,s})\Ns{CMB}$&-5.9&17.3&-5.8&-42.0&-3.0&2.2&12.9&18.8&11.7&-2.8&48.9\R
$(\si_{b_{d\,s}})\Ns{CMB}$&0.9&1.7&4.5&16.8&10.7&8.0&20.5&9.6&15.6&9.7&195.9\R
$(\chi^2/\nu)\Ns{CMB}$&15.0&3.1&1.6&1.0&0.8&0.7&0.6&0.6&0.7&0.9&0.6\R
$(Q_s)\Ns{CMB}$&0.000&0.000&0.000&0.332&1.000&1.000&1.000&1.000&1.000&0.817&0.999\R
$(H_{d\,s})\Ns{LG}$&79.1&86.6&94.6&97.4&95.9&97.9&97.0&99.7&100.3&98.5&95.9\R
$(\si_{H_{d\,s}})\Ns{LG}$&1.5&0.6&0.7&0.6&0.6&0.7&0.8&0.8&1.0&0.7&1.9\R
$(\be_s)\Ns{LG}$&15.9&20.0&14.2&14.9&8.6&4.9&4.7&1.5&2.0&4.7&3.6\R
$(\si_{\be\,s})\Ns{LG}$&1.7&1.1&1.2&0.8&0.9&1.0&1.4&1.3&1.8&1.4&3.1\R
$(\ell_{d\,s})\Ns{LG}$&324.6&52.7&50.3&87.2&85.3&86.3&107.0&265.4&353.4&348.1&111.3\R
$(\si_{\ell_{d\,s}})\Ns{LG}$&12.7&3.9&8.3&4.9&10.1&32.9&27.9&115.6&217.1&24.2&76.0\R
$(b_{d\,s})\Ns{LG}$&-30.5&-34.8&-55.4&-36.3&-44.2&-58.7&-42.5&-51.0&-76.2&-38.1&-17.8\R
$(\si_{b_{d\,s}})\Ns{LG}$&6.4&2.2&5.5&4.3&8.0&18.5&20.7&59.3&57.3&14.3&36.5\R
$(\chi^2/\nu)\Ns{LG}$&7.8&3.6&1.6&1.1&0.8&0.7&0.6&0.6&0.7&0.9&0.6\R
$(Q_s)\Ns{LG}$&0.000&0.000&0.000&0.158&1.000&1.000&1.000&1.000&1.000&0.882&1.000\\
\noalign{\smallskip}\tableline\noalign{\smallskip}
Shell $s$&1'&2'&3'&4'&5'&6'&7'&8'&9'&10'&11\\
$N_s$&321&513&553&893&681&485&343&273&164&206&91\R
$r_s$ ($h^{-1}$Mpc)&6.25&18.75&31.25&43.75&56.25&68.75&81.25&93.75&106.25&118.75&156.25\R
$\mr_s$ ($h^{-1}$Mpc)&12.26&23.46&37.61&49.11&61.74&73.92&87.15&99.12&111.95&131.49&182.59\R
$(H_{d\,s})\Ns{CMB}$&87.5&90.4&93.8&97.5&96.1&98.2&98.5&100.5&97.2&99.3&95.8\R
$(\si_{H_{d\,s}})\Ns{CMB}$&0.7&0.6&0.7&0.6&0.7&0.7&0.9&0.9&1.0&0.9&1.9\R
$(\be_s)\Ns{CMB}$&39.5&16.2&8.5&3.0&5.8&5.2&5.0&8.9&6.7&5.4&0.9\R
$(\si_{\be\,s})\Ns{CMB}$&1.2&1.1&1.2&1.0&1.3&1.6&1.6&1.7&2.1&1.5&2.9\R
$(\ell_{d\,s})\Ns{CMB}$&294.9&296.1&307.3&273.9&276.2&270.1&289.3&279.7&314.8&308.8&147.2\R
$(\si_{\ell_{d\,s}})\Ns{CMB}$&2.6&4.6&10.1&20.2&12.8&17.8&21.0&12.0&17.0&20.7&389.1\R
$(b_{d\,s})\Ns{CMB}$&17.9&15.2&-13.4&-16.6&14.9&-0.1&26.8&7.0&4.3&2.1&48.9\R
$(\si_{b_{d\,s}})\Ns{CMB}$&1.2&3.5&7.0&15.4&11.4&10.2&16.9&8.4&13.9&14.2&195.9\R
$(\chi^2/\nu)\Ns{CMB}$&3.9&2.1&1.4&0.9&0.8&0.6&0.6&0.7&1.0&0.6&0.6\R
$(Q_s)\Ns{CMB}$&0.000&0.000&0.000&0.997&1.000&1.000&1.000&1.000&0.333&1.000&0.999\R
$(H_{d\,s})\Ns{LG}$&88.2&89.4&94.2&97.4&96.2&98.2&98.4&100.5&97.3&99.4&95.9\R
$(\si_{H_{d\,s}})\Ns{LG}$&0.8&0.7&0.7&0.5&0.7&0.7&0.9&0.9&1.0&0.9&1.9\R
$(\be_s)\Ns{LG}$&20.9&15.9&15.1&11.8&5.4&5.1&2.8&3.7&5.0&3.6&3.6\R
$(\si_{\be\,s})\Ns{LG}$&1.6&1.1&1.1&0.8&1.1&1.0&1.5&1.6&2.2&1.6&3.1\R
$(\ell_{d\,s})\Ns{LG}$&42.9&51.9&64.0&94.5&94.1&103.7&70.4&286.4&0.0&360.0&111.3\R
$(\si_{\ell_{d\,s}})\Ns{LG}$&4.5&4.8&6.0&6.3&17.2&33.9&43.3&35.4&27.1&33.9&76.0\R
$(b_{d\,s})\Ns{LG}$&-35.5&-31.8&-43.4&-38.0&-38.8&-57.8&-33.8&-37.0&-29.5&-36.7&-17.8\R
$(\si_{b_{d\,s}})\Ns{LG}$&2.7&3.7&5.1&5.1&14.9&17.1&31.4&23.1&16.0&24.4&36.5\R
$(\chi^2/\nu)\Ns{LG}$&4.3&2.7&1.4&0.9&0.8&0.6&0.6&0.7&1.0&0.6&0.6\R
$(Q_s)\Ns{LG}$&0.000&0.000&0.000&0.988&1.000&1.000&1.000&1.000&0.404&1.000&1.000
\enddata
\caption{Least squares fit of dipole Hubble law (\ref{di}) in radial
shells in CMB and LG frames, for the same two choices of shells given in
Table~\ref{shell}. We tabulate $N_s$, $r_s$, $\mr_s$; the best fit dipole
Hubble constant, $H_{d\,s}$ (units $h\kmsMpc$), dipole slope $\be_s$ (units
$h\kmsMpc$), the galactic longitude $\ell_{d\,s}$ and latitude $b_{d\,s}$ of
the dipole apex, and their respective standard deviations $\si_{H_{d\,s}}$,
$\si_{\be\,s}$, $\si_{b_{d\,s}}$ and $\si_{\ell_{d\,s}}$. We also tabulate the
reduced $\chi^2$ (for $\nu=N_s-4$) and goodness of fit
probability $Q_s$ in each case.}
\label{dshell}\end{table*}\endgroup

\def\R{\\\noalign{\vskip-1pt}}\bigtable{*}{lrrrrrrrrrrr}
\multispan{12}\hfil CMB frame\hfil\\ \noalign{\smallskip}
$r_o$\hfil&$N$\hfil&$H_d$\hfil&$\si_{H_d}$\hfil&$\be$\hfil&$\si_\be$\hfil&
$\ld$\hfil&$\si_{\ell_d}$\hfil&$\bd$\hfil&$\si_{b_d}$\hfil&$\chi^2/\nu$\hfil&
$Q$\hfil\\ \noalign{\smallskip}
15&4358&96.0&0.2&6.1&0.4&300.2&4.3&6.1&3.1&1.18&0.000\R
20&4158&97.1&0.2&5.8&0.4&289.6&4.7&1.0&3.5&1.04&0.032\R
25&3937&97.6&0.2&5.5&0.5&290.4&5.2&0.7&3.8&0.92&1.000\R
30&3742&97.6&0.2&4.8&0.5&291.7&6.0&2.9&4.4&0.87&1.000\R
35&3538&97.8&0.3&4.4&0.5&292.0&6.7&3.3&4.9&0.83&1.000\R
40&3308&97.8&0.3&4.1&0.5&288.7&7.4&4.8&5.5&0.78&1.000\R
45&3055&98.1&0.3&4.5&0.5&283.4&7.1&7.8&5.3&0.76&1.000\R
50&2692&97.9&0.3&5.4&0.6&289.0&6.5&8.3&4.6&0.73&1.000\R
55&2328&98.1&0.3&5.6&0.6&286.2&6.6&9.2&4.7&0.71&1.000\R
60&2008&98.3&0.3&5.3&0.7&291.5&7.4&7.3&5.2&0.71&1.000\R
65&1741&98.5&0.3&5.5&0.7&289.0&7.8&7.8&5.3&0.69&1.000\R
70&1520&98.6&0.4&5.4&0.7&289.1&8.1&6.8&5.5&0.68&1.000\R
75&1311&98.6&0.4&5.3&0.8&289.8&9.0&11.3&6.5&0.69&1.000\R
80&1124&98.7&0.4&5.6&0.8&291.8&8.9&9.2&6.4&0.72&1.000\R
85&965&99.0&0.4&5.8&0.8&291.1&9.2&7.8&6.5&0.72&1.000\R
90&833&99.1&0.5&5.7&0.9&292.9&9.9&8.1&7.1&0.74&1.000\R
95&704&99.1&0.5&6.0&0.9&297.8&10.1&3.3&7.0&0.75&1.000\R
100&593&98.9&0.5&5.5&1.0&303.3&12.0&4.6&8.4&0.79&1.000\R
105&480&98.2&0.6&5.7&1.1&308.8&13.0&6.6&9.3&0.79&1.000\R
110&401&98.3&0.6&5.1&1.2&311.0&16.3&1.6&10.7&0.80&0.998\R
115&343&98.3&0.7&5.1&1.3&312.4&18.0&3.5&11.9&0.86&0.975\R
120&287&98.5&0.8&4.6&1.4&319.1&22.2&7.0&14.5&0.62&1.000\\
\noalign{\smallskip}
\tableline\noalign{\smallskip}\multispan{12}\hfil LG frame\hfil\\
\noalign{\smallskip}
$r_o$\hfil&$N$\hfil&$H_d$\hfil&$\si_{H_d}$\hfil&$\be$\hfil&$\si_\be$\hfil&
$\ld$\hfil&$\si_{\ell_d}$\hfil&$\bd$\hfil&$\si_{b_d}$\hfil&$\chi^2/\nu$\hfil&
$Q$\hfil\\ \noalign{\smallskip}
15&4358&94.7&0.2&11.4&0.4&68.3&2.7&-38.0&2.0&1.39&0.000\R
20&4158&96.1&0.2&9.4&0.4&78.0&3.4&-36.2&2.6&1.18&0.000\R
25&3937&97.1&0.2&8.0&0.4&80.8&4.4&-40.2&3.2&0.97&0.906\R
30&3742&97.2&0.2&7.7&0.4&83.1&4.6&-38.2&3.4&0.93&0.999\R
35&3538&97.5&0.3&7.4&0.4&82.8&4.9&-38.1&3.7&0.89&1.000\R
40&3308&97.7&0.3&6.9&0.4&85.6&5.4&-38.6&4.0&0.84&1.000\R
45&3055&98.1&0.3&5.6&0.4&89.4&7.5&-42.7&5.3&0.79&1.000\R
50&2692&97.8&0.3&4.3&0.5&68.4&12.6&-52.2&7.6&0.75&1.000\R
55&2328&98.0&0.3&3.5&0.5&55.1&21.4&-60.7&10.2&0.71&1.000\R
60&2008&98.2&0.3&3.6&0.5&45.9&20.1&-57.8&10.1&0.72&1.000\R
65&1741&98.4&0.3&3.3&0.6&31.3&26.5&-61.9&11.7&0.69&1.000\R
70&1520&98.5&0.4&3.2&0.6&20.6&29.4&-62.9&12.6&0.68&1.000\R
75&1311&98.6&0.4&2.6&0.7&17.0&31.2&-56.6&15.7&0.68&1.000\R
80&1124&98.7&0.4&2.8&0.8&0.2&28.3&-53.3&15.1&0.71&1.000\R
85&965&99.0&0.4&2.7&0.8&345.6&29.4&-52.5&15.8&0.71&1.000\R
90&833&99.1&0.5&2.7&0.9&347.1&28.1&-48.2&16.2&0.73&1.000\R
95&704&99.0&0.5&3.3&1.0&343.9&23.3&-44.8&13.5&0.74&1.000\R
100&593&98.8&0.5&3.4&1.1&355.9&23.8&-41.2&14.6&0.78&1.000\R
105&480&98.2&0.6&3.6&1.2&357.9&21.8&-32.9&14.7&0.77&1.000\R
110&401&98.3&0.6&3.8&1.2&3.7&25.4&-40.1&17.2&0.79&0.999\R
115&343&98.4&0.7&3.7&1.4&4.3&25.6&-36.1&19.2&0.83&0.989\R
120&287&98.6&0.8&3.5&1.6&18.1&25.5&-29.9&22.3&0.61&1.000
\enddata
\caption{Least squares fit of dipole Hubble law (\ref{di}) using
data outside a radius $r>r_o$, as $r_o$ is varied, in both CMB and LG frames.
We tabulate $r_o$, $N(r>r_o)$, the best fit dipole Hubble constant, $H_d$
(units $h\kmsMpc$), dipole slope $\be$ (units $h\kmsMpc$), the galactic
longitude $\ld$ and latitude $\bd$ of the dipole apex, and their respective
standard deviations $\si_{H_d}$, $\si_\be$, $\si_{b_d}$ and $\si_{\ell_d}$.
We also tabulate the reduced $\chi^2$ (for $\nu=N-4$) and goodness of fit
probability $Q$ in each case.}
\label{oshell}\end{table*}\endgroup

More robust statistical results might be obtained if in place
of (\ref{di}) we were to fit an alternative dipole law
\beq\frac r{cz}=\frac1{H_d}-\frac{\be'}{H_d^2}\cos\ph'\,,\label{di2}\eeq
since the uncertainties (\ref{sial}) in $H_i^{-1}=r_i/(cz_i)$ are
more directly related to measurement uncertainties than the uncertainties
(\ref{sih}) in $H_i$. However, while the two laws will agree if $|\be|\ll H_d$,
in general the relationship between (\ref{di}) and (\ref{di2}) is not
linear, so that $\be\ne\be'$ and $\ph\ne\ph'$ when $\be/H_d$ can be of order
10\%, as is typical for the data here. Since we have plotted the variation
of $H_\al$ in Figs.~\ref{wholesky}--\ref{lgcutsky}, rather than $H_\al^{-1}$,
the only way we can expect to obtain angular agreement of the dipoles is
to use (\ref{di}). Furthermore, fitting (\ref{di}) should give results
for the dipole in angular agreement with bulk flows found in the
standard peculiar velocity framework.

We have performed the analysis in the
CMB and LG frames in two ways: (i) in each of the radial shells defined in
Table~\ref{shell}; and (ii) with data aggregated into inner and outer shells
split at a radius $r_o$, as it is varied. In this way we have counterparts
for both the analysis of section~\ref{rad} and of section~\ref{avar}.
The results, by radial shell, are given in Table~\ref{dshell}. In
Table~\ref{oshell} we show the equivalent results for all data outside a
cutoff $r>r_o$ as $r_o$ is varied.

As in the case of the spherical averages the goodness of fit of the
linear relation is poor in the first few radial shells, whose radius is smaller
than the typical largest voids. We have checked that using the reduced
datasets of Appendix~\ref{nonl} leads to a goodness of fit close to $1.0$
without substantially changing any of the conclusions.

We note that the values of $H_d$ are smaller than those found
in Sec.~\ref{rad}. This is a direct consequence of fitting the law
(\ref{di}), rather than the alternative (\ref{di2}), and agrees with
our observation in the case of the Gaussian window averages that when IV
weightings are used a fit to (\ref{Htp})--(\ref{sih}) rather than
(\ref{Hal})--(\ref{sial}) gives lower mean values of $H_\al$. We have checked
that if (\ref{di2}) is used in place of (\ref{di}) then the values of
$H_d$ agree within $1\si$ with the values of $H_s$ in Table~\ref{shell} for
those shells with $Q_s>0.2$ in both fits. Thus it is the relative value
$\be/H_d$ which is of most interest, rather than the absolute value of $\be$,
in Tables~\ref{dshell} and \ref{oshell}.

\section{Statistical significance of results: Monte Carlo simulations}
\label{mc}
To check that there are no systematic effects introduced by incomplete
sky coverage in outer shells or the nonlinear nature of the Hubble law in
inner shells, within each shell we have performed $12\times10^6$ random
shuffles of the angular positions of the data points with respect to their
distances and redshifts, and have recomputed the dipoles. Such a procedure
will not significantly change the value of the spherically averaged Hubble
constant in each shell, but would lead to a nonzero value of the dipole
slope $\be$ if the dipole in any shell originated from insufficient sky cover,
for example.

The results of the Monte Carlo analysis are presented in Table~\ref{ranshell}.
We find in particular that in each of $10^6$ shuffles per shell the weighted
mean Cartesian projections of the dipole slope, $\be_x$, $\be_y$ and $\be_z$
are zero to within $0.01\,\si$ for all shells other than shell 1 which is not
used in drawing statistical conclusions. For the outer 7 shells (primed
or unprimed) we find $\be_s/\si_{\be\,s}<0.005$ for the random reshuffles,
consistent with no dipole to high accuracy. In particular, although there are
fewer data points in the outermost three shells we can be confident that any
conclusions drawn from the dipole analysis are not due to unaccounted
systematics.

\def\R{\\\noalign{\vskip-.2pt}}\bigtable{*}{lrrrrrrrrrrr}
Shell $s$&1&2&3&4&5&6&7&8&9&10&11\\
$(\be_{x\,s})\Ns{CMB}$&-0.561&0.002&0.004&0.002&-0.002&0.000&-0.001&-0.006&-0.006&0.001&0.004\R
$(\si_{\be_x\,s})\Ns{CMB}$&55.8&4.3&3.1&1.5&1.2&1.3&1.6&1.9&2.3&2.1&3.5\R
$(\be_{y\,s})\Ns{CMB}$&2.533&-0.002&-0.011&0.000&0.001&0.003&0.002&0.007&0.002&-0.004&-0.017\R
$(\si_{\be_y\,s})\Ns{CMB}$&46.4&4.1&2.8&1.5&1.3&1.3&1.6&1.7&2.1&2.1&2.7\R
$(\be_{z\,s})\Ns{CMB}$&-6.212&0.000&0.006&0.000&0.001&0.002&0.000&-0.001&-0.004&-0.001&0.002\R
$(\si_{\be_z\,s})\Ns{CMB}$&36.3&3.2&2.4&1.3&1.0&0.9&1.3&1.5&1.7&1.5&2.2\R
$(\be_{\,s})\Ns{CMB}$&6.732&0.003&0.013&0.002&0.002&0.003&0.002&0.010&0.008&0.004&0.018\R
$(\si_{\be\,s})\Ns{CMB}$&55.6&6.3&4.4&1.7&1.8&1.6&2.2&2.7&3.2&2.5&3.6\R
$(P\ns{ran})\Ns{CMB}$&0.000&0.000&0.000&0.435&0.001&0.000&0.093&0.000&0.034&0.003&0.983\R
$(\be_{x\,s})\Ns{LG }$&-0.012&-0.028&0.000&0.018&-0.001&-0.003&-0.001&-0.003&-0.003&0.001&0.001\R
$(\si_{\be_x\,s})\Ns{LG }$&13.7&4.2&2.7&3.1&1.5&1.3&1.7&1.7&1.9&2.0&3.7\R
$(\be_{y\,s})\Ns{LG }$&0.322&0.053&-0.001&0.009&0.000&-0.003&0.002&0.003&0.001&-0.004&-0.008\R
$(\si_{\be_y\,s})\Ns{LG }$&11.4&4.1&2.4&3.2&1.5&1.4&1.7&1.6&1.8&2.0&2.9\R
$(\be_{z\,s})\Ns{LG }$&-0.478&-0.019&-0.001&0.002&0.000&-0.007&-0.001&-0.001&-0.002&-0.001&0.000\R
$(\si_{\be_z\,s})\Ns{LG }$&8.4&3.2&2.0&2.7&1.2&1.0&1.3&1.3&1.4&1.5&2.4\R
$(\be_{\,s})\Ns{LG }$&0.576&0.064&0.001&0.020&0.001&0.008&0.002&0.004&0.003&0.004&0.008\R
$(\si_{\be\,s})\Ns{LG }$&13.6&6.3&3.3&4.5&1.9&1.8&2.6&2.6&2.8&2.5&3.2\R
$(P\ns{ran})\Ns{LG}$&0.000&0.000&0.000&0.000&0.000&0.000&0.002&0.718&0.624&0.040&0.691\\
\noalign{\smallskip}\tableline\noalign{\smallskip}
Shell $s$&1'&2'&3'&4'&5'&6'&7'&8'&9'&10'&11\\
$(\be_{x\,s})\Ns{CMB}$&-0.003&-0.006&0.001&0.001&0.002&-0.001&-0.004&-0.003&-0.005&0.002&0.002\R
$(\si_{\be_x\,s})\Ns{CMB}$&8.0&3.3&2.5&1.1&1.5&1.4&1.6&2.1&3.2&2.0&3.5\R
$(\be_{y\,s})\Ns{CMB}$&0.012&0.016&0.002&0.000&0.001&0.002&0.000&0.010&0.001&-0.006&-0.015\R
$(\si_{\be_y\,s})\Ns{CMB}$&7.2&3.1&2.4&1.2&1.6&1.3&1.7&1.9&2.9&2.1&2.7\R
$(\be_{z\,s})\Ns{CMB}$&0.000&-0.008&-0.004&0.001&0.000&-0.001&-0.003&-0.003&-0.004&0.001&0.000\R
$(\si_{\be_z\,s})\Ns{CMB}$&4.9&2.7&1.9&1.0&1.2&1.0&1.3&1.6&2.3&1.5&2.2\R
$(\be_{\,s})\Ns{CMB}$&0.012&0.019&0.004&0.001&0.002&0.002&0.005&0.011&0.006&0.006&0.015\R
$(\si_{\be\,s})\Ns{CMB}$&8.9&4.8&3.1&1.5&2.2&2.0&2.2&2.7&4.5&2.8&3.2\R
$(P\ns{ran})\Ns{CMB}$&0.000&0.000&0.000&0.112&0.000&0.003&0.008&0.000&0.115&0.008&0.983\R
$(\be_{x\,s})\Ns{LG }$&-0.019&-0.016&0.003&0.002&-0.002&0.001&-0.002&-0.002&-0.006&0.001&0.000\R
$(\si_{\be_x\,s})\Ns{LG }$&5.2&4.4&3.4&1.9&1.5&1.5&1.7&1.8&3.0&1.8&3.7\R
$(\be_{y\,s})\Ns{LG }$&0.052&0.040&0.026&0.001&-0.001&-0.005&-0.001&0.008&-0.001&-0.004&-0.005\R
$(\si_{\be_y\,s})\Ns{LG }$&4.7&4.2&3.3&2.0&1.6&1.5&1.7&1.6&2.7&1.9&2.9\R
$(\be_{z\,s})\Ns{LG }$&-0.041&-0.015&-0.016&0.005&-0.002&0.000&-0.005&-0.002&-0.006&0.001&-0.001\R
$(\si_{\be_z\,s})\Ns{LG }$&3.2&3.6&2.7&1.7&1.2&1.1&1.3&1.4&2.1&1.4&2.4\R
$(\be_{\,s})\Ns{LG }$&0.069&0.046&0.030&0.006&0.003&0.006&0.005&0.008&0.009&0.004&0.005\R
$(\si_{\be\,s})\Ns{LG }$&6.8&6.5&4.5&2.7&2.3&1.8&2.1&2.3&3.8&2.5&3.3\R
$(P\ns{ran})\Ns{LG}$&0.000&0.000&0.000&0.000&0.000&0.000&0.255&0.080&0.422&0.146&0.691
\enddata
\caption{Monte Carlo analysis of randomized angular variations by radial
shell in CMB and LG frames. The analysis of Table~\ref{dshell} is reperformed
with random shuffling of angular coordinates of the data points in radial
shells in each rest frame.
For each shell, and each rest frame, we tabulate the weighted mean Cartesian
projections $\be_x$, $\be_y$, $\be_z$ (defined in Appendix~\ref{dipest}), the
combined dipole magnitude, $\be$, and their weighted standard deviations. We
also tabulate the probability, $P\ns{ran}$, that a random shuffle in each shell
produced a better fit to the linear dipole law (\ref{di}) than the actual
data. The results are based on $10^6$ shuffles in each shell, and
12 independent runs.}
\label{ranshell}\end{table*}\endgroup

The standard deviations $\si_{\be\,s}$ obtained from the Monte Carlo analysis
in Table~\ref{ranshell} are generally somewhat larger than the values in
in Table~\ref{dshell}, and provide a better estimate of uncertainty which
accounts for systematic and nonlinear effects, particularly in the innermost
shells.

In order to assess statistical confidence we have computed the $\chi^2$
statistic for each random shuffle, and have determined the probability,
$P\ns{ran}$, that a random reshuffle of the angular positions of the
data points yields a better linear dipole fit than that of the actual data
determined in Table~\ref{dshell}. The statistic $1-P\ns{ran}$ then determines
the statistical confidence that we have for a non-zero dipole in each shell.
For example, in the outermost control shell 98.3\% of random angular
reshuffles in the CMB frame and 69.1\% of reshuffles in the LG frame give
a better fit than the data, meaning that there is no evidence for a dipole in
that shell in either frame, consistent with the angular uncertainties already
noted. Since the same conclusion applies to both frames,
there are insufficient data points to establish the existence of a
dipole, and much more data beyond $150\hm$ is required to determine whether
a dipole is genuinely absent.

The values of $P\ns{ran}$ lead to a robust interpretation of
Fig.~\ref{sdipfit}. Tables~\ref{dshell} and \ref{ranshell} show that in the
unprimed shells\footnote{$\vbox to15pt{\vfill}$The conclusions drawn from the primed shells are
similar, with the proviso that in shell 9' the CMB frame dipole is only found
at the 88.5\% confidence level. Shell 9' has only 164 points, and the goodness
of fit for a dipole Hubble law is only 0.132 and 0.404 in the CMB and LG
frames, as compared to 0.992 and 0.996 respectively for the monopole Hubble
law. Evidently the number of points in the outer shells needs to be of the
order of at least 200 in order to make statements about the dipoles at the
90\% confidence level in at least one frame. Fortunately this is true of the
unprimed shells.} there is a significant dipole at the 90\% confidence level in
all shells up to $s=10$ apart from shell 4, with $\mr_4=44.5\hm$, where as
has been noted the CMB frame also provides a better fit to a monopole Hubble
law. In shell 4, 43.5\% of random angular shuffles in the CMB frame produce
a better dipole fit, meaning that the dipole is found with only 56.5\%
confidence. By contrast, in the LG frame the dipole is found at a confidence
level of more than 99.999\% in shell 4. The difference is therefore a frame
effect.

In the unprimed shells in the LG frame there is a significant dipole at
the 90\% confidence level in all shells $s\le7$ and also in shell $10$. In
shells 8 and 9 the probability of a random angular shuffle giving a better fit
than the data is greater than 50\%, consistent with the values of $\be$ close
to zero shown in Fig.~\ref{sdipfit}. By contrast a dipole is found in the
CMB frame in shells 8 and 9 respectively at the levels of 99.98\% and 96.6\%
confidence. Thus we can be confident that the absence of the dipole in the LG
frame in shells 8 and 9 is a genuine frame effect, rather than being due to
small number statistics.

In shell 10, whose inner boundary $r_{10}=112.5\hm$ is close to the BAO scale,
there is a dipole at the level of 96\% confidence in the LG frame and 99.7\%
in the CMB frame. Since shells 8 and 9 lack a significant dipole in the LG
frame, it would appear that it is the feature in this shell that is driving the
appearance of the residual LG frame dipole for the largest radii shown in the
smoothed value of $\be$ shown in Fig.~\ref{dipfit}. Furthermore, since there
is only a LG frame dipole at the 85.4\% confidence level in shell 10', whose
inner boundary is at $r_s=118.75\hm$, it would appear that the data in the
closer portions of shell 10 contains the relevant structures. A great deal more
data should be added to the analysis before we draw any firm conclusions about
features in this shell, however.

\end{document}